\documentclass[a4paper,fleqn,usenatbib]{mnras}

\usepackage[T1]{fontenc}

\DeclareRobustCommand{\VAN}[3]{#2}
\let\VANthebibliography\thebibliography
\def\thebibliography{\DeclareRobustCommand{\VAN}[3]{##3}\VANthebibliography}

\def\psj{\ref@jnl{Planet.~Sci. J.}}  

\usepackage{graphicx}	
\usepackage{amsmath}	
\usepackage{amssymb}	
\usepackage{gensymb}	
\usepackage[compatibility=false]{caption}
\usepackage{subcaption}

\newcommand{\gcm}{~$\mathrm{g~cm}^{-3}$}

\newcommand{\kgs}{kg~s$^{-1}$}
\newcommand{\QD}{$Q_\mathrm{D}^\star$}
\newcommand{\ar}{$\alpha_\mathrm{r}$}
\newcommand{\kgy}{kg~yr$^{-1}$}
\graphicspath{{figures/}{figures/method/}{figures/toy/}{figures/frag_results/}{figures/dust_results/}{figures/discussion/}}
\usepackage{newtxtext,newtxmath}

\title[Comet fragmentation]{Comet fragmentation as a source of the zodiacal cloud}

\author[J. K. Rigley et al.]{
Jessica K. Rigley,$^{1}$\thanks{E-mail: jkr40@ast.cam.ac.uk}
Mark C. Wyatt,$^{1}$
\\
$^{1}$Institute of Astronomy, University of Cambridge, Madingley Road, Cambridge, CB3 0HA, UK\\
}

\date{Accepted XXX. Received YYY; in original form ZZZ}

\pubyear{2021}

\begin{document}
\label{firstpage}
\pagerange{\pageref{firstpage}--\pageref{lastpage}}
\maketitle

\begin{abstract}
Models of the zodiacal cloud’s thermal emission and sporadic meteoroids suggest Jupiter-family comets (JFCs) as the dominant source of interplanetary dust. However, comet sublimation is insufficient to sustain the quantity of dust presently in the inner solar system, suggesting that spontaneous disruptions of JFCs may supply the zodiacal cloud. We present a model for the dust produced in comet fragmentations and its evolution. Using results from dynamical simulations, the model follows individual comets drawn from a size distribution as they evolve and undergo recurrent splitting events. The resulting dust is followed with a kinetic model which accounts for the effects of collisional evolution, Poynting-Robertson drag, and radiation pressure. This allows to model the evolution of both the size distribution and radial profile of dust, and we demonstrate the importance of including collisions (both as a source and sink of dust) in zodiacal cloud models. With physically-motivated free parameters this model provides a good fit to zodiacal cloud observables, supporting comet fragmentation as the plausibly dominant dust source. The model implies that dust in the present zodiacal cloud likely originated primarily from disruptions of $\sim 50$ km comets, since larger comets are ejected before losing all their mass. Thus much of the dust seen today was likely deposited as larger grains $\sim 0.1$ Myr in the past. The model also finds the dust level to vary stochastically; e.g., every $\sim 50$ Myr large ($> 100$ km) comets with long dynamical lifetimes inside Jupiter cause dust spikes with order of magnitude increases in zodiacal light brightness lasting $\sim 1$ Myr. If exozodiacal dust is cometary in origin, our model suggests it should be similarly variable.
\end{abstract}



\section{Introduction}

The zodiacal cloud consists of a diffuse, interplanetary dust complex which permeates throughout the inner solar system. Thermal emission and scattered light from this dust is seen as the zodiacal light. The first all-sky map of the zodiacal emission was produced by IRAS \citep{Hauser84,Sykes88}, which observed the thermal emission in four infrared bands. This lead to the discovery of structure in the zodiacal cloud. Dust bands \citep{Dermott84}, populations of dust that are prominent at the same ecliptic latitude (inferred to be from dust with the same proper inclination), have been linked to collisions of specific asteroid families \citep[see, e.g.][]{Nesvorny03,Nesvorny06,Nesvorny08,EspyKehoe15}. IRAS also discovered narrow trails of dust \citep{Sykes86,Sykes92}, associated with the orbits of comets, believed to be debris ejected from the comet. This implies that both asteroids and comets must contribute to the interplanetary dust complex. \par 
Many studies have tried to constrain the relative contributions of different sources to the interplanetary dust cloud. Comets are believed to be the dominant source, with asteroids contributing at most 30 per cent. Although interstellar dust should be present, it is believed to be a very small contribution, and dominate only at the smallest, submicron, grain sizes \citep[e.g.][]{Landgraf00,Kruger10}. While some dust will migrate in from the Kuiper belt, its contribution in the inner solar system is likely extremely low due to the fact that not much dust will migrate past Jupiter \citep[e.g.][]{Moro-Martin03}. Most of the evidence favouring comets as the dominant source comes from the distribution of 25~$\mu$m emission seen as a function of ecliptic latitude by IRAS, which describes the vertical distribution of the zodiacal cloud. Typically comets have higher inclinations than asteroids, and will therefore produce broader latitudinal profiles. \citet{Liou95} required a combination of 1/4 to 1/3 asteroidal dust and 3/4 to 2/3 cometary dust to reproduce the latitudinal profile. On the other hand, \citet{Durda97} modelled the collisional evolution of the main asteroid belt. By using the ratio of emission from the asteroid families to the main belt and the fraction of thermal emission which comes from the dust bands, they concluded that asteroidal dust is responsible for at least 1/3 of the zodiacal cloud. More recent models find comets contribute a much higher fraction of interplanetary dust. For example, the modelling of \citet{Nesvorny10} found a contribution of $> 90$ per cent from comets was required to fit the vertical distribution of thermal emission seen by IRAS, with asteroids contributing less than 10 percent. \citet{Rowan-Robinson13} simultaneously modelled the infrared emission from IRAS and COBE empirically, and required contributions of 70, 22, and 7.5 per cent respectively from comets, asteroids, and interstellar dust. Another constraint comes from the Earth's resonant ring, in which dust particles are trapped in mean motion resonances with Earth. This exhibits a leading-trailing brightness asymmetry, in which the zodiacal cloud is always brighter behind the Earth than ahead of it on its orbit \citep{Dermott94}.  Using infrared observations from AKARI, \citet{Ueda17} concluded that cometary dust must dominate to fit the leading-trailing brightness asymmetry, with asteroidal dust contributing less than 10 percent of the infrared emission. Further, comparison of the optical properties of the zodiacal light with those of different minor bodies suggests more than 90 per cent originates from either comets or D-type asteroids \citep{Yang15}. \par
Sporadic meteoroids provide constraints on the types of comets supplying interplanetary dust. These are meteoroids which are not associated with any meteoroid streams. Comets can be broadly categorised based on their orbits. Short-period comets (SPCs) have periods $<200$~yr, while long-period comets (LPCs) have longer periods, and originate in the Oort cloud. SPCs can be further divided into Jupiter-family comets (JFCs), which are dominated by their interactions with Jupiter, and have Tisserand parameters with respect to Jupiter $2 \lesssim T_\mathrm{J} \lesssim 3$, and Halley-type comets (HTCs), which have Tisserand parameters $T_\mathrm{J} < 2$. The impact velocities and orbital elements of the sporadic meteoroid complex, suggest SPCs must be the dominant source of sporadic meteoroids \citep{Wiegert09}. Different types of comets are linked to meteoroids in different parts of the sky \citep[e.g.][]{Pokorny14}, though JFCs dominate the helion and anti-helion sources, which contain most of the mass flux \citep{Nesvorny11_ZC}. \par
While the structure of the zodiacal cloud is best explained by a cometary source, comet sublimation is insufficient to sustain the quantity of dust presently in the inner solar system \citep{Nesvorny10}. However, cometary sublimation is not the only source of mass loss from comets. 
Many comets  have been observed to spontaneously disrupt, with more splittings observed at lower perihelion distances \citep[see, e.g.][]{Fernandez05}. While tidal splitting due to close encounters with a planet or the Sun is one possible cause of splitting, this accounts for very few of the observed cases. Other possible mechanisms are rotational spin-up due to asymmetric outgassing, thermal stress due to variable distances from the Sun, or internal gas pressure build-up due to sublimation of sub-surface volatiles \citep[for a review see][]{Boehnhardt04}. \citet{DiSisto09} showed that dynamical simulations of bodies originating in the trans-Neptunian region could not fully reproduce the observed orbital distribution of JFCs, requiring a mechanism which could limit the physical lifetime of comets. They therefore developed a model for the frequency and mass loss of comet fragmentation, fitted to the observed distribution of JFCs. Similarly, \citet{Nesvorny17} modelled the origin of ecliptic comets (ECs), and showed that to match the observed distribution they needed to limit the number of perihelion passages comets were active for, with larger bodies requiring more passages than smaller bodies. Observations also suggest comets should fragment frequently, with \citet{Chen94} finding a lower limit for the rate a given comet fragments of 0.01 /yr. Given the likely high frequency of fragmentation events and their ability to cause much greater mass loss than cometary activity, comet fragmentation may dominate the input to the interplanetary dust complex. \par
Numerical models of the dust in the zodiacal cloud can be classified into two types: empirical and dynamical. Empirical models describe the 3D structure of the zodiacal cloud along with the size distributions of the dust. The parameters describing these distributions may have a basis in the underlying physics, but are ultimately fitted to be able to reproduce certain observations. Some empirical models of the zodiacal cloud are \citet{Grun85,Divine93,Kelsall98,Rowan-Robinson13}. \par
Dynamical models of the zodiacal cloud use N-body integrators to follow the orbital evolution of individual dust particles from their source to their ultimate loss \citep[e.g.][]{Liou95,Wiegert09,Nesvorny10,Nesvorny11_ZC,Pokorny14,Ueda17,Soja19,Moorhead20}. Some of these consider dust from comets, while others compare dust of different cometary types with asteroidal dust. In all cases, the initial orbits of particles are determined by those of their parent bodies. A dynamical approach is useful for following dynamical interactions with planets, and allows inclusion of the effects of radiation pressure, solar wind and Poynting-Robertson (P-R) drag. However, since individual particles are followed, only a simplified collisional prescription can be used. Either particles are removed after their collisional lifetime has elapsed, or a stochastic prescription based on collisional lifetimes determines when to remove particles. The production of smaller grains in collisions is neglected, such that dynamical models are limited in their ability to model the size or radial distribution of dust consistently. The production of collisional fragments (and subsequent disruption of these fragments) supplies smaller grain sizes, and is important in order to follow the size distribution of particles. When modelling meteoroids, only including collisions as a loss mechanism may be a valid approximation, as for larger ($\gtrsim 1$~mm) dust particles this will be the net effect of collisions. However, when considering smaller particles which contribute to the zodiacal light and thermal emission, it is important to include the supply of smaller particles from disruption of larger grains. \par
We propose to use a different approach in using a kinetic model, which follows the evolution of a population of particles in a phase space of mass and orbital elements. Such models have found much use in the study of extrasolar debris discs \citep[e.g.][]{Krivov05,Krivov06,vLieshout14}, but we are only aware of one use in the context of the zodiacal cloud \citep{Napier01}. Kinetic models incorporate the effects of radiation pressure, P-R drag, and solar wind, along with collisional evolution. While such models allow the size distribution to be modelled self-consistently, using a statistical approach requires a simplified prescription of the effect of dynamical interactions with planets. \citet{Napier01} modelled dust produced by comets on Encke-like orbits and followed the distribution of dust with semimajor axis, eccentricity, and particle mass. We improve on this model by using a more realistic size distribution of comets, N-body simulations of cometary dynamics and a more physical prescription for mass loss from comets by spontaneous fragmentation. Further, in \citet{Napier01} collisions were only included as removal mechanism; here we include the full effects of collisional evolution, including the production of smaller grains, when modelling interplanetary dust. This allows us to produce a self consistent model for the size distribution, whereas dynamical models must either approximate that a single size dominates, or make assumptions about that distribution (e.g. by using a power law with parameters that are fit to observations). \par
A further limitation of some models is that given our much better knowledge of interplanetary dust near Earth, most models focus on the dust at 1~au. The radial distribution is typically only included empirically \citep[e.g.][]{Rowan-Robinson13}, although ESA's IMEM2 \citep{Soja19} and NASA's MEM 3 \citep{Moorhead20} are dynamical models which consider the radial distribution. Our use of a kinetic model allows us to study the radial distribution of dust, taking into account the effect of collisions on the distribution. \par
Using a kinetic model not only allows us to model the size and spatial distribution of the dust self-consistently, but also addresses issues such as the stochasticity of dust production in the zodiacal cloud. Asteroidal input should be stochastic due to collisional evolution \citep{Durda97,Dermott01}. As far as we are aware, only \citet{Napier01} has previously studied the stochasticity of a cometary input to the zodiacal cloud. While most comets seen today are smaller than $\sim$15~km, bodies in the Kuiper belt, the source of JFCs, can be as large as hundreds of km, though they are far fewer. Thus, it is possible that occasionally in the history of the solar system, large bodies could be scattered inwards and deposit large amounts of dust in the interplanetary dust complex. For example, it is hypothesised that the Taurid complex, a collection of asteroids and comets with similar orbits \citep{Ferrin21}, originated from a series of fragmentations of a large progenitor comet $\gtrsim 100$~km in size tens of thousands of yr ago \citep[e.g.][]{Clube84,Napier19}. Any cometary contribution to interplanetary dust will be highly variable over long timescales depending on the sizes of comets which are scattered in. Studying the potentially stochastic nature of a cometary source is therefore important for understanding the history of the zodiacal cloud. \par
The final way we aim to improve on previous models is by using a physically-motivated mechanism for the production of dust. We apply a physical prescription for individual comet fragmentations, rather than placing dust on cometary orbits randomly. \citet{Marboeuf16} modelled the thermo-physical evolution of comets in the context of (exo-)zodiacal dust produced by comet sublimation, but as discussed earlier that is not thought to be the dominant mass loss mechanism from comets. \citet{Nesvorny10,Nesvorny11_ZC} model the production of zodiacal dust via comet fragmentation, but simply release dust grains from comets once they reach a critical pericentre, as opposed to modelling individual, recurrent events. We use a more physical model of comet fragmentation which has been fitted to observations of JFCs such that we can model the evolution of individual comets as they fragment repeatedly. The model will also be able to follow the stochasticity of that fragmentation. \par
To summarise, in this paper we develop a model for mass input to the zodiacal cloud from comet fragmentation based on realistic cometary dynamics, with a self-consistent model for the evolution of the dust produced by comets as a result of mutual collisions and P-R drag. Dynamical effects are included with a simplified prescription. We aim to show whether comet fragmentation can produce a viable model of the zodiacal cloud in terms of the spatial and size distribution of dust. We also investigate the variability of any cometary source of the zodiacal cloud due to stochasticity relating to inward scattering of comets and its implications for the zodiacal cloud's history. \par 
Our model of comet fragmentation is given in Section~\ref{sec:frag_model}, and the model of the dust produced by these comets is presented in Section~\ref{sec:dust}. Fitting of the model parameters to observational constraints is discussed in Section~\ref{sec:fitting}. Our results are given in Section~\ref{sec:results} and discussed in Section~\ref{sec:discussion}. Section~\ref{sec:comparison} compares our model to previous zodiacal cloud models. Finally, we give our conclusions in Section~\ref{sec:conc}. \par

\section{Comet model}
\label{sec:frag_model}
To determine the potential contribution of comet fragmentation to the zodiacal cloud, we model the mass input from fragmentation events within a population of comets. That population is created by starting with N-body simulations of the dynamical evolution of solar system comets over 100~Myr. We clone particles from the N-body simulations in time (to simulate the continual injection of comets), with each cloned particle representing a size distribution of comets. Then each particle in the size distribution is followed as it bounces around the inner solar system, randomly undergoing fragmentation events which produce dust and reduce the particle's size.

\subsection{N-body data}
\label{subsec:nbody}
JFCs are comets with short orbital periods and relatively low inclinations. We define JFCs to have periods $P<20$~yr and a Tisserand parameter with respect to Jupiter of $2 < T_J < 3$ as in \citet{Nesvorny17}. They are believed to originate in the scattered trans-Neptunian disc, from which some bodies are randomly scattered inside Neptune's orbit, then into the inner solar system \citep[e.g.][]{Duncan97}. \par
We apply a fragmentation model to JFCs as they evolve with trajectories from the CASE2 simulation of \citet{Nesvorny17}. \citeauthor{Nesvorny17} followed the evolution of objects from the trans-Neptunian region to the inner planetary system over 1~Gyr to model the origin and evolution of JFCs. Interactions with the giant planets are included, but terrestrial planets are not. Their data give the orbital elements of comets with pericentre distances $q < 5.2$~au at 100~yr intervals. The number of particles in the N-body simulations is relatively low (21,548), and spread over 1~Gyr. We therefore assume that each N-body particle is representative of a size distribution of comets, which is described in Section~\ref{subsec:sizedist_com}. Additionally, we assume the time a particle is scattered in is unimportant, and clone each N-body particle in time so that the same particle is introduced every 12,000~yr (see Section~\ref{subsec:fragmethod}). \par

\begin{figure}
	\centering
	\includegraphics[width=\linewidth]{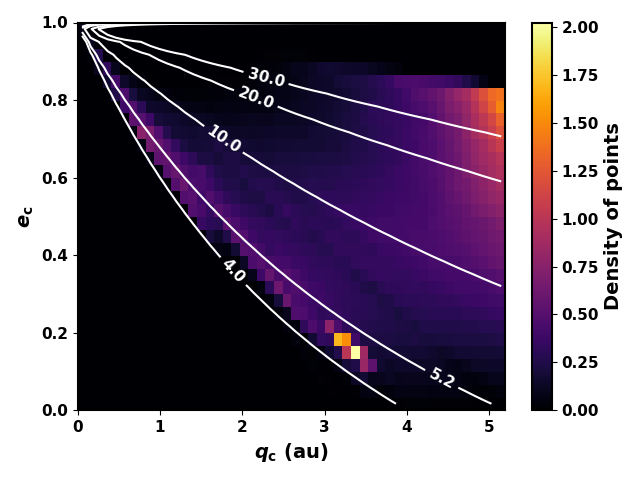}
	\caption{Histogram showing the locations of N-body data points in pericentre-eccentricity space, from the simulations of \citet{Nesvorny17}, averaged over all times. Contours show lines of constant apocentre in au. }
	\label{fig:comet_qe}
\end{figure}
The orbital elements of the N-body data points in pericentre-eccentricity space are shown in Figure~\ref{fig:comet_qe} as the density of comets in each pericentre-eccentricity bin, averaged over the full time span. The orbital elements are only recorded once the bodies reach $q < 5.2$~au, so generally bodies start at 5.2~au and move inwards. The density is therefore highest at pericentres closest to 5.2~au, as some comets may be scattered outside Jupiter again before reaching very low pericentres. Note that comets may fully disrupt before reaching the innermost regions, such that the distribution of mass deposited by fragmentation, and indeed the distribution of comets, may not match that of the parent N-body particles. The peak in the density of points at $q \sim 3.2$~au and $e \sim 0.15$ is due to one particular body which spends a long time ($\sim 40$~Myr) in the inner system without being scattered by Jupiter. This illustrates how individual comets can have a significant effect on the distribution through rare but long-lived dynamical pathways. The simulations do not include the terrestrial planets, and so the only way comets can reach the inner regions is by scattering off the outer planets, which means their apocentre must be close to or beyond the giant planets. Therefore a dearth of JFCs with apocentres $\lesssim 4$~au is seen. \par
Comets will bounce around the phase space (Figure~\ref{fig:comet_qe}) as they evolve, with the amount of dust produced at each location determined randomly depending on the likelihood of fragmentation events. The mass produced also depends on the initial size of the comet: larger comets have more mass to lose, whereas smaller comets are likely to deplete all of their mass before their dynamical lifetime has elapsed. It is therefore important to follow the evolution of individual comets of different sizes and the fragmentations they undergo to determine the mass input into the zodiacal cloud. \par

\subsection{Size distribution}
\label{subsec:sizedist_com}
Each cloned particle is representative of a size distribution of comets which could be scattered in from the Kuiper belt. We consider comets of radii ranging from 0.1 to 1000~km, placing them into 40 logarithmic size bins. Each time an N-body particle is cloned, these size bins are filled by choosing random numbers from a Poisson distribution, with the mean values in each bin given by the size distribution described in this section and Table~\ref{tab:sizedist}. \par
Many attempts have been made to characterise the size distribution of JFCs by converting observed absolute nuclear magnitudes $H_N$ to nuclear radii $R_\mathrm{N}$. Most observations cover the range of radii $1 \lesssim R_\mathrm{N} \lesssim 10$~km. For a cumulative size distribution (CSD), defined as $N_\mathrm{R}(>R_\mathrm{N}) \propto R_\mathrm{N}^{-\gamma}$, a range of slopes have been found, $1.6 \leq \gamma \leq 2.7$  \citep{Weissman03,Lamy04,Tancredi06,Snodgrass11,Fernandez13,Belton14}. Here we choose to use a slope $\gamma = 2.0$ in this size range, which is also in agreement with observations of Jupiter Trojans, thought to have the same source as JFCs. For example, \citet{Yoshida17} found a cumulative slope of $1.84 \pm 0.05$ for Jupiter Trojans in the size range $1 \lesssim R \lesssim 10$~km. \par
For small comets with $R_\mathrm{N} \lesssim 1$~km, the size distribution is seen to turn over to a shallower slope, measured by \citet{Fernandez06} to be $\gamma = 1.25$. It is difficult to pinpoint the exact size this turnover occurs at. It has been shown by both \citet{Meech04_size} and \citet{Samarasinha07} that this is not purely an observational effect due to smaller comets being more difficult to observe, but a result either of the inherent parent distribution or the evolution of comets as they are scattered inwards from the Kuiper belt - perhaps smaller comets are more susceptible to erosion by physical effects such as sublimation and fragmentation. In their recent analysis of cratering on Charon and Arrokoth, \citet{Morbidelli21} found that bodies $\lesssim 1$~km in the Kuiper belt have a slope of $\gamma = 1.2$, which could suggest that the shallow slope for small JFCs may be a result of the primordial distribution of their source in the scattered disc.  \par
For JFCs larger than $\sim10$~km, observations are very few, so instead we turn to the size distribution of their parent population. Using the size distribution of the primordial trans-Neptunian disc from Figure~14 of \citet{Nesvorny17}, we assume a slope of $\gamma = 5.0$ for the range $50 \leq R \leq 150$~km and $\gamma = 2.5$ for $150 \leq R \leq R_\mathrm{max}$~km. This is based upon \citet{Nesvorny16}, including their requirement for 1000-4000 Pluto-sized objects in the primordial planetesimal disc. We set an upper limit of $R_\mathrm{max} = 1000$~km in our model. \par 
An overview of the differential size distribution slopes, $\alpha$, used for our input comets when a particle is cloned is given in Table~\ref{tab:sizedist}. These are defined such that the differential size distribution of comets at a given size scales as
\begin{equation}
\label{eq:dNdR}
N(R) = dN/dR \propto R^{-\alpha},
\end{equation} 
meaning that $\alpha = \gamma + 1$ in terms of the slope of the cumulative size distributions given in the literature. \par
Densities of comets have large uncertainties, as the mass can only be measured indirectly. We assume the comet nuclei to have a bulk density of 0.6\gcm, in agreement with the most likely value suggested by \citet{Weissman08}. \par

\begin{table}
	\centering
	\caption{Slopes of the differential size distribution of JFC nuclei used in our model as input to the inner solar system.}
	\label{tab:sizedist}
	\begin{tabular}{c | c}
		\hline
		Size range (km) & Slope, $\alpha$ \\
		\hline
		$0.1 \leq R \leq 1$ & 2.25 \\
		$1 \leq R \leq 50$ & 3.0 \\
		$50 \leq R \leq 150$ & 6.0 \\
		$150 \leq R \leq 1000$ & 3.5 \\
		\hline
	\end{tabular}
\end{table}

We normalise the mean size distribution of comets when cloning a particle using the mass in comets of radii $1 \leq R \leq 10$~km. Note that each particle may receive more or less than the mean due to the way the population of each size bin is assigned stochastically based on this distribution. This mass input is a free parameter which is fitted to the number of active visible comets in the given size range. The most complete catalogue of JFCs is \citet{Tancredi00,Tancredi06}, who have estimated radii for 58 JFCs in the given size range. This is a lower limit on the number of active visible comets in this range with pericentres $< 2.5$~au, as many observed comets do not have estimated radii. In our model, we consider comets to be 'active' for the first 12,000~yr inside 2.5~au based on \citet{Levison97}. We tuned the mass input to fit on average 58 active visible comets inside 2.5~au, which gave a mass input of $8.08 \times 10^{19}$~g of comets in the range $1 \leq R \leq 10$~km every 12,000~yr. \par
\begin{figure}
	\centering
	\includegraphics[width=\linewidth]{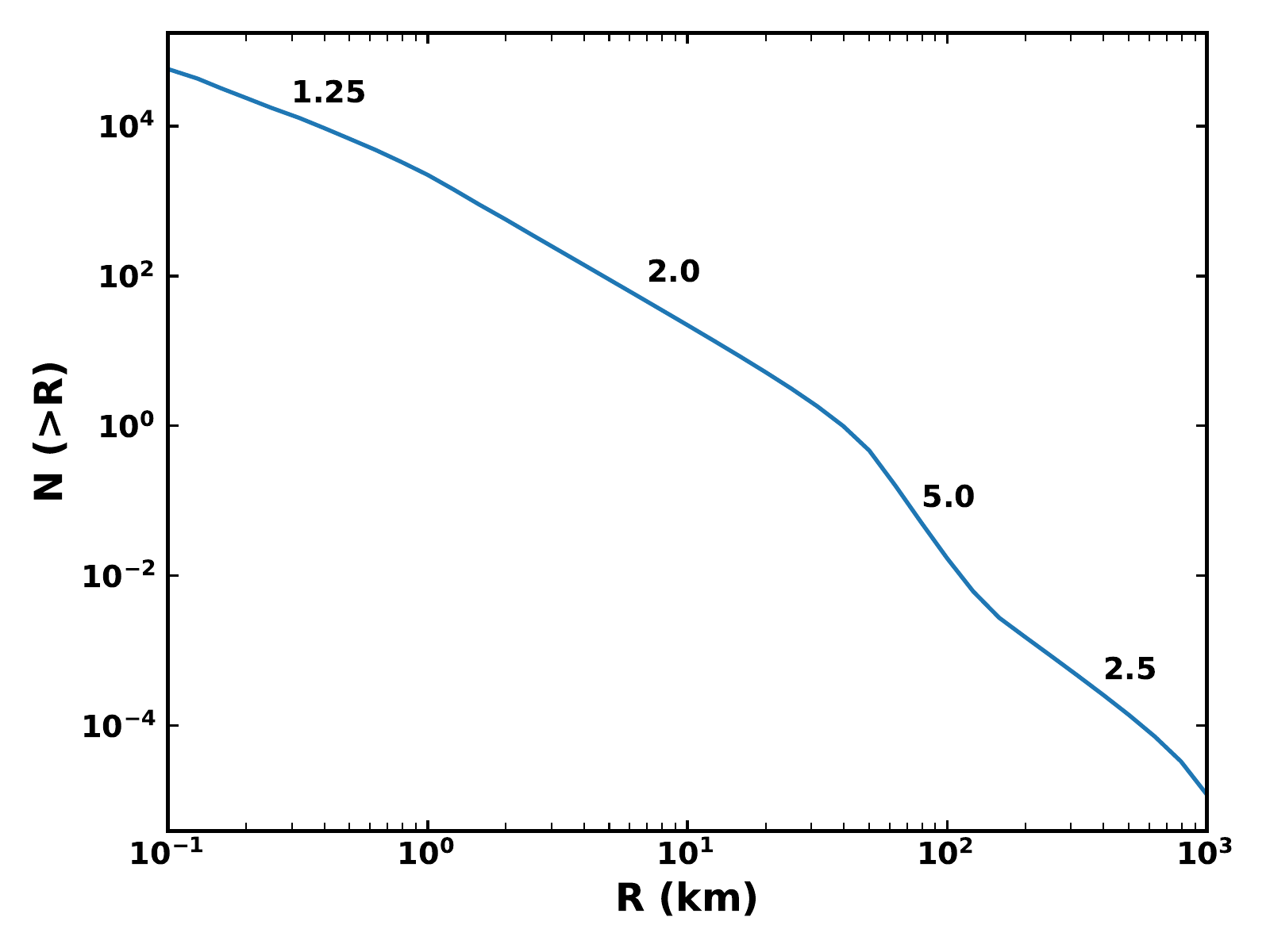}
	\caption{Cumulative size distribution of comet nuclei radii which is input into our simulations each time the N-body particles are cloned. The cumulative slope of the size distribution in each region is labelled above the curve, related to the differential slope (Table~\ref{tab:sizedist}) as $\alpha - 1$. This size distribution has been normalised to have a mass of $8.08\times 10^{19}$~g in comets of sizes $1 \leq R \leq 10$~km, and is shared between all N-body particles each time cloning is done.}
	\label{fig:input_size}
\end{figure}

The final input size distribution of comets which is used every time particles are cloned is shown in Figure~\ref{fig:input_size} as the cumulative size distribution. The slopes of the cumulative distribution in each region are given above the line. \par                                                                                                          

\begin{figure}
	\centering
	\includegraphics[width=\linewidth]{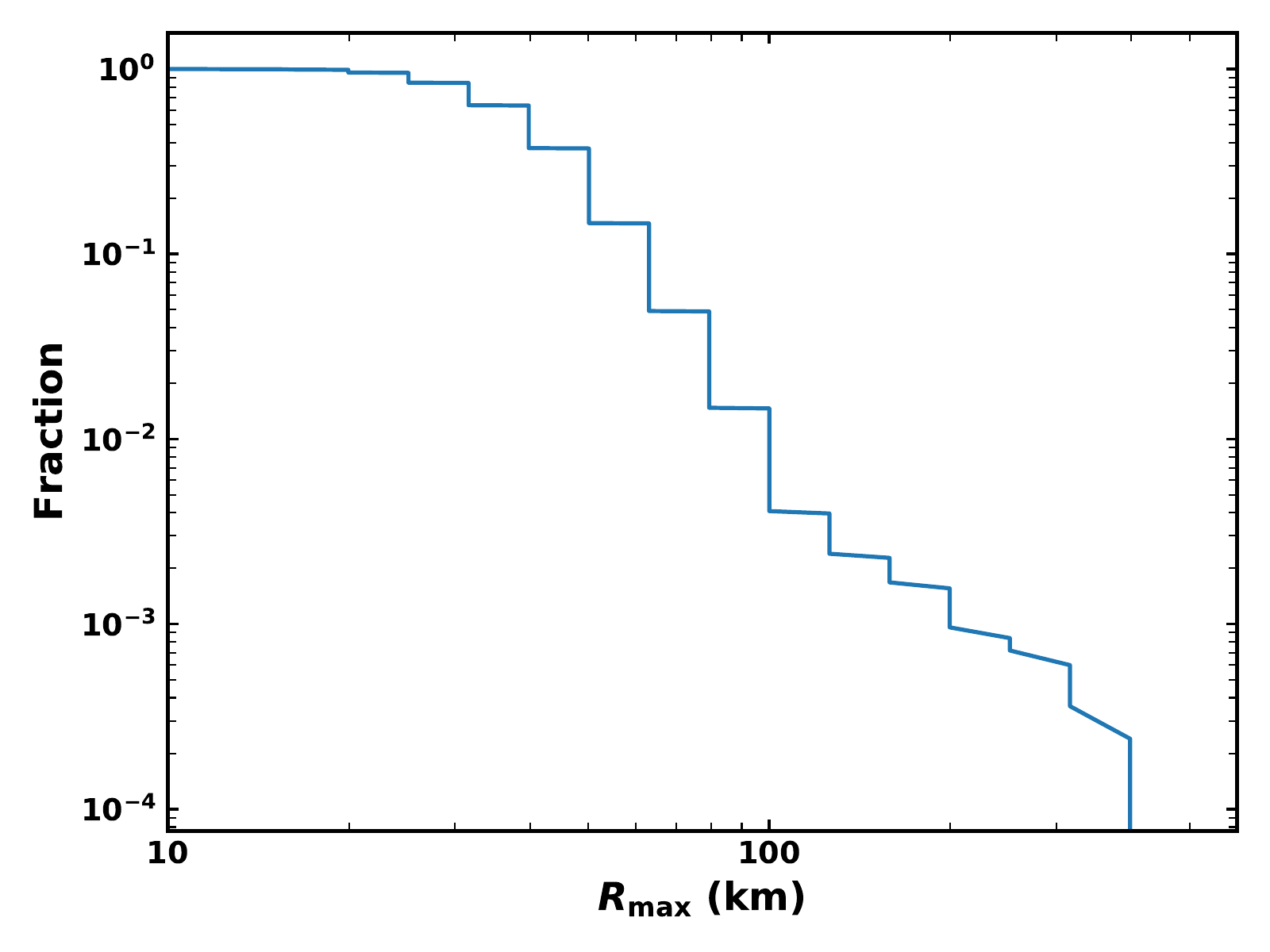}
	\caption{Fraction of cloning time steps for which the largest comet present amongst the randomly drawn size distributions of all N-body particles is larger than $R_\mathrm{max}$.}
	\label{fig:maxR}
\end{figure}

The steepness of the size distribution means that it is rare for bodies $R \gtrsim 100$~km to be scattered into the inner solar system. Figure~\ref{fig:maxR} shows the distribution of the largest comet size which is present amongst all 21,548 of the N-body particles each time cloning is done (every 12,000~yr). The largest comet seen in the whole simulation is 501~km, with a total of 34 out of 8334 cloning steps (0.4 per cent) containing a comet with $R \geq 125$~km. Most commonly, the largest comet present will be in the range $\sim$30-60~km. The size distribution will always contain many comets a few km in size. The largest comet present in a given cloning step ranges from 16 to 501~km. Given the steep dependence of mass on radius, in the rare cases very large (>100~km) comets are present, they may dominate the mass distribution if they lose a significant fraction of their mass, and it is therefore important to study the effects of such events. \par

\subsection{Cloning}
\label{subsec:fragmethod}
We simulate the fragmentation of comets for a total of 100~Myr. This total run time is limited due to computational issues, though is sufficient time for the dust distribution to come to a quasi-steady state. Due to the low number of N-body particles, they are cloned every 12,000~yr, assuming that the time a body is scattered inwards is unimportant. Ideally cloning would happen more frequently if it were feasible computationally, but the time resolution of our output is also limited due to computational resources. The exact frequency of cloning is not too important, so long as it is frequent enough to give good statistics. Given the canonical lifetime of JFCs of 12,000~yr, this should be frequent enough to study the variation in the zodiacal cloud. \par 
Each cloned N-body particle is assumed to represent a size distribution of comets. Every time a particle is cloned, the number of comets in each size bin is randomly drawn from a Poisson distribution, with a mean given by the size distribution of Section~\ref{subsec:sizedist_com}. Each individual comet in this distribution is followed as it evolves, calculating the probability that a fragmentation event occurs each orbit, as described in Section~\ref{subsec:frag}. Once all comets have been followed, this gives a mass input into the zodiacal cloud as a function of time, pericentre, and eccentricity. \par
Observations of comet splittings show that often mass goes to fragments which are tens or hundreds of metres in size \citep[see e.g.][for a review]{Fernandez09}. However, fragments are often seen to disappear on relatively short timescales, varying from days to months or a few yr. We assume that a fraction of the mass a comet loses as it fragments is inputted into the interplanetary region as dust grains with a range of sizes, with the rest of the mass lost going into larger fragments that are assumed to follow the same dynamical evolution as the parent comet. The fraction of mass which goes to dust is a free parameter of our model, $\epsilon$. Dust grains are placed onto the relevant orbits depending on their size, taking into account radiation pressure (see Section~\ref{subsec:dust_sizedist}). The evolution of this dust is followed with a kinetic code that follows the evolution of particles due to collisions and drag (Section~\ref{sec:dust}). \par

\subsection{Fragmentation}
\label{subsec:frag}
Since little is known about the exact mechanism of comet fragmentation \citep[for a review see][]{Boehnhardt04}, we make no assumptions about which of the possibilities is best, and just apply the prescription given below for the mass loss and occurrence rate. \par
We use the model of \citet{DiSisto09} to simulate the splitting of comets. This is a dynamical-physical model which is fitted to the distributions of orbital elements of observed JFCs in order to determine the frequency and mass loss of fragmentation events. \par
The probability in the model that a comet fragments in a given orbit is given by
\begin{equation}
\label{eq:fprob}
f = f_0 (q/q_0)^{-\beta},
\end{equation}
where $q_0 = 0.5$~au, and $f_0$ and $\beta$ are free parameters of the model. When a comet does split, its mass loss is some fraction $s$ of its original mass,
\begin{equation}
\label{eq:dM}
\Delta M = sM,
\end{equation}
where the fraction $s(R)$ of mass lost is,
\begin{equation}
\label{eq:sR}
s(R) = \frac{s_0}{R/R_0},
\end{equation}
where $R$ is the comet radius in km, $R_0 = 10$~km and $s_0$ is a free parameter of the model. \citet{DiSisto09} fit the free parameters of their splitting model to the orbital distributions of observed JFCs, and give four best fit models. Here we choose to use their model 2, which has $\beta = 1$, $f_0 = 1/3$, and $s_0 = 0.007$. The general trend is that the best fit models have a mass loss per event (and therefore $s_0$) that is lower when the frequency of splitting ($f_0$) is higher, which is why they produce comparably good fits to the observed comet population. \par
Each individual comet is followed as it evolves along its dynamical path. For each 100~yr timestep, the number of orbits with the given orbital elements is found. For each orbit, a random number in the range [0,1) is chosen, and compared to the fragmentation probability, $f$, (equation~\ref{eq:fprob}). If $f$ is higher than the random number, a fragmentation event is assumed to occur. Otherwise nothing happens. \par
If a fragmentation event does occur, the fraction of mass lost, $s$, is calculated using equation~\ref{eq:sR}. The mass which is lost, $sM$, is then deposited in the corresponding pericentre-eccentricity bin, and distributed in dust grains as described in Section~\ref{sec:dust}. The mass of the comet is reduced by a factor $(1 - s)$ such that the radius will shrink by a factor $(1 - s)^{\frac{1}{3}}$, and the comet's size decreases after each fragmentation. Eventually the comet mass may reach zero; when this happens the comet is assumed to have fully disrupted, and the evolution of the comet is stopped. There are thus two possible end states for a comet: either the comet is lost dynamically (which almost always means it is scattered outwards) after its dynamical lifetime ends with a nonzero mass, or all of its mass is lost in fragmentation events. \par

\subsection{Outcomes of fragmentation}
\label{subsec:frag_results}
\begin{figure}
	\centering
	\includegraphics[width=\linewidth]{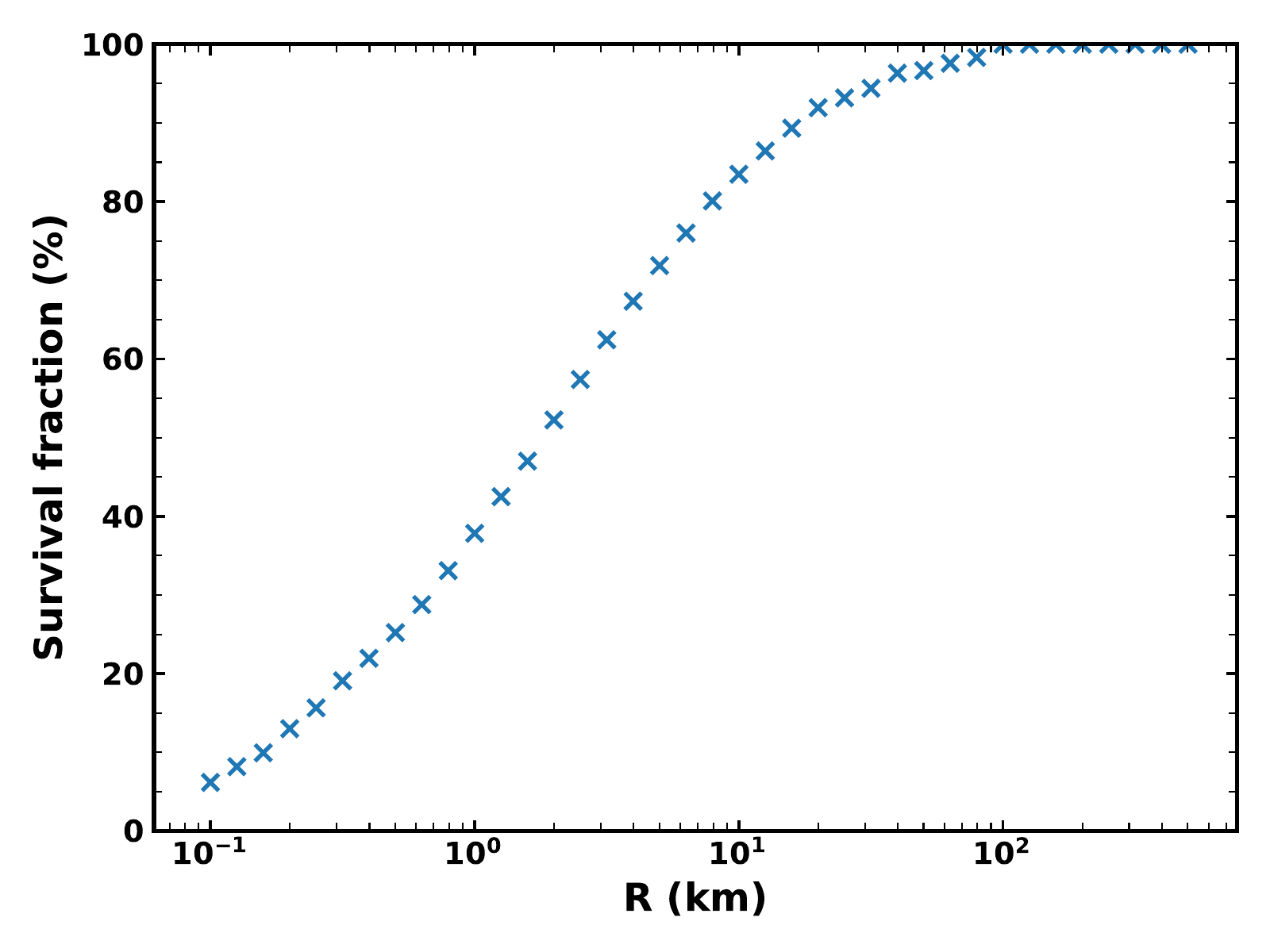}
	\caption{Fraction of comets which survive their dynamical lifetime without fully disrupting as a function of initial comet radius, $R$.}
	\label{fig:survival_frac}
\end{figure}

Models fitted to observations of JFCs \citep{DiSisto09,Nesvorny17} have suggested the need for shorter active lifetimes of comets than the canonical 12,000~yr found by \citet{Levison97}, with a potential increase of lifetime with size. As highlighted by \citet{DiSisto09}, this is a natural outcome of spontaneous fragmentation. Whether a comet survives its dynamical lifetime without fully disrupting depends on two things. First, the initial size of the comet: larger comets have more mass and therefore can survive more splitting events. It also depends on what dynamical path the comet is on. For example, some of the bodies in the simulations of \citet{Nesvorny17} only spend a few hundred~yr inside Jupiter's orbit before being scattered outwards again, such that they may not have sufficient time to disrupt. The fraction of comets of each size which survive their dynamical lifetime, rather than fully disrupting, are shown in Figure~\ref{fig:survival_frac}. As expected, the general trend is higher survival fractions for larger comet nuclei, saturating at $R \sim 100$~km. Small comets disrupt much more rapidly, such that generally all of their mass will be input into the zodiacal cloud. Conversely, larger comets do not lose all of their mass, and therefore may not necessarily dominate the input to the zodiacal cloud. Overall, we found that 13 per cent of comets survived, while 87 per cent fully disrupted. \par
\begin{figure}
	\centering
	\begin{subfigure}[b]{\linewidth}
		\centering
		\includegraphics[width=\linewidth]{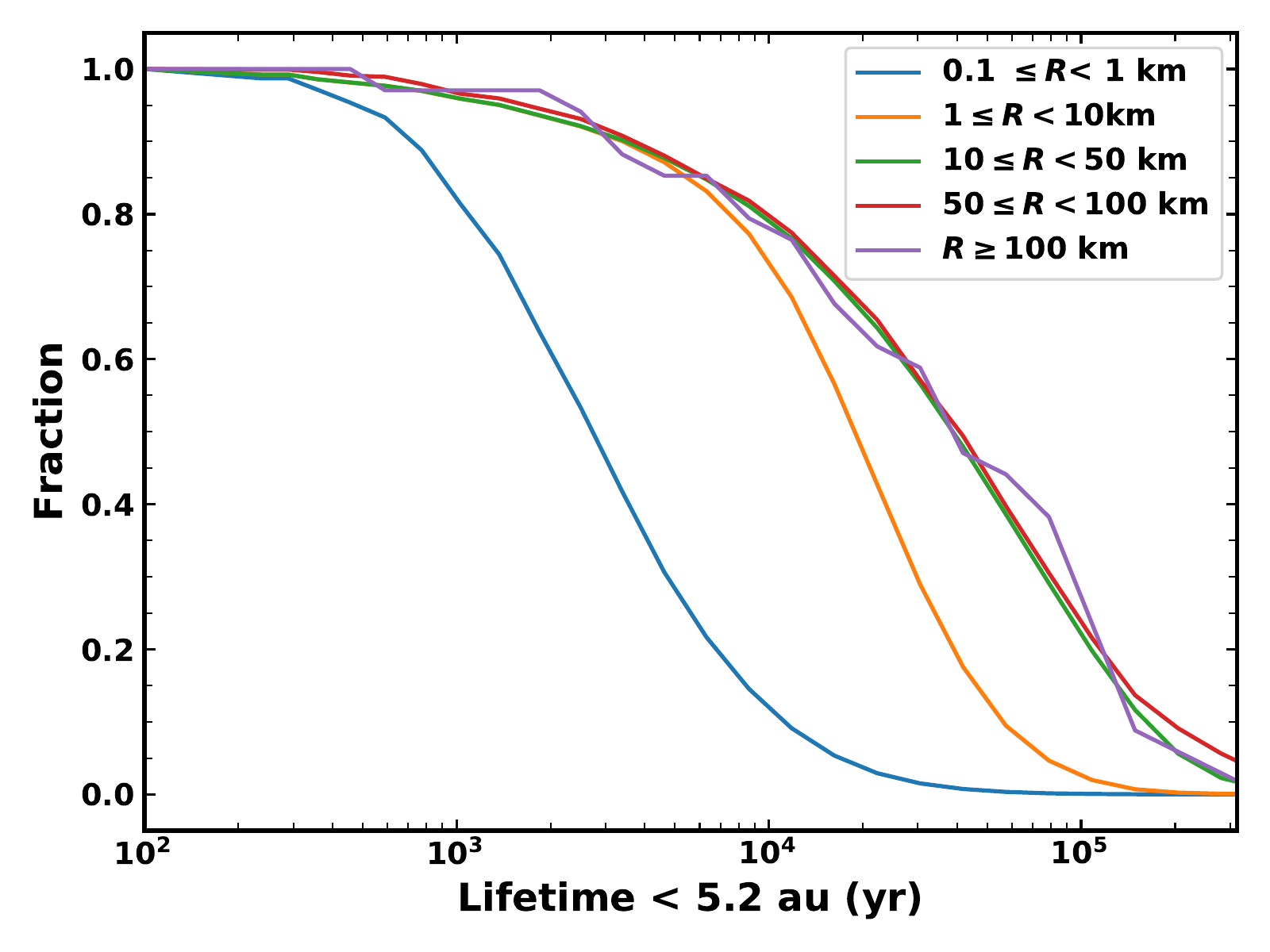}
	\end{subfigure}
	\begin{subfigure}[b]{\linewidth}
		\centering
		\includegraphics[width=\linewidth]{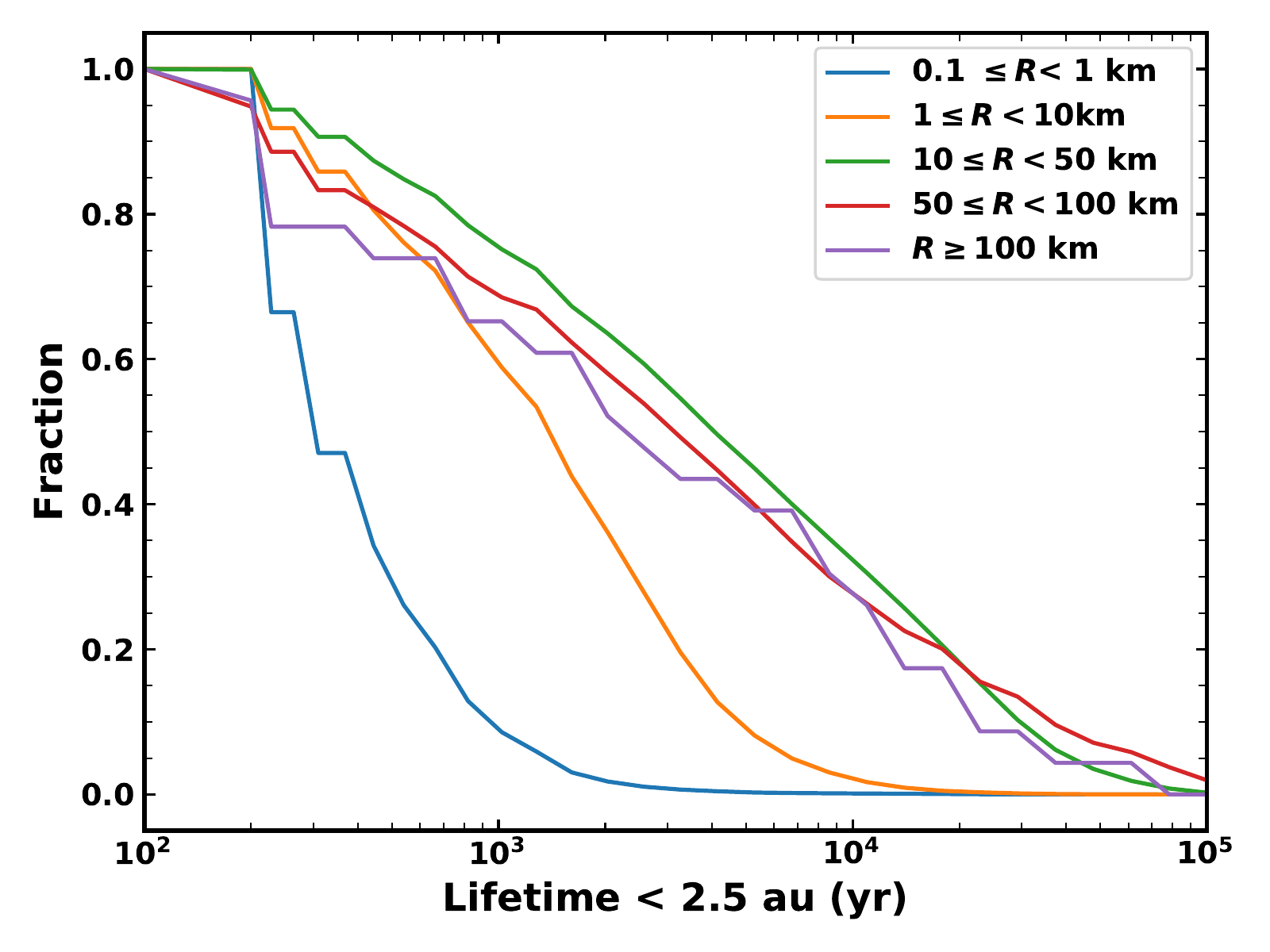}
	\end{subfigure}
	\caption{Cumulative distributions of the lifetime each individual comet survives with $q < 5.2$~au (top) and $q < 2.5$~au (bottom) as a function of initial comet radius.}
	\label{fig:lifetimes}
\end{figure}

Figure~\ref{fig:lifetimes} (top) shows the distribution of lifetimes individual comets have inside Jupiter's orbit in different size ranges. As expected, larger comets have longer lifetimes, with sub-km comets in particular having far shorter lifetimes than other sizes. Comets with radii $> 10$~km tend to survive their dynamical lifetime, such that the distributions for comets $> 10$~km in size generally match the distribution of dynamical lifetimes in the N-body data, although there is some fluctuation for $R > 100$~km comets due to the small number of dynamical paths they sample. Figure~\ref{fig:lifetimes} also suggests that some comets survive for much longer than expected, with a non-negligible fraction of the large comets surviving for over 100,000~yr. The median dynamical lifetime of bodies from the N-body data is 40,200~yr, with a range of 100~yr to 57~Myr. \par

\begin{figure}
	\centering
	\includegraphics[width=\linewidth]{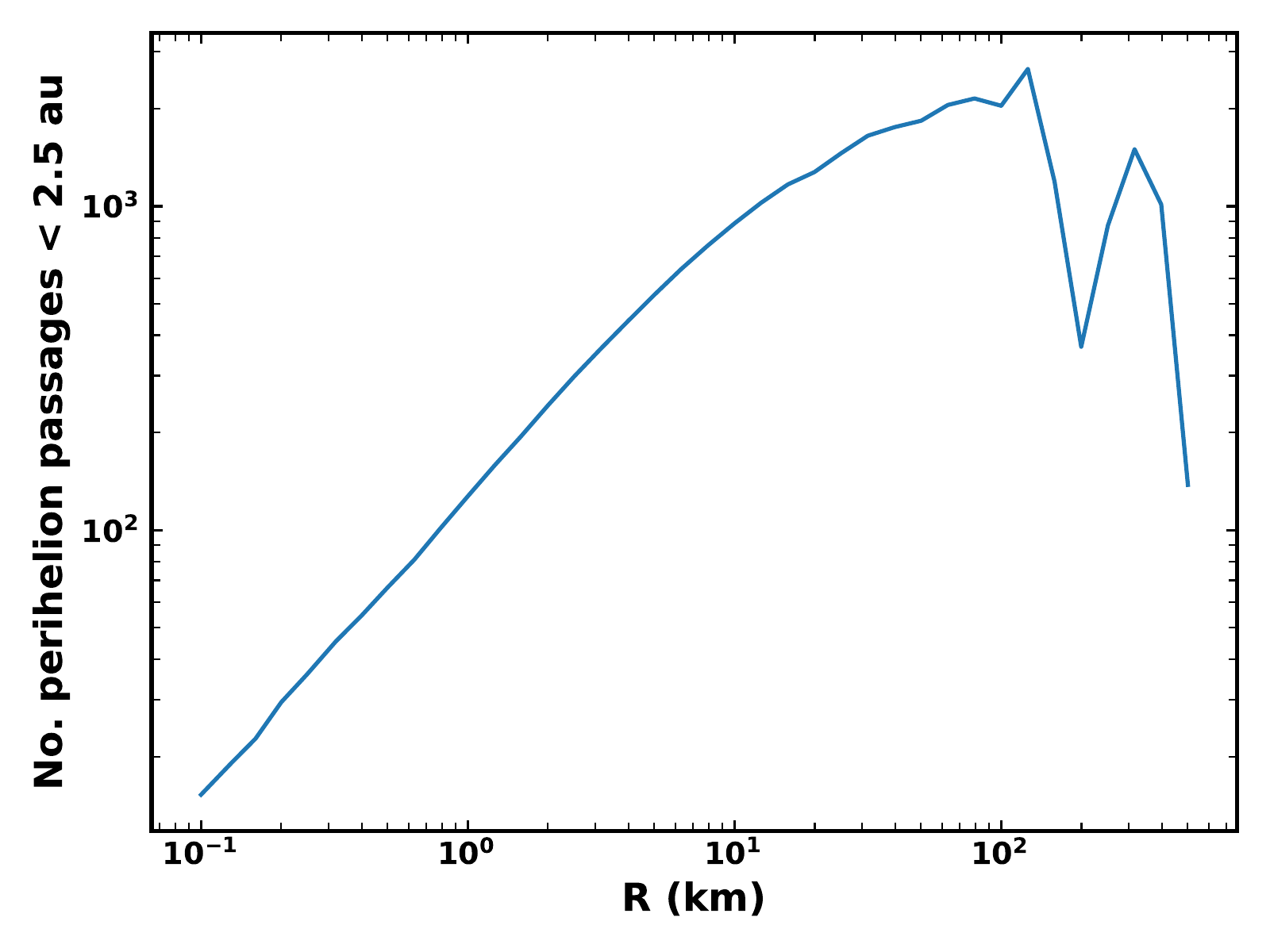}
	\caption{The mean number of perihelion passages comets survive for with $q < 2.5$~au, as a function of the initial comet radius.}
	\label{fig:nperi25}
\end{figure}

The canonical result is that the active lifetime of a comet is 12,000~yr \citep{Levison97}. This applies to comets which are 'visible', defined as those with pericentres $<2.5$~au. We find that 18 per cent of our comets reach $q < 2.5$~au. Not all comets will reach small pericentres because they either fully disrupt or get scattered outwards before this point: 63 per cent of the N-body particles reach $< 2.5$~au at some point, suggesting that the main factor is that small comets fully disrupt before reaching small pericentres. Only 6 per cent of these comets survive their dynamical lifetime - they will fragment more frequently due to the lower pericentre (see equation~\ref{eq:fprob}). We show the lifetimes with $q < 2.5$~au in Figure~\ref{fig:lifetimes} (bottom). Once more larger comets have longer lifetimes than km-sized and sub-km comets. This plot makes it appear that 10-50~km comets are longer-lived than $\geq 100$~km comets inside 2.5~au. However, the distributions of these largest comets are likely affected by small number statistics, with only 123 comets larger than 100~km. Out of the comets with $R \geq 100$~km, 12 per cent are on dynamical paths with a single timestep (i.e. 100~yr) with $q < 2.5$~au. We find that 0.2 per cent of comets that reach inside 2.5 au survive there for longer than 12,000~yr. This is because the size distribution is dominated by sub-km comets, which lose all of their mass rapidly, whereas comets larger than $\sim 10$~km are able to survive for longer than 12,000~yr. Therefore, it is reasonable that some of our larger comets survive to continue fragmenting past the 'active' lifetime. It may be that they stop sublimating after this time as they run out of volatiles, or due to the build-up of a surface layer, but can continue to fragment spontaneously while dormant. \par
Previous JFC models have considered the lifetime of comets in terms of the number of times they pass perihelion with $q < 2.5$~au. In Figure~\ref{fig:nperi25} we therefore show the mean number of times comets of a given size pass pericentre at $< 2.5$~au. This has a strong size dependence, as larger comets have more mass to lose and therefore survive for longer. However, it starts to turn over at $\sim 70$~km as comets no longer lose all of their mass in fragmentations, such that the limiting factor becomes the dynamical lifetime of comets at $<2.5$ au. In particular, a dip is seen at $>100$~km due to the small numbers of comets sampled at these sizes, such that individual dynamical paths become important. We find that 1-10~km JFCs should survive hundreds of perihelion passages, while $>10$~km comets should survive $>1000$ passages. This is broadly consistent with the model of \citet{Nesvorny17}, which found that $\sim 500$ perihelion passages is needed to fit the inclination distribution of JFCs, which are mostly a few km in size, but $\sim 3000$ passages are needed to fit the number of $>10$~km comets. \par
We also compared the rate of comet splitting in our model with observations. The mean rate of comet splitting for visible comets ($q < 2.5$~au) was 0.01~yr$^{-1}$ per comet, which is consistent with the lower limit of 0.01~yr$^{-1}$ per comet found by \citet{Chen94}. However, including all comets, the average rate decreases due to the drop in fragmentation probability with pericentre (equation~\ref{eq:fprob}). \par
\begin{figure}
	\centering
	\includegraphics[width=\linewidth]{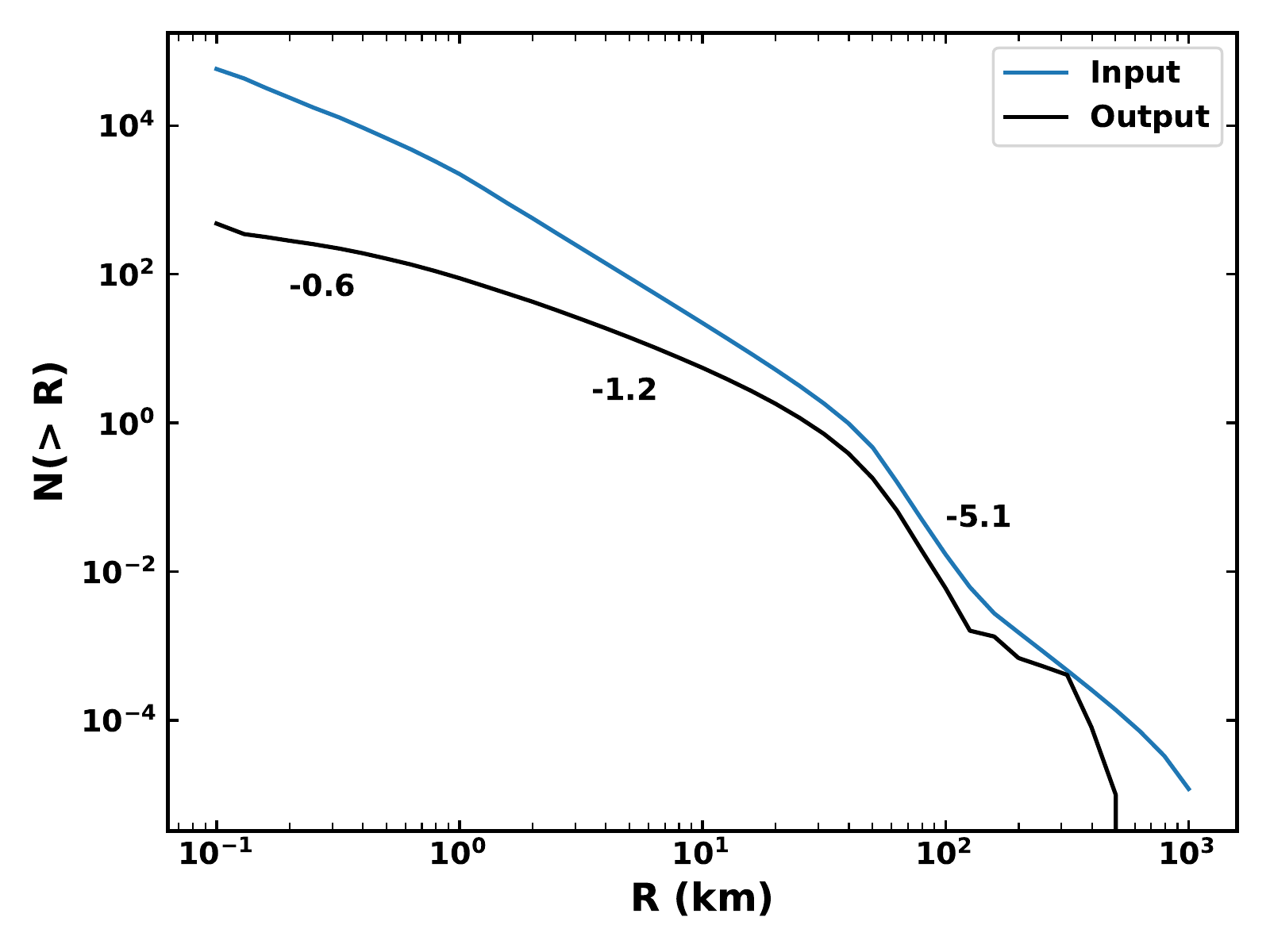}
	\caption{Cumulative size distribution (CSD) of visible comets ($q < 2.5$ au) which is present on average in a 100~yr period (black) compared with the initial distribution of comets which is input (blue). The slopes of the CSD of visible comets in each region are labelled by the curve.}
	\label{fig:CSD_com}
\end{figure}

As comets undergo fragmentations their radii shrink, such that the size distribution of comets changes from our input distribution (Figure~\ref{fig:input_size}). The average cumulative size distribution of visible comets ($q < 2.5$~au) in a 100~yr period is shown in Figure~\ref{fig:CSD_com}, taking into account the change in comet size as mass is lost through fragmentation. The shorter lifetimes of comets due to fragmentation causes the slopes of the size distribution to become shallower than our input size distribution. For sub-km comets, the CSD slope found by fitting a power law to this size range goes from -1.25 to -0.6. For $1 \leq R \leq 10$~km comets, the slope goes from -2.0 to -1.2. The slope of $50 \leq R \leq 200$~km comets is relatively unchanged, however, going from -5.0 to -5.1. This change in slope may suggest that if fragmentation is the significant mass loss mechanism for JFCs, the slope of the size distribution of Kuiper belt objects from which they originate should be steeper than the observed distribution of JFCs at smaller sizes. Estimates of the Kuiper belt size distribution suggest that its slope is similar to the observed JFC distribution at these smaller sizes (Section~\ref{subsec:sizedist_com}), although the uncertainties in these size distributions can be quite large. For example, the slope in the sub-km Kuiper belt size distribution was measured to be in the range -1.0 to -1.2 \citep{Morbidelli21}, and the sub-km JFCs were measured as $-1.25 \pm 0.3$ \citep{Fernandez06}. One possible resolution, if these size distributions are in fact identical, is that the prescription for the size dependence of fragmentation used in our model should be changed. \citet{DiSisto09} assumed that the fraction of mass lost in a splitting event is proportional to $1/R$ (equation~\ref{eq:sR}) based on the escape velocity from the comet nucleus being proportional to its radius. This means that sub-km comets will only survive one or two events, while larger comets almost never fully disrupt. A weaker size dependence would cause the size distribution of comets produced by the model to be closer to their input distribution. The size dependence is therefore potentially another free parameter of the fragmentation model which should be explored. \par

\begin{figure}
	\centering
	\includegraphics[width=\linewidth]{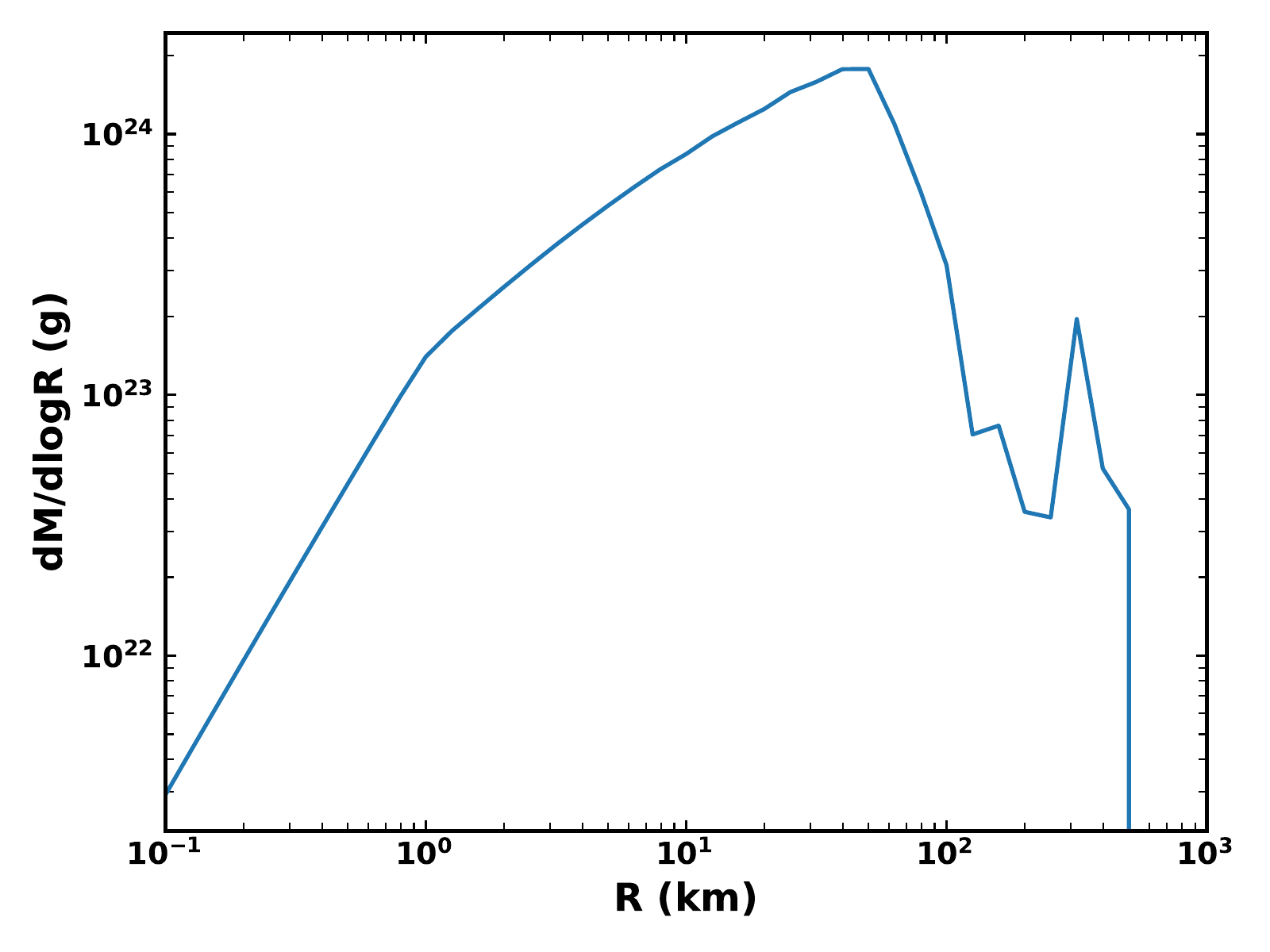}
	\caption{Distribution of mass produced by fragmentation of comets with different initial sizes over the whole 100~Myr simulation. This mass will be distributed over a range of sizes, from dust up to m-size fragments, such that a fraction of this will supply the zodiacal cloud. }
	\label{fig:dm_dlogR}
\end{figure}

To investigate what sizes of comet should dominate the mass input to the zodiacal cloud, Figure~\ref{fig:dm_dlogR} shows the total mass lost by comets due to fragmentation over 100~Myr vs. the initial size of the comet which produced the mass. It should be noted that this is the mass lost by comets in fragmentations, but only a fraction of this will supply the zodiacal cloud. We assume that only a fraction of the mass lost in a fragmentation becomes dust (see Section~\ref{subsec:parameters}), and larger dust grains may be dominated by dynamical interactions and follow an evolution that sticks with the parent comet (Section~\ref{subsec:timescales}). Figure~\ref{fig:dm_dlogR} shows that the total mass input is dominated by comets around 50~km in size. This is likely due to a balance between larger comets having more mass to potentially lose, and  larger comets not losing all of their mass before being scattered out of the inner solar system. The fraction of mass lost by a comet in a splitting is inversely proportional to its size (equation~\ref{eq:sR}), such that a very small comet could lose all of its mass in a single event, while larger comets require many splittings to lose their mass. Further, the nature of our input size distribution of comets (Table~\ref{tab:sizedist}) means that very few $>100$~km comets are scattered in throughout the simulation, whereas $\sim 50$~km comets are present half the time. In terms of the mass in comets, the steep negative slope for 50-150~km means that the second break in the size distribution at 50~km is where the mass in comets peaks. Since most $>10$~km comets survive their dynamical lifetime (Figure~\ref{fig:survival_frac}), the comet size which dominates the input to the zodiacal cloud is determined by what fraction of their mass large comets lose before the end of their dynamical lifetime. The size distribution is such that the larger fractional mass loss for $\sim 10$~km comets compared to 50~km comets is not sufficient to overcome the lower mass in such comets, which is why the mass input is dominated by $\sim 50$~km comets. The smallest comets ($<1$~km) do not contribute much mass because, although there are many of them and they will fully disrupt, losing all of their mass, the size distribution is such that most of the mass is in larger comets.  \par

\begin{figure}
	\centering
	\includegraphics[width=\linewidth]{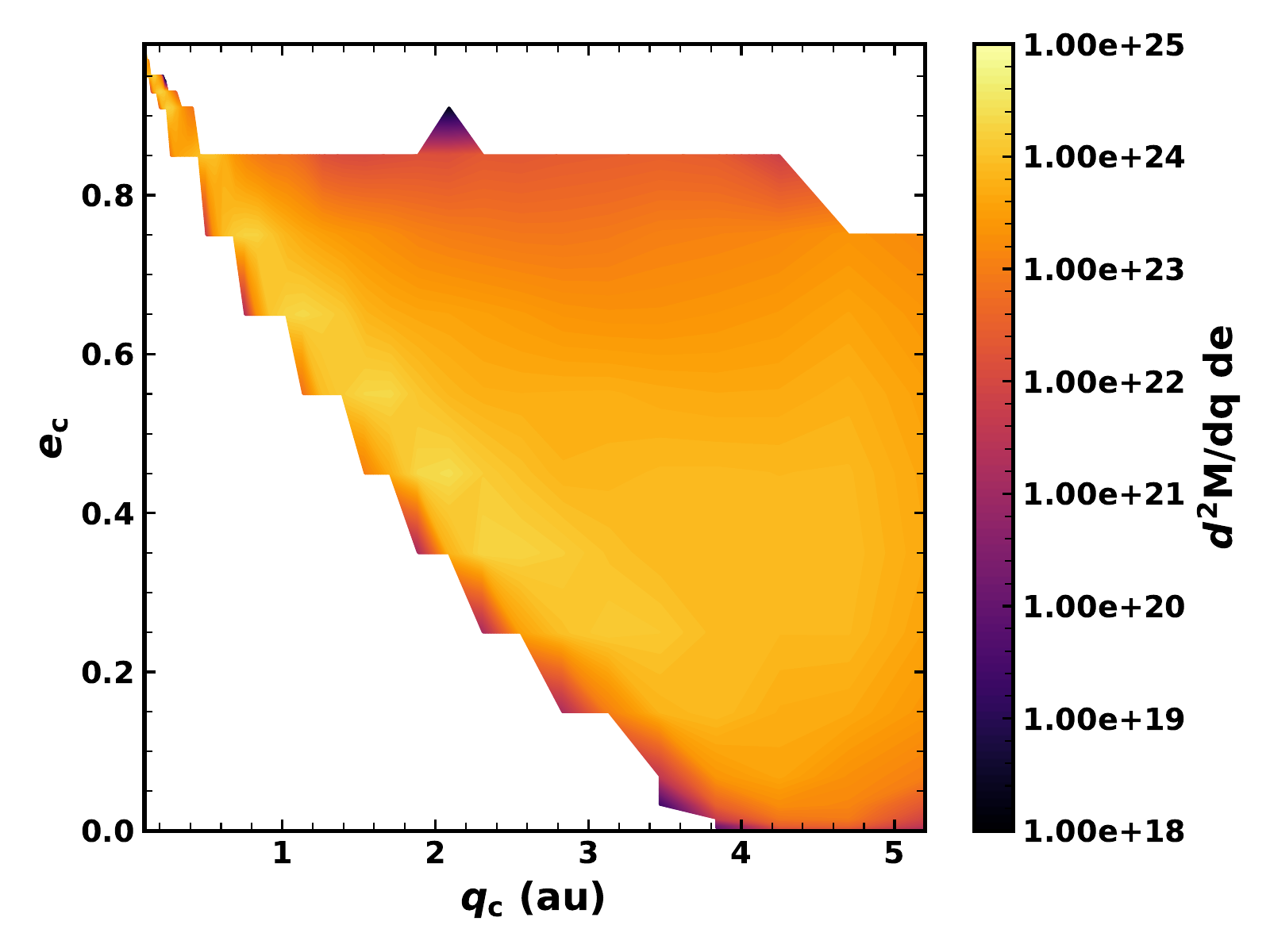}
	\caption{Total mass lost by comets in fragmentation events as a function of pericentre and eccentricity, summed over 100~Myr. A fraction of this will go to dust grains, which will be put on different orbits due to radiation pressure.}
	\label{fig:mloss}
\end{figure}

Given that comets have finite mass, and their fragmentation probability depends on pericentre, the distribution of mass produced by comet fragmentations will not match their distribution in pericentre-eccentricity space (Figure~\ref{fig:comet_qe}). Figure~\ref{fig:mloss} shows the mass lost by comets as a function of pericentre and eccentricity. Comets bounce around in the phase space, though not all will reach $< 2.5$~au. Conversely, the likelihood of fragmentation increases as pericentre decreases. Therefore, the production of dust peaks at the lowest pericentres, where comets fragment frequently and lose a lot of mass if they reach such low pericentres. The distribution of mass lost (Figure~\ref{fig:mloss}) looks similar to the distribution of N-body data points (Figure~\ref{fig:comet_qe}), weighted towards smaller pericentres due to the fragmentation probability decreasing with pericentre. It should be noted that the orbits of dust grains are affected by radiation pressure, and some grains will be removed dynamically, so the distribution of dust input into the zodiacal cloud will differ from Figure~\ref{fig:mloss}. In particular, smaller grains will be put onto higher eccentricity orbits.  \par

\section{Dust model}
\label{sec:dust}

\subsection{Input size distribution and dust properties}
\label{subsec:dust_sizedist}
We assume that some fraction of the mass lost in the comet splittings of our fragmentation model (Section~\ref{sec:frag_model}) becomes dust. This mass is distributed into particles with a range of sizes via a piece-wise power law size distribution. \par
The size distribution of dust produced in the comae of comets has been measured by several spacecraft flybys, finding various slopes over different size ranges \citep[e.g.][]{McDonnell93,Horz06,Economou13}. Flybys of comet 1P/Halley showed that the slope varies with both particle mass and time \citep{McDonnell93}. \par
More recently, various instruments on the Rosetta mission observed the coma of comet 67P/Churyumov-Gerasimenko. \citet{Rotundi15} found the dust had a differential slope $\alpha \sim -2$ for $<$ mm-sized grains, and a slope of $\alpha = -4$ for grains larger than mm-sized. \citet{Fulle16b} found the size distribution of smaller ($<1$~mm) grains varied with time: before perihelion the slope was -2, while after perihelion it was -3.7. However, \citet{Moreno16} used ground-based images from the VLT to show that a slope of -3 is needed to fit the dust tail, disagreeing with in situ measurements. Further, \citet{Soja15} found a slope of -3.7 for large ($> 100~\mu$m) particles from Spitzer observations of the dust trail of 67P. This suggests that the size distribution varies with time and particle size, and may differ between the coma and the tail. \par
Dust measured in comae by flybys likely originates from sublimation. Since we are concerned with the products of comet fragmentation, we instead choose to focus on the debris trails, which may be linked to the break-up of comets rather than just sublimation. Indeed, some of the dust seen near a comet is placed on unbound orbits, and so does not remain in the system. \citet{Reach07} observed the debris trails of 27 JFCs with Spitzer, and found three populations of particles with different size distribution slopes. The breaks in the size distribution occurred at $D$ of $100~\mu$m and $500~\mu$m. The differential size distribution slopes resulting from the mass distribution of \citet{Reach07} are given in Table~\ref{tab:input_dist}. Notice that for this distribution, the mass will be dominated by the largest grains, while the cross-sectional area will be dominated by grains near the second break in the distribution, at $D \sim 0.5$~mm. Assuming the dust in comet trails is linked to comet fragmentation, we therefore choose this distribution for the mass produced in our model. The lower limit of our size distribution is set by the fact that the smallest grains will be blown out on hyperbolic orbits by radiation pressure. For grains released from circular orbits, $D_{\mathrm{bl}} \sim 1.2~\mu$m, but the blowout limit depends on the eccentricity of the parent body, the assumed composition of dust grains, and where around the orbit grains are released, as discussed later in this subsection. Submicron interplanetary dust grains are believed to be primarily of interstellar origin \citep[e.g.][]{Landgraf00}, and we therefore do not try to model such grains. The maximum grain size is chosen to be $D_\mathrm{max} = 2$~cm; the effects of this parameter are discussed in Section~\ref{subsec:freepars}. \par

\begin{table}
	\centering
	\caption{Slopes of the differential size distribution of dust grains produced in comet splittings in our model.}
	\label{tab:input_dist}
	\begin{tabular}{c | c}
		\hline
		Size range & Slope, $\alpha$ \\
		\hline
		$D_\mathrm{bl} \leq D \leq 100~\mu$m & 3.25 \\
		$100 \leq D \leq 500~\mu$m & 1.0 \\
		$500~\mu$m$ \leq D \leq 2$~cm & 3.25 \\
		\hline
	\end{tabular}
\end{table}

Cometary dust is typically thought to be composed of fluffy, porous grains containing ices, though they are often approximated to be compact and spherical. Measurements from the Grain Impact Analyzer and Dust Accumulator (GIADA) of the Rosetta mission found a bulk density range of $1.9 \pm 1.1$\gcm~ for spherical grains in the size range $50~\mu$m $\lesssim D \lesssim 0.5$~mm \citep{Rotundi15}. \citet{Fulle16a} derived a density of $0.795^{+0.84}_{-0.065}$\gcm~ for compact $\sim$mm-sized particles of porous icy dust also from GIADA. We assume cometary fragments to have a bulk density of 1.9\gcm~based on \citet{Rotundi15}. \par 
Radiation pressure means that the orbits of dust created in the break-up of a comet on an orbit with a given pericentre and eccentricity depend on where around the orbit the break-up occurs. Thus the model needs to make an assumption about where around the orbit mass is lost in order to determine the orbits grains are placed on. For instance, most mass loss from sublimation occurs close to perihelion. Comet splittings have been observed even at large distances from the Sun. For example, splitting beyond 100~au was suggested for Comet C/1970 K1 by \citet{SekaninaChodas02}, and the progenitor of the Kreutz sungrazer system is believed to have fragmented near aphelion \citep{Sekanina21}. There is evidence that splitting should occur all around the orbit \citep[e.g.][]{Sekanina82,Sekanina97,Sekanina99}, although it could be argued that some mechanisms may cause fragmentation to be more likely closer to perihelion due to their temperature dependence. \par
\begin{figure}
	\centering
	\includegraphics[width=\linewidth]{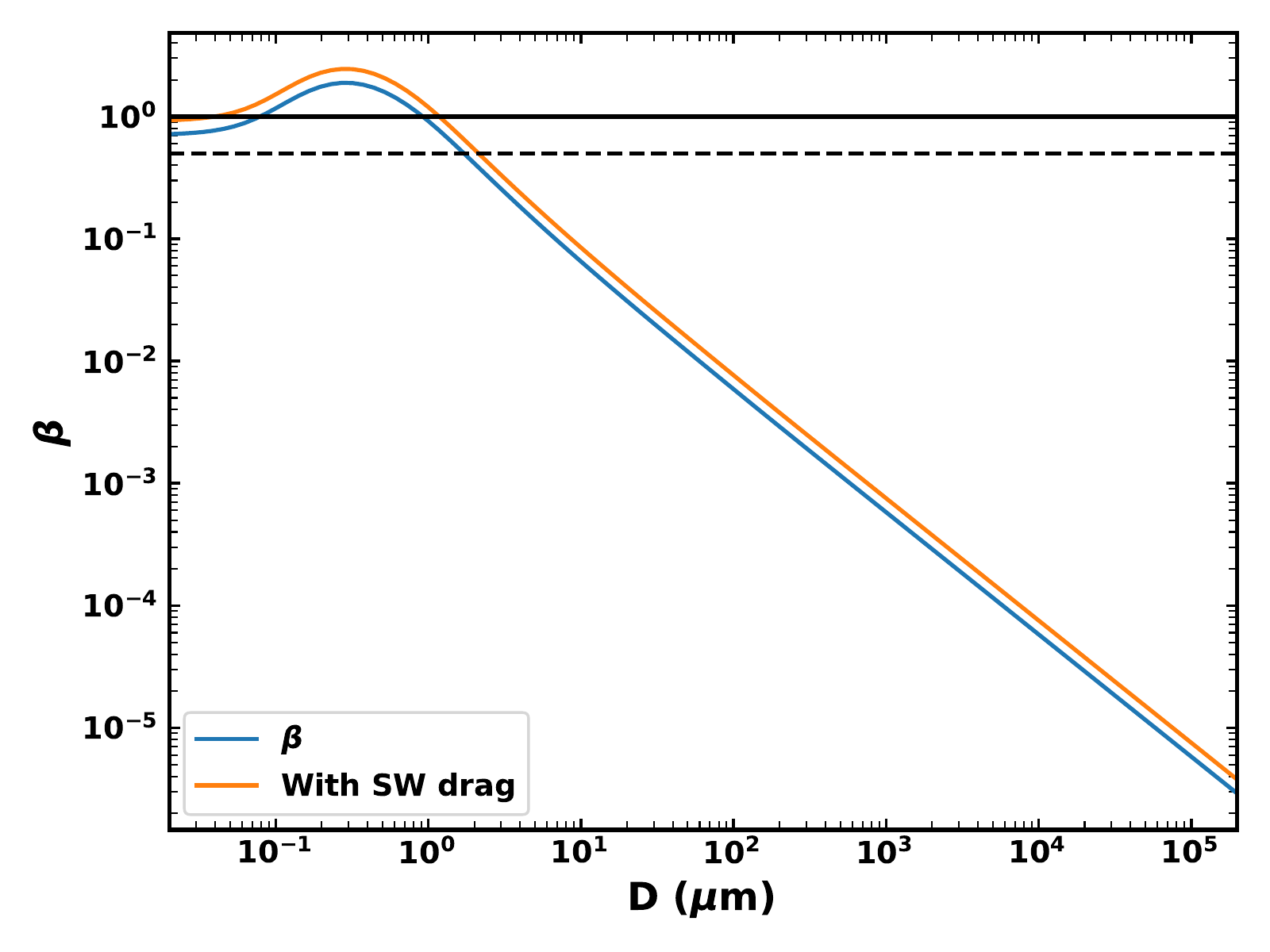}
	\caption{The ratio of radiation pressure to gravity, $\beta$, for grains of different sizes (blue) with a composition which is 1/3 silicate to 2/3 organic material with a porosity of 20 per cent. The orange line shows $\beta$ after being multiplied by a factor 1.3 to include the solar wind drag. The dashed and solid horizontal lines show $\beta = 0.5$ and 1 respectively. Grains released from circular orbits will be put onto hyperbolic orbits for $\beta > 0.5$, while any grains with $\beta > 1$ will be blown out.}
	\label{fig:beta}
\end{figure}

We assume that each comet splitting occurs at a random location around the orbit, choosing a random mean anomaly for each event. The true anomaly $f$ and heliocentric distance $r$ at which a fragmentation takes place can then be found using Kepler's equation. The orbits of dust released by a comet depend on the ratio of radiation pressure to gravity acting on the particle, which is given by 
\begin{equation}
\label{eq:beta}
\beta = \frac{3 L_{\star}Q_\mathrm{pr}}{8\pi G M_\star c D \rho},
\end{equation}
where $D$ is the particle diameter, $L_\star$ is the stellar luminosity, $M_\star$ is stellar mass, $\rho$ is the bulk density of the particle, and $c$ is the speed of light. $Q_\mathrm{pr}$ is the radiation pressure efficiency averaged over the stellar spectrum. Then the orbital elements $q_\mathrm{d}$ and $e_\mathrm{d}$ of particles released from a comet with semimajor axis $a_\mathrm{c}$ and eccentricity $e_\mathrm{c}$ are determined from $\beta$ as follows:
\begin{equation}
\label{eq:enew}
e_\mathrm{d}^2 = \frac{e_\mathrm{c}^2 + \beta^2 + 2e_\mathrm{c} \beta \cos f}{(1 - \beta)^2}
\end{equation}
and
\begin{equation}
\label{eq:qnew}
q_\mathrm{d} = \frac{a_\mathrm{c} (1 - \beta)(1-e_\mathrm{d})}{(1 - 2\beta a_\mathrm{c} / r)},
\end{equation}
where $f$ is the true anomaly of the parent comet at the time of fragmentation, and $r$ is the heliocentric distance its fragmentation occurs at. For our purposes we assume zero ejection velocity of particles from the comet when it fragments. The ratio of radiation pressure to gravity, $\beta$, is plotted in Figure~\ref{fig:beta}. We calculated the optical properties of dust grains using the method of \citet{Wyatt02}, which is based on the core-mantle model of \citet{Li97}. In order to fit the density of 1.9\gcm~used in our model, grains were assumed to have a volume fraction of 1/3 silicate to 2/3 organic refractory material, with a porosity of 20 per cent. The smallest grains will be put onto hyperbolic orbits ($e_\mathrm{d} > 1$) by radiation pressure, and are therefore rapidly ejected from the solar system. For grains released at pericentre ($f = 0$), this is given by $\beta > \frac{1}{2} (1 - e_\mathrm{c})$, while for grains released at apocentre ($f = \pi$), this is given by $\beta > \frac{1}{2} (1 + e_\mathrm{c})$. The mean value of $e_\mathrm{c}$ is 0.45. \par

\subsection{Timescales}
\label{subsec:timescales}
Dust grains in the inner solar system will be subject to radiation pressure, mutual collisions, P-R drag, and dynamical interactions with the planetary system. However, it is not possible to fully model all of these effects simultaneously. We therefore find the dominant physical process acting on a given debris particle by comparing the timescales for dynamical interactions, P-R drag, and collisions. Where the P-R drag or collision timescales are shortest the dust is put into a code which follows the evolution of debris in a kinetic model that accounts for drag and collisions. If the dynamical lifetime is shortest, the dust is assumed to stick with its parent comet and be lost on the dynamical timescale. \par
One limitation of this approach is that the model ignores the possibility of dynamical interactions with the planets during the drag and collision-dominated phase. This is a necessary approximation, and for example does not allow for the possibility that dust becomes trapped in mean-motion resonances with planets (such as the Earth's resonant ring), or migrates into a region where the scattering timescales once more become dominant. The secular resonances at 2~au may also be important, increasing particle eccentricities and inclinations, which would influence their accretion onto Earth. Smaller particles will migrate through resonances faster than larger particles, such that larger particles would be affected more significantly. However, we expect this approximation to allow the model to reproduce the broad structure of the zodiacal cloud, but not detailed structures such as the resonant ring. \par

\subsubsection{Dynamics}
JFCs and grains released from them are subject to close encounters and dynamical interactions with Jupiter. Dynamical interactions will dominate the motion of the largest fragments, which are less affected by radiation pressure and P-R drag. When a particle is released from a comet, we define its dynamical lifetime to be the remaining time the parent comet has left with $q < 5.2$~au.

\subsubsection{Poynting-Robertson Drag}
The tangential component of radiation pressure, known as P-R drag, circularises the orbits of bodies and causes them to spiral in towards the star as they lose angular momentum \citep[see, e.g.][]{Wyatt50,Burns79}. The strength of effect P-R drag has on a body depends on the ratio of radiation pressure to gravity, $\beta$ (equation~\ref{eq:beta}). The inverse dependence of $\beta$ on particle size means that P-R drag is strongest for the smallest particles. We define the P-R drag timescale to be the time for drag to reduce the aphelion of the particle to below 4~au, such that the particle is effectively dynamically decoupled from Jupiter. Given that the combination of orbital elements 
\begin{equation}
\label{eq:PRdrag_C}
C_0 = Q(1 - e) e^{-4/5},
\end{equation}
where $Q$ is the aphelion and $e$ is the eccentricity, is constant throughout evolution due to P-R drag, we can find the corresponding eccentricity for an aphelion of 4~au based on the initial orbital elements. Since P-R drag decreases both the eccentricity and aphelion monotonically, we can then find the time taken for the particle to reach an aphelion of 4~au by finding the time to reach the corresponding eccentricity, using an equation for $de/dt$ in terms of only $e$ and constants, as in the method of \citet{Wyatt50}. \par
In order to take into account the effect of solar wind drag, we assume it has a strength 30 per cent that of P-R drag \citep[e.g.][]{Gustafson94,Minato06}. We therefore multiply the values of $\beta$ by a factor 1.3 to incorporate the solar wind into our model, effectively reducing P-R timescales (see Figure~\ref{fig:beta}).

\begin{figure*}
	\centering
	\begin{subfigure}[b]{0.48\textwidth}
		\centering
		\includegraphics[width=\textwidth]{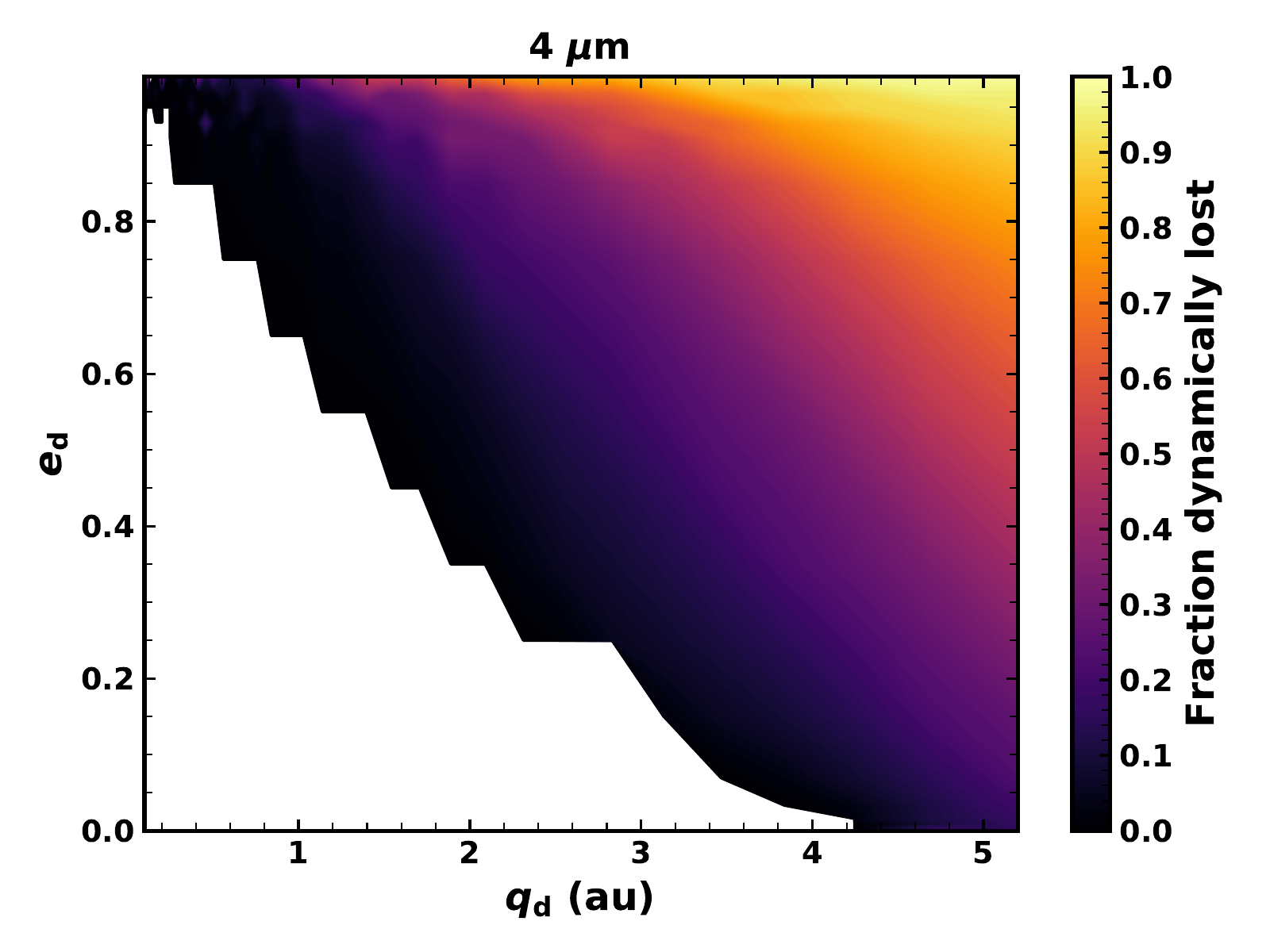}
	\end{subfigure}
	\hfill
	\begin{subfigure}[b]{0.48\textwidth}
		\centering
		\includegraphics[width=\textwidth]{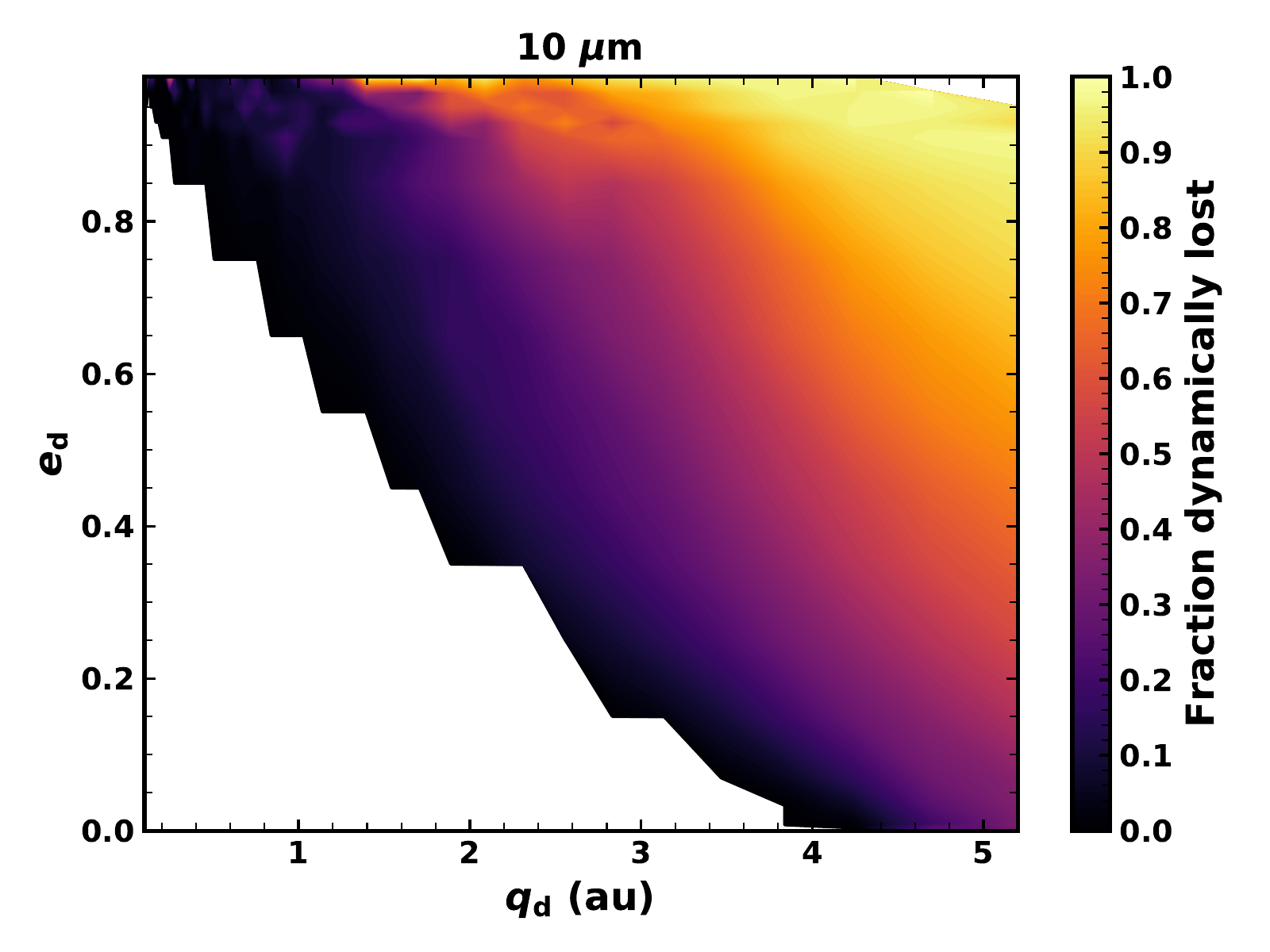}
	\end{subfigure}
	\begin{subfigure}[b]{0.48\textwidth}
		\centering
		\includegraphics[width=\textwidth]{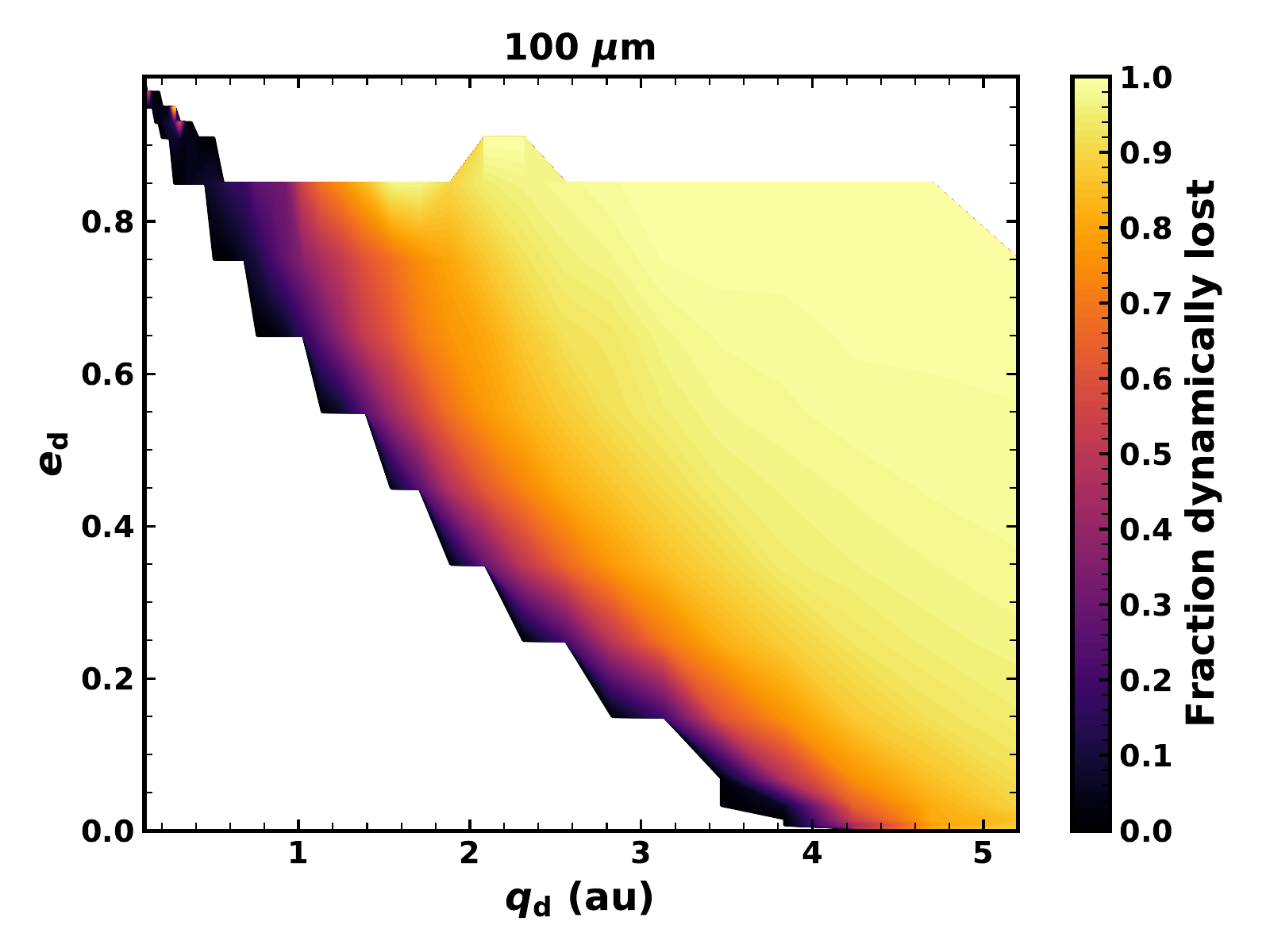}
	\end{subfigure}
	\hfill
	\begin{subfigure}[b]{0.48\textwidth}
		\centering
		\includegraphics[width=\textwidth]{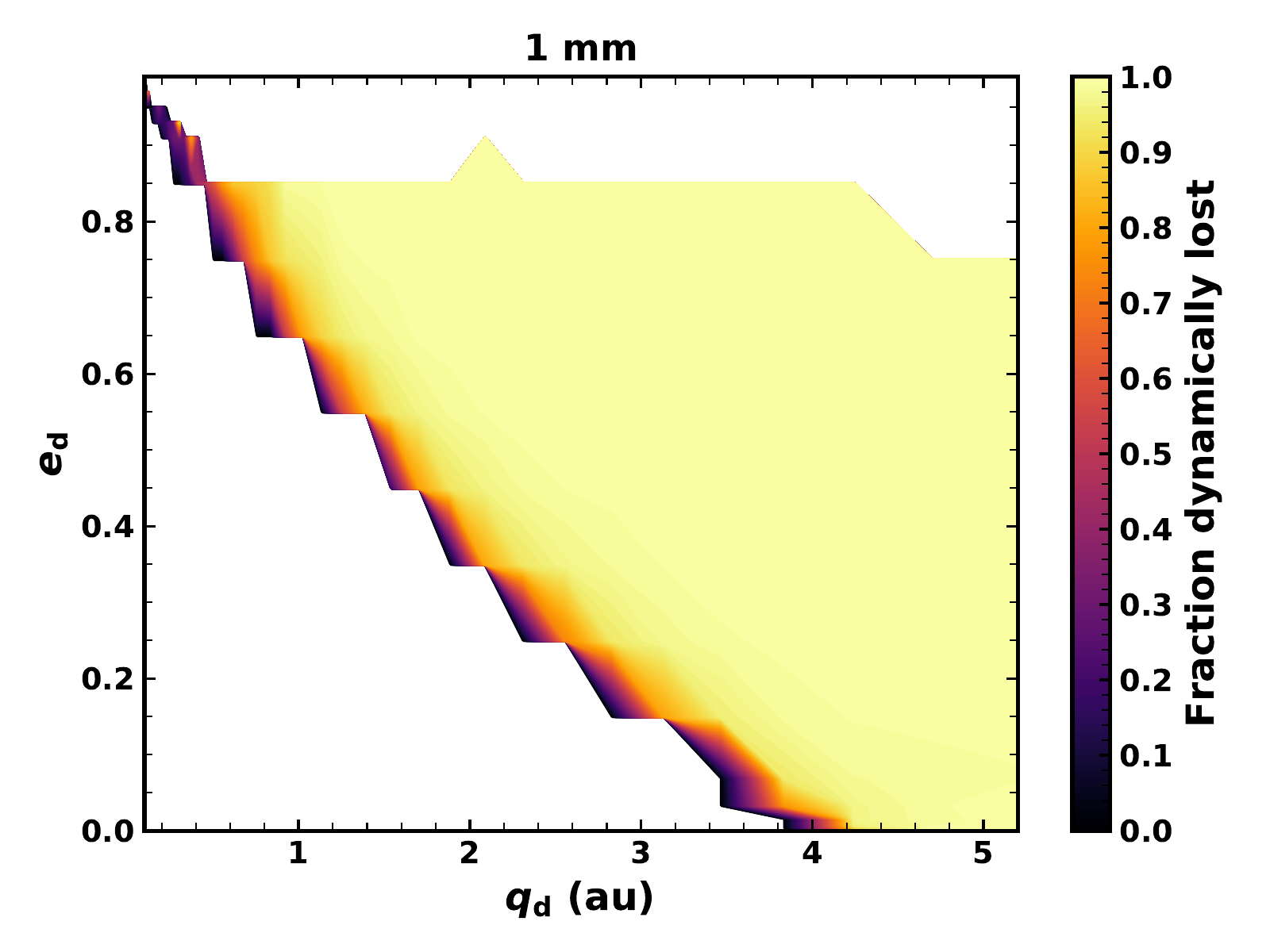}
	\end{subfigure}
	\begin{subfigure}[b]{0.48\textwidth}
		\centering
		\includegraphics[width=\textwidth]{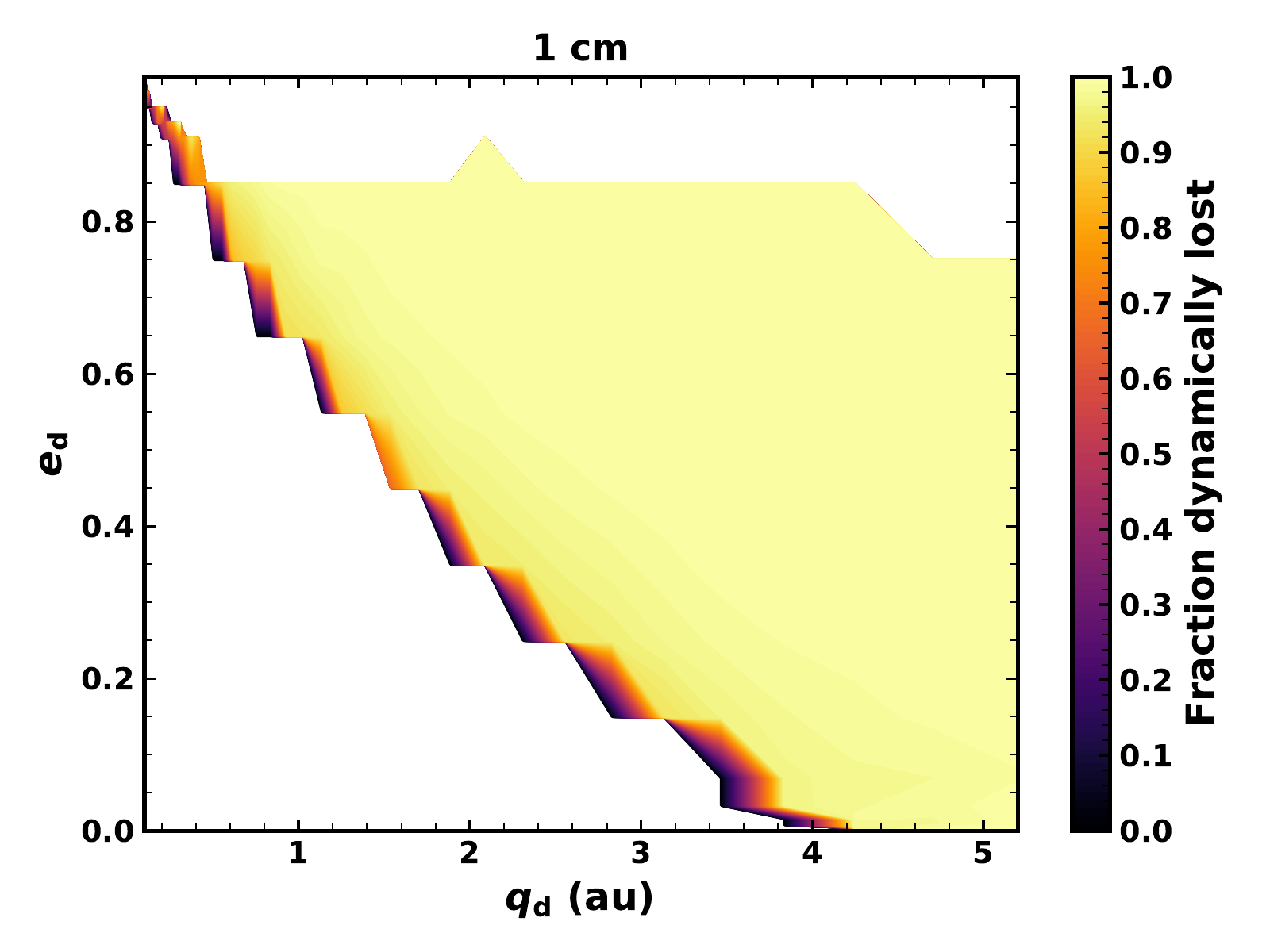}
	\end{subfigure}
	\hfill
	\begin{subfigure}{0.48\textwidth}
		\centering
	\end{subfigure}
	\caption{Fraction of grains of diameter $4~\mu$m, $10~\mu$m, $100~\mu$m, 1mm, and 1cm which are dominated by dynamical interactions with Jupiter, as a function of pericentre $q_\mathrm{d}$ and eccentricity $e_\mathrm{d}$. Grains for which the dynamical timescale is shorter than those of collisions and P-R drag are assumed to be lost on the dynamical timescale, and are therefore not followed by our kinetic model.}
	\label{fig:timescales}
\end{figure*}

\subsubsection{Mutual collisions}
\label{eq:collt}
We calculate the mean time between mutual destructive collisions using the method of \citet{vLieshout14}, further discussed in Section~\ref{subsec:evol_frags}. This involves binning the particles in terms of their size, pericentre, and eccentricity, and taking into account the overlap of different orbits in order to calculate the rate of collisions between grains of different sizes/orbits. Collision rates are scaled based on the population of each bin. Summing over all sizes of impactors which can destroy target particles of a given size gives the rate of catastrophic collisions; its inverse is the mean collisional lifetime.

\subsubsection{Effect on size distribution}
Every time a splitting event occurs, we distribute the lost mass in a size distribution as described in Section~\ref{subsec:dust_sizedist}. Grains for which the dynamical timescale is shortest are assumed to be dominated by their interaction with the planets (mostly Jupiter), such that we do not further include them in our calculations. Ideally we would continue to follow these grains for their dynamical lifetime, however this proved too computationally expensive. Hence the approximation that dynamically-dominated grains are lost is made, although such particles will likely contribute to the zodiacal light in part before they are scattered outwards by Jupiter. The effect of these 'lost' grains is discussed further in Section~\ref{subsub:lostgrains}. \par 
The fraction of different-sized grains which are dominated by dynamics as a function of pericentre and eccentricity is shown in Figure~\ref{fig:timescales}. The dynamical lifetime will depend on which comet dust is released from, and the collision lifetimes vary with time based on how much dust is present. Therefore, this is an average over all times and all comets. This also shows where in $q-e$ space particles of different sizes are produced, which differs from the distribution of comets (Figure~\ref{fig:comet_qe}) due to radiation pressure (see equations~\ref{eq:enew} and~\ref{eq:qnew}). Smaller grains are put onto higher eccentricity orbits by radiation pressure, while mm-cm size grains follow the same orbits as their parent comets. Collisions sometimes dominate for the largest grains which are very close in, or at times when the density of dust is high, but in general P-R drag dominates for the smallest dust grains and those closer to the Sun, while dynamical interactions dominate the largest grains and those which are further out. The fractions of the total cross-sectional area of dust dominated by drag, collisions, and dynamics are 10.2, 0.5, and 89 per cent respectively. It should be noted that while collisions are not usually dominant when a grain is released from a comet, this does not mean that collisions will not become important later in the evolution. For example, as dust migrates inwards, collisions become more destructive due to increased velocities. While we remove the dynamically-dominated grains, their effect is discussed further in Section~\ref{subsub:lostgrains}. \par
\begin{figure}
	\centering
	\includegraphics[width=\linewidth]{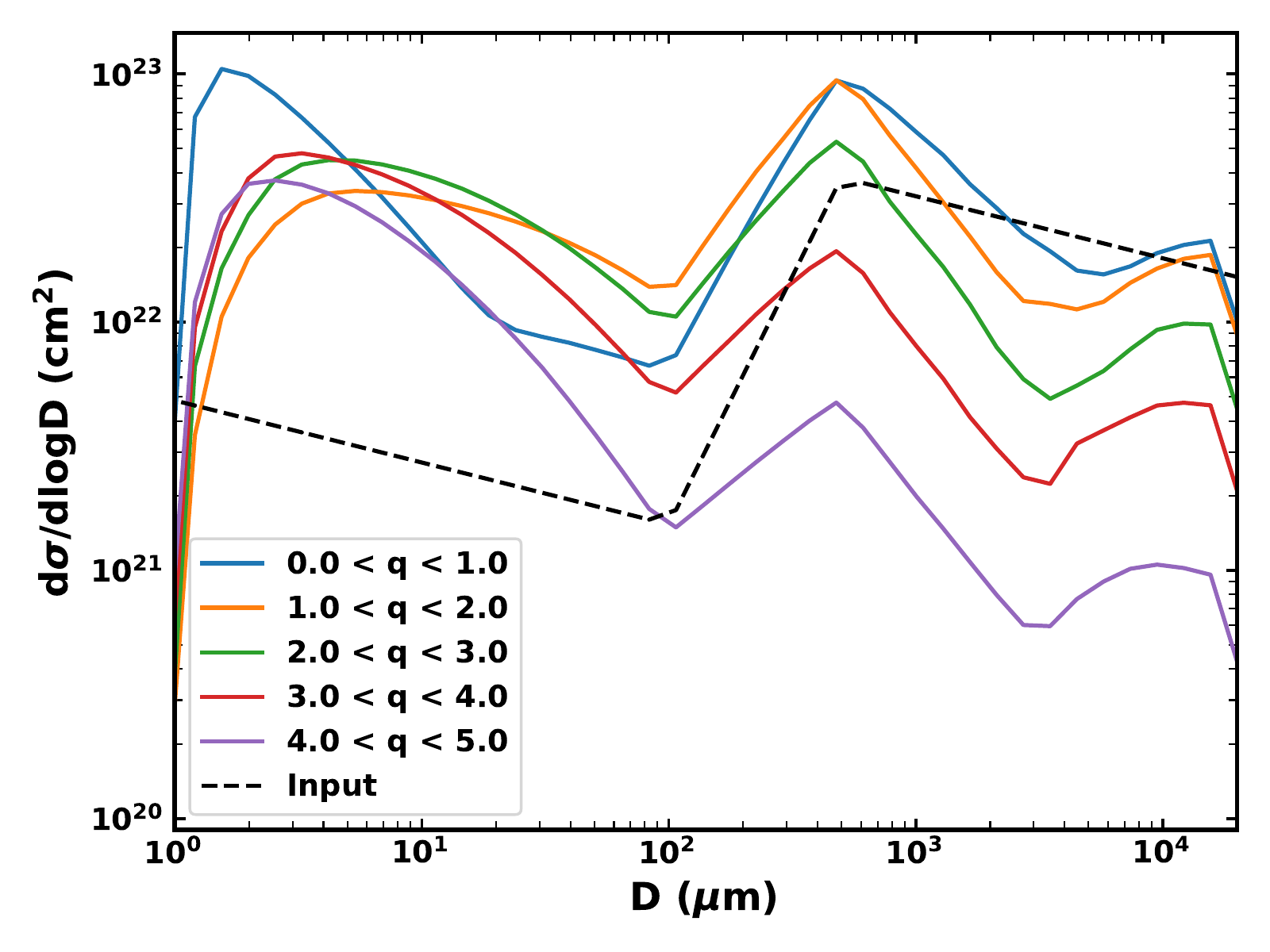}
	\caption{The distribution of cross-sectional area per decade of grain size input into our dust model (Section~\ref{subsec:evol_frags}) within different ranges of pericentres, summed over the whole 100~Myr. Also shown is the size distribution produced by comets based on dust trails \citep[dashed black]{Reach07} with arbitrary scaling. The size distribution input to our dust model is modified by removal of grains which are believed to be dominated by dynamical interactions. }
	\label{fig:dsig_dlogD_in}
\end{figure}

The dependence of the drag and collision timescales on grain size affects the size distribution input into our dust model. Figure~\ref{fig:dsig_dlogD_in} shows this as the distribution of cross-sectional area of grains per size decade input into the model, once 'dynamical' grains have been removed, summed over all time. The distribution of cross-sectional area produced by comets is also shown with arbitrary scaling, to highlight the effect of removing dynamically-dominated grains on the shape of the distribution. Due to the fact that larger grains are very weakly affected by radiation forces, and therefore preferentially removed from the model due to dynamics dominating their evolution, our original input size distribution is shifted towards smaller grain sizes. In particular this effect is more prominent at larger pericentre distances, where P-R drag timescales are longer, such that most large grains are removed dynamically. Hence the input size distribution is close to the distribution we assume is produced by comets (Table~\ref{tab:input_dist}) for grains which are close in ($q \lesssim 1~$au), whereas the size distribution of grains further out is much more dominated by the smaller grains. In all cases two peaks are seen in the cross-sectional area distribution due to the three-slope nature of the original power law: one at the smallest grain sizes, and another around where the second break in the size distribution is at $D\sim 0.5$~mm. These are unchanged by the physical processes, but the relative magnitude of the peak at 0.5 mm decreases for input at larger pericentres due to the loss of large grains. \par

\subsection{Collisional evolution}
\label{subsec:evol_frags}
After using the relevant timescales to determine which particles are lost to dynamics and which evolve due to collisions and drag, we input the drag- and collision-dominated particles into the numerical model of \citet{vLieshout14}. This is a kinetic model which follows the distribution of particles in the phase space described by orbital elements and particle size. This includes the effects of mutual collisions and P-R drag on a population of particles, using a statistical method based on \citet{Krivov05,Krivov06} to find the spatial and size distribution of particles. Particles are distributed in phase space bins in terms of their pericentre $q$, eccentricity $e$, and mass $m$; other orbital elements are averaged over under the assumption that the disc is axisymmetric. A uniform distribution of inclinations is assumed. The population of each bin changes with time according to the rates of collisions and migration due to P-R drag. Starting from no mass being present, dust is added as it is produced in our comet model (Section~\ref{sec:frag_model}). We follow the evolution of the mass produced by comet fragmentations over 100~Myr to find the radial profile and size distribution of the zodiacal cloud which would result from the outcome of comet fragmentation. \par
Only catastrophic collisions are considered by the model. The outcome of a collision is determined by the specific energy $Q$ relative to the critical specific energy, \QD, of the target. This is defined as the energy per unit mass of a collision for which the largest fragment has half the mass of the target. When two particles collide destructively, their mass is redistributed amongst smaller size fragments according to a redistribution function, which is a power law $n_\mathrm{r}(D) \propto D^{-\alpha_\mathrm{r}}$, where $D$ is particle diameter. The maximum fragment mass scales with the specific energy of the collision as -1.24. These fragments are placed onto orbits determined by radiation pressure in a similar manner to equations~\ref{eq:enew} and \ref{eq:qnew}, but including radiation pressure on the disrupted particles too. \par

\subsection{Model parameters}
\label{subsec:parameters}
For our phase space grid we use 30 logarithmic bins of pericentre from 0.1 to 5.2 au. Grain sizes are distributed into 30 logarithmic bins from diameters of 0.1~$\mu$m to 2~cm. For eccentricity, 9 logarithmic bins from $2\times 10^{-4}$ to 0.1 are used for low eccentricity grains. For higher eccentricity, there are 8 linear bins from 0.1 to 0.9, then 5 more linear bins up to $e = 1$ for the highest eccentricities. \par
As discussed in Section~\ref{subsec:dust_sizedist}, we assume the dust to be porous with a density of 1.9\gcm. For small ($<100$ m-size) bodies, collisional strength \QD~is dominated by the material strength, and decreases with particle size. For dust grains self-gravity will be negligible. The strength of grains can therefore be parametrized by a single power law, \QD$~=~Q_0(\frac{s}{\mathrm{cm}})^{-a}$, in this regime. While the collisional strength of various materials has been studied in the literature, most laboratory experiments are performed with particles $\gtrsim 10$~cm in size, and simulations focus mostly on larger sizes. Therefore the collisional strength for dust ($<$ cm-size) is poorly constrained, and we must extrapolate from simulations of $\geq$~cm-size objects. \citet{Benz99} used SPH simulations, and found basalt should have a slope $\sim -0.37$, while ice should have a slope $\sim -0.4$ and lower strength overall. \citet{Jutzi10} simulated collisional destruction of both porous and non-porous bodies, and found that in the strength regime porous bodies (such as pumice) are stronger than non-porous bodies (such as basalt), with similar dependencies on grain size to \citet{Benz99}. Cometary material is believed to be porous, so as a starting point we used the prescription of \citet{Jutzi10} for porous materials, with $Q_0 = 7.0\times 10^7$~erg/g and $a = 0.43$, but both $Q_0$ and $a$ are considered as free parameters. \par
The numerical model assumes a uniform distribution of inclinations. In principle, inclination could be added as another dimension of the phase space grid, but this would increase computational time, which is already a limiting factor. Based on the fact that $> 95$ per cent of JFCs should lie within this range \citep[see Figure 8 of][]{Nesvorny17}, we use a maximum inclination of $30 \degree$ to approximate the inclination distribution. P-R drag does not affect the inclinations of particles, and collisions should not have a major effect either. However, it should be noted that \citet{Nesvorny10} showed from their dynamical model that the inclinations of JFC particles will be increased by interactions with Jupiter after their release from comets. \par
Another free parameter of the model is the slope of the size distribution of fragments produced in collisions, \ar. The canonical value for this is \ar~$= 3.5$, based on the slope of a collisional cascade with constant collisional strength \citep{Dohnanyi69}. Laboratory experiments of catastrophic impacts suggest a range of $2.5 \lesssim~$\ar$~\lesssim 4.0$ \citep{Fujiwara86}. For values of \ar~$>4$, the total mass will be dominated by the smallest particles, whereas for \ar~$<3$, the cross-sectional area will be dominated by the largest grains. \par
The final free parameter of our model is $\epsilon$, the fraction of mass lost by a comet in a fragmentation event which goes to dust. This is fitted in Section~\ref{sec:fitting} to match the absolute value of geometrical optical depth to the present-day zodiacal cloud. \par
 
\section{Model fitting}
\label{sec:fitting}
We compare our model with observables of the zodiacal cloud in order to fit four free parameters: the size distribution of collisional fragments, \ar; the normalisation $Q_0$ and slope $a$ of the collisional strength, \QD; and the fraction of mass lost in fragmentations which becomes dust, $\epsilon$. These are chosen based on finding a model which can best fit the present-day zodiacal cloud at some point in time. \par

\subsection{Observational constraints}
\label{subsec:obs_constraints}
We aimed to fit both the radial profile of geometrical optical depth, equivalent to the surface density of cross-sectional area, and the size distribution of interplanetary dust. The structure of the zodiacal cloud has been characterised in a lot of depth using COBE/DIRBE \citep{Kelsall98}. The DIRBE model has different parametrisations for various components of the zodiacal cloud, but the dominant structure is the smooth cloud, which has a fan-like structure, with a density which decreases with heliocentric distance. Integration of the density of the smooth cloud vertically gives an optical depth of the zodiacal cloud at 1~au of $7.12\times10^{-8}$. \par \citet{Kelsall98} measure a radial power law slope of $-1.34 \pm 0.022$ with DIRBE for the volume density of cross-sectional area. This is in agreement with other measurements of the radial structure of the zodiacal cloud. Photometry on Helios 1 and 2 found the spatial density of zodiacal light particles to vary with a slope -1.3 in the range $0.3 \leq r \leq1$~au \citep{Leinert81}. Meanwhile, \citet{Hanner76} fit a power law to Pioneer 10 observations of the zodiacal light at $> 1$~au, and found the best fit was either a single power law with a slope $\sim -1$, or a power law with a slope of -1.5 with additional enhancement in the asteroid belt. Both models had a cut off at 3.3 au, outside which the zodiacal light is no longer visible over the background. We use these measurements to fit both the absolute value of the geometrical optical depth and its radial slope. Since geometrical optical depth is the volume density of cross-sectional area integrated vertically, if the number density has a radial dependence $n(r) \propto r^{-\nu}$, the geometrical optical depth should have a radial dependence $\tau(r) \propto r^{1 - \nu}$ (for our assumption about the inclination distribution, which means that the scale height is proportional to $r$). Therefore, we want to fit to a radial slope for the geometrical optical depth of $\sim -0.34$ between 1 and 3~au. \par
Finally, we consider the size distribution of zodiacal dust, focussing on the grain size which dominates the cross-sectional area and therefore the zodiacal light emission. At present, the size distribution of particles in the interplanetary dust cloud is best known at 1~au. The most comprehensive model of the size distribution is that of \citet{Grun85}, which combined measurements of different particle sizes based on in situ measurements from Pioneer 8 and 9, Pegasus, and HEOS-2 along with lunar microcraters to produce an empirical model for the size distribution of interplanetary dust particles (IDPs) near Earth. The model of \citet{Grun85} has a peak in the cross-sectional area distribution $d\sigma / d\log D$ at $D \approx 60~\mu$m. \citet{Love93} measured the flux of particles onto a plate on the LDEF satellite near Earth. Converting their flux to a distribution of cross-sectional area gives a peak at $D \approx 140~\mu$m. We therefore want our distribution of cross-sectional area at 1~au to peak at particle sizes of $\gtrsim 60~\mu$m. \par

\subsection{Fitting}
\label{subsec:fitting}
The free parameters of our model are fitted to three observables: the absolute value of geometrical optical depth at 1~au, $\tau$(1~au), the slope of that optical depth between 1 and 3~au, and the grain size which dominates the cross-sectional area. It is part of the stochastic nature of our model that these variables will vary over time depending on what comets are scattered in and how much they fragment. Therefore, our aim in fitting this model to the zodiacal cloud is simply to find for what parameters can it pass through the correct values of all three observables simultaneously at some time. While a range of parameters can give reasonable results, here we try to find the best combination. This is not to say that we have developed a comprehensive model for the zodiacal cloud: we have made some important approximations about the inclinations of particles and the effects of dynamical interactions with Jupiter. Further, other sources (other types of comets and asteroids) should contribute a small amount to our current zodiacal cloud. Here we are simply trying to show the feasibility of a physical comet fragmentation model.  \par
As expected, since $\epsilon$ determines the total mass input into our dust model, the primary effect of increasing $\epsilon$ is an increase in the absolute value of the geometrical optical depth. However, increasing the overall number of particles will also increase collision rates, which are proportional to the number density of particles. This shifts the size distribution to smaller grain sizes, as increased collision rates cause the destruction of larger grains and increased production of smaller fragments. The relationship between $\epsilon$ and $\tau$(1~au) is not linear, but it is monotonic, and so we can simply adjust the efficiency to match the absolute optical depth $\tau$(1~au). \par 
Dust migrating inwards due to P-R drag is expected to have a flat radial profile. Collisions will be more destructive closer in due to higher collision velocities, which would give a positive radial slope. However, the fact that our source is extended, with comets fragmenting at a range of heliocentric distances, produces a negative radial slope as seen for the zodiacal cloud \citep[see ][]{Leinert83}. With the canonical values of our free parameters described in Section~\ref{subsec:parameters}, the slope is too steep at all times, with a maximum value of -0.45. Therefore, \ar~and \QD~must be altered to improve this slope.  \par 
The redistribution function \ar~describes the distribution of mass produced in disruptive collisions. The range of potential values is $2.5 <~$\ar$~<4$, with our initial value \ar$~= 3.5$. Decreasing \ar~shifts the mass produced in collisions to larger sizes, which also shifts the overall size distribution to larger particles. It also causes an increase in the overall optical depth and a decrease in the radial slope. \par
The collisional strength \QD~has two parameters which can be varied: the absolute value $Q_0$, and the dependence on particle size $a$. Increasing the absolute value $Q_0$ makes particles of all sizes more difficult to disrupt via collisions, increasing their collision timescales. The peak in cross-sectional area should occur at a grain size for which the collision and P-R drag timescales are the same. Therefore, increasing $Q_0$ and thus the collision timescales means this occurs at a larger grain size, such that the cross-sectional area peaks for larger grains. Increasing $Q_0$ also causes a slight decrease in the radial slope. \par
The other free parameter is $a$, the slope of the power law of \QD. Increasing $a$ makes smaller particles more difficult to disrupt. This shifts the mass to larger grains overall, as grains $<$cm-size will be lost to collisions less frequently. Again, the reduced collision rates cause a slight decrease in the radial slope. \par
The main difficulty in fitting the zodiacal cloud was the radial slope. With our canonical model, the maximum slope was -0.45, which was too steep to fit the observed slope of -0.34. In order to increase the radial slope, we increased \ar~to 3.75. However, this shifts the size distribution to smaller grain sizes such that the size distribution fit was poor. We therefore had to increase $a$ to 0.9 to shift the size distribution to larger grains. Finally, we decreased $Q_0$ to $2 \times 10^7$~erg/g. Fitting to the absolute optical depth, we found an efficiency of 5 per cent. In one representative run, this gave us a best fit of a slope of -0.34, optical depth at 1~au of $7.1 \times 10^{-8}$, and a peak of cross-sectional area at $60~\mu$m at 66.7~Myr, with another good fit of -0.34, $7.3 \times 10^{-8}$ and 57~$\mu$m at 37.4~Myr. While the model matches the observed values on two occasions during this run, the observables are highly variable due to stochasticity, as discussed in Section~\ref{subsec:stochasticity}.  \par
The model has several free parameters, and it could be argued that there are alternative ways the model could be parametrized while fitting the observables. However, while we do not claim to have a unique model, it does allow to link the zodiacal cloud to its origin in the dynamical and physical evolution of comets, using a physically plausible model. \par

\section{Results}
\label{sec:results}

\subsection{Mass input to the zodiacal cloud}
\label{subsec:min_zc}
\begin{figure}
	\centering
	\includegraphics[width=\linewidth]{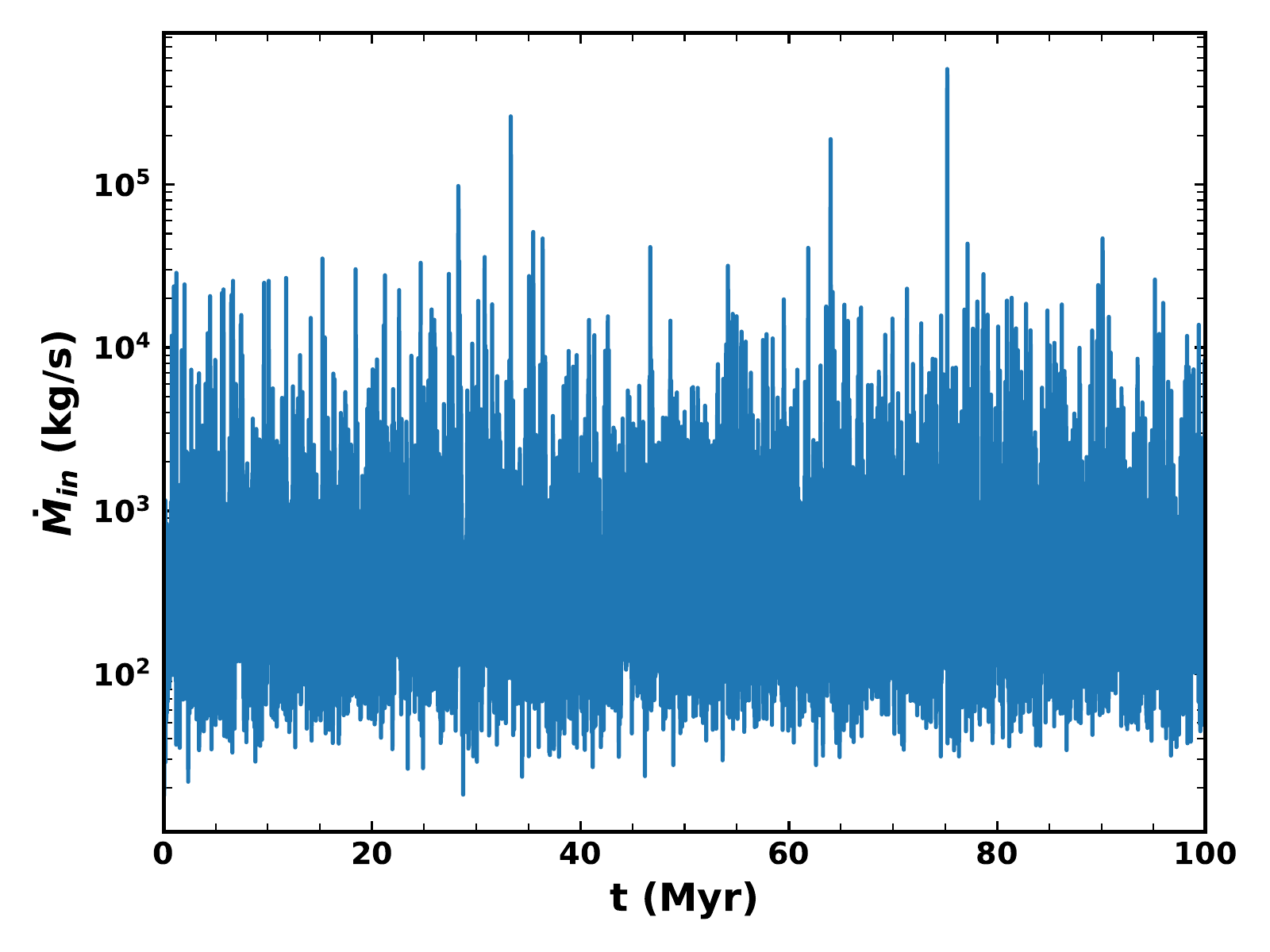}
	\caption{Total mass input rate into our dust model from comet fragmentation as a function of time after removing dynamically-dominated grains and assuming a fraction $\epsilon = 5$ per cent of mass produced in a comet fragmentation becomes dust.}
	\label{fig:Minrate_tot}
\end{figure}

Given the stochastic nature of what comets scatter in as part of our model (and what path they take), the amount of mass being produced by comet splittings is stochastic and highly variable. Figure~\ref{fig:Minrate_tot} shows the total mass input rate to our dust model as a function of time. This takes into account the 'loss' of grains dominated by dynamical interactions and our efficiency parameter of 5 per cent. The mass input to the zodiacal cloud then ranges from 18 to $5.1 \times 10^5$~\kgs, with a mean value of 990~\kgs and a median of 300~\kgs. At the two times our model best fits the zodiacal cloud, the mass input rate is 6,240 and 11,100~\kgs~(i.e. these are epochs of higher than average mass input). Spikes of around two orders of magnitude are seen, which can be linked to the presence of very large comets, highlighting the importance of the stochastic element of our model. This can be compared to previous estimates of the amount of mass required to sustain the zodiacal cloud. Based on their model of Helios 1/2 data, \citet{Leinert83} required a mass input of 600-1000\kgs~to sustain the zodiacal cloud in steady state, while \citet{Nesvorny10} required a slightly higher mass input of 1000-1500\kgs, though did not fully take into account loss of mass through collisions. However, \citet{Nesvorny11_ZC} suggested a much higher rate of $\sim$10,000\kgs~was needed due to the fact that they found grains released closer to the Sun had shorter collisional lifetimes. Our mean mass input rate is thus comparable to previous estimates. \par
\begin{figure}
	\centering
	\includegraphics[width=\linewidth]{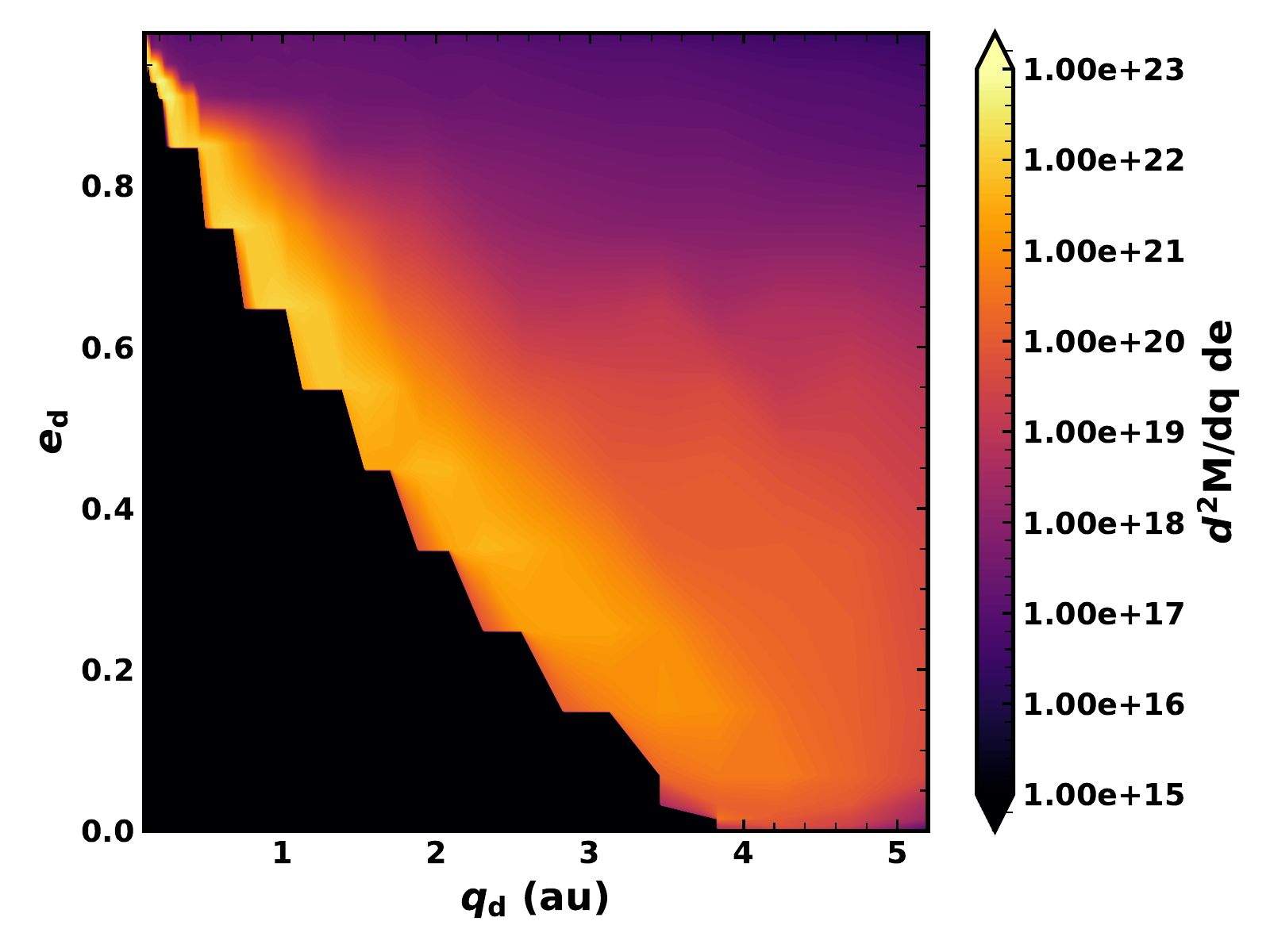}
	\caption{Distribution of mass input into our dust model in terms of pericentre and eccentricity, assuming $\epsilon = 5$ per cent of mass lost in comet fragmentations goes to dust and removing grains dominated by dynamical interactions with Jupiter, summed over the whole 100 Myr.}
	\label{fig:dm_dlogqe}
\end{figure}

The total mass input to our dust model distributed in pericentre-eccentricity space is shown in Figure~\ref{fig:dm_dlogqe}. Most mass is inputted at moderately high eccentricities due to the high eccentricities of the comets. More mass is inputted at lower pericentres due to a combination of the higher rates of splitting events closer in and our removal of grains dominated by dynamical interactions, which are more important at larger pericentres. There is a lower-bound in pericentre-eccentricity space which corresponds to an apocentre of 4~au; this is based on the orbital distribution of the parent comets (Figure~\ref{fig:comet_qe}). \par
\begin{figure}
	\centering
	\includegraphics[width=\linewidth]{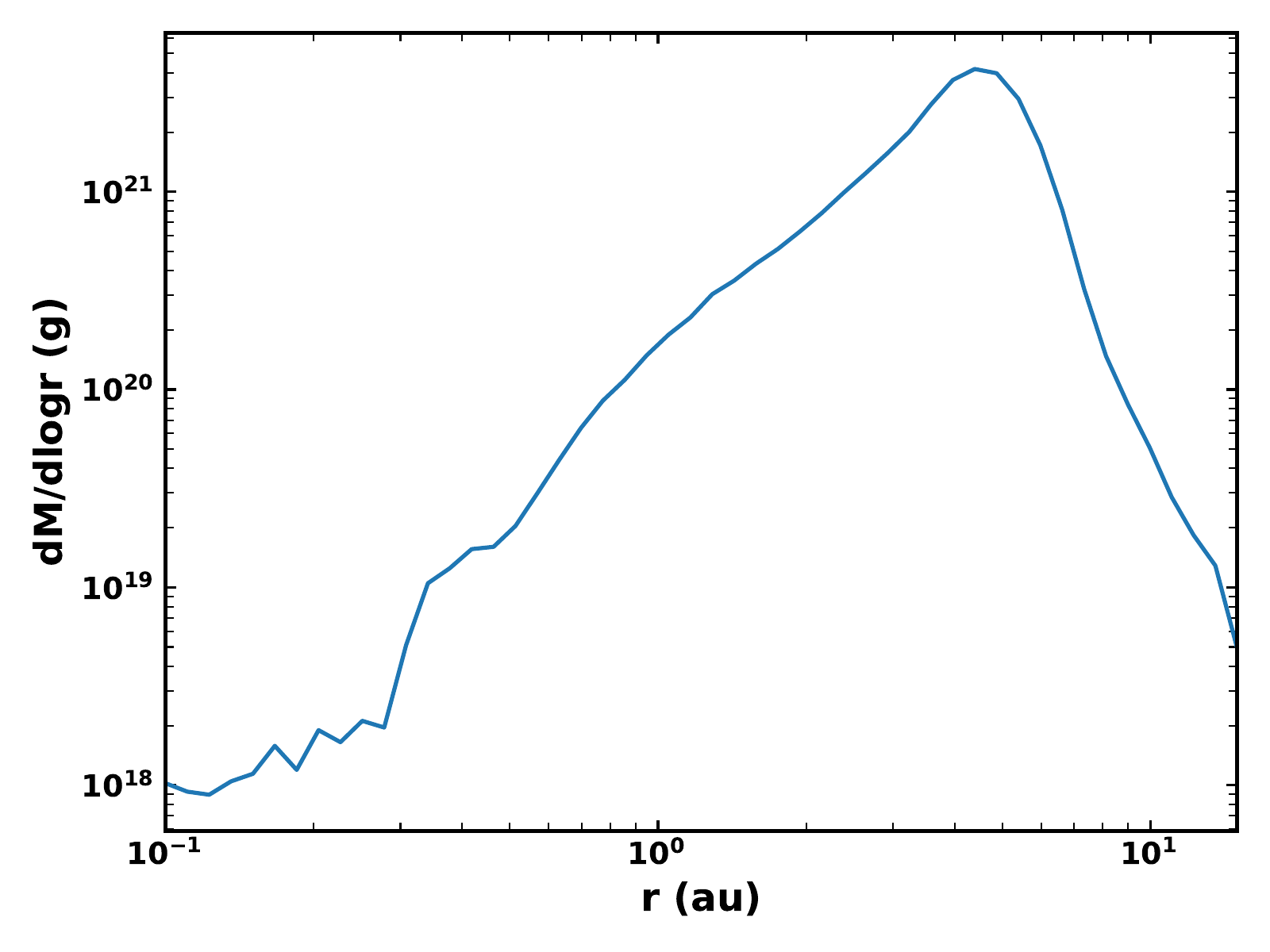}
	\caption{Mass input into our dust model as a function of heliocentric distance after weighting by $\epsilon = 5$ per cent and removing dynamically-dominated grains, summed over 100 Myr. Mass is distributed equally around the orbit in terms of mean anomaly for each combination of pericentre and eccentricity.}
	\label{fig:dMdlogr}
\end{figure}

The mass input from our fragmentation model as a function of heliocentric distance is shown in Figure~\ref{fig:dMdlogr}, which was found by distributing the mass equally around orbits for each ($q$, $e$) bin. The mass input peaks at 4.5~au, with a sharp drop-off further from the Sun. This is due to a balance between the fact that comet fragmentation is more likely closer to the Sun, and that comets move inwards from 5.2~au, such that some may fully disrupt before getting too close to the Sun. Further, the removal of the largest grains, which dominate the mass, is much more effective further out, where drag timescales are longer, which will shift the mass input towards smaller heliocentric distances. The eccentricity of cometary orbits means that a comet on a given orbit can produce dust at a range of heliocentric distances depending where around the orbit it fragments. The comets act as a distributed source of dust, with a mass input which is concentrated inside Jupiter's orbit, but continues outside Jupiter. \par
\begin{figure}
	\centering
	\includegraphics[width=\linewidth]{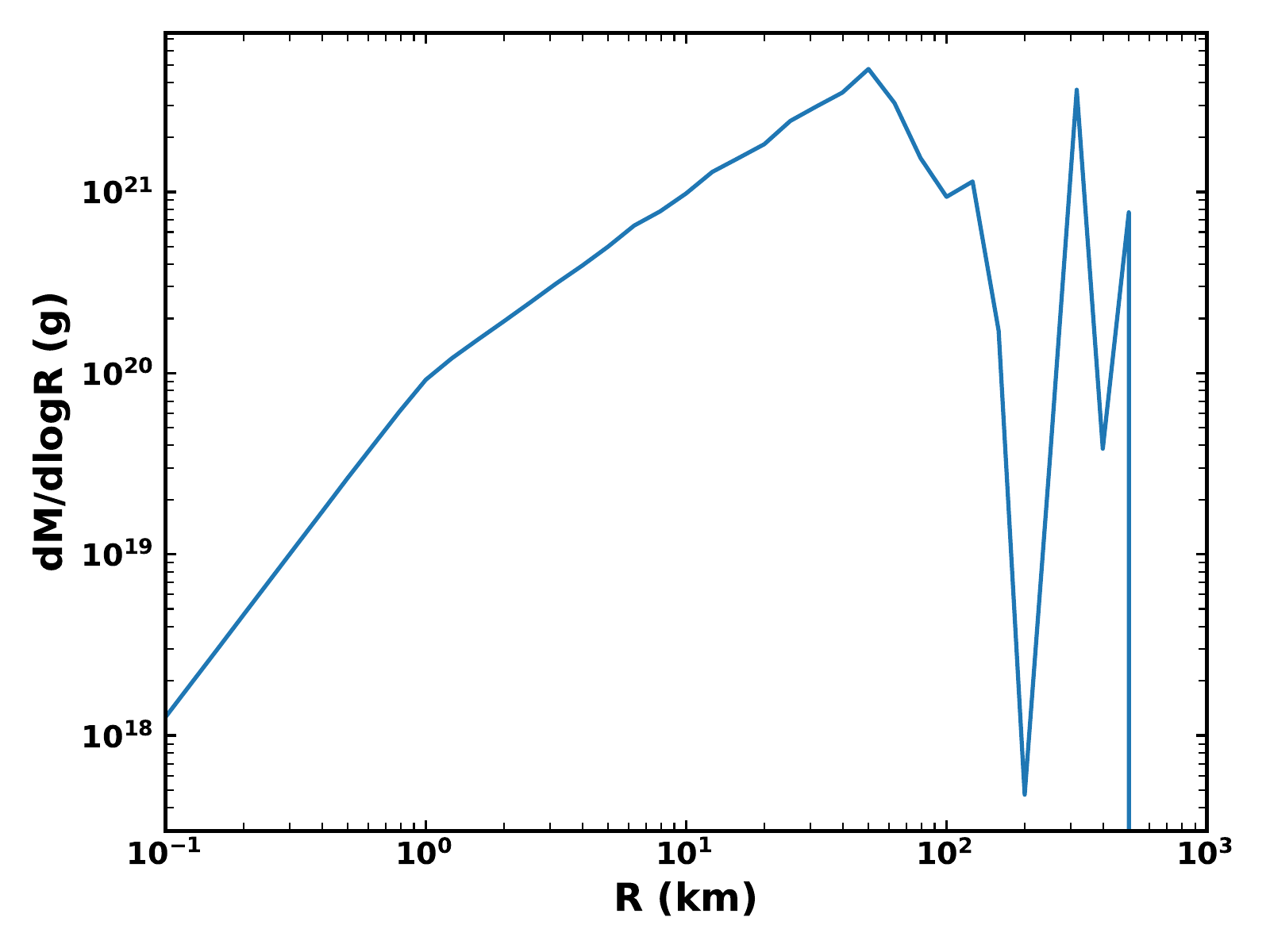}
	\caption{Mass of dust produced by comets of different initial sizes over 100~Myr which supplies the zodiacal cloud (i.e. excluding the dusty fragments that are dominated by dynamical evolution). The same as Figure~\ref{fig:dm_dlogR}, but weighted by a factor of $\epsilon = 5$~per cent, and removing dynamically-dominated dust grains. }
	\label{fig:dm_dlogR_nodyn}
\end{figure}

In Figure~\ref{fig:dm_dlogR} we showed which sizes of comet produced the most mass in fragmentations. However, this is slightly different from how much dust each size of comet produces which supplies the zodiacal cloud. Figure~\ref{fig:dm_dlogR_nodyn} shows the distribution of dust which is inputted into our kinetic model as a function of the initial size of the comet which produced it. This is essentially Figure~\ref{fig:dm_dlogR}, scaled by our efficiency $\epsilon$ of 5 per cent, and removing the dust which is assumed to stay with its parent comet. In both cases the mass input is dominated by comets of initial size $\sim 50$~km, and other than the absolute value, the distribution is very similar for $R \lesssim 50$~km. However, the contribution from $R > 200$~km comets is much more significant after the removal of dynamically-dominated grains. This is likely because these comets do not fully disrupt, such that they have longer lifetimes, and are more likely to survive to reach lower pericentres, where dust is dominated by drag and collisions. The sharp drop at $R \sim 200$~km is probably because the break in our input comet size distribution at $R = 150$~km is a minimum in terms of the mass in comets. Therefore, even including dynamical interactions, our conclusion remains that the mass input to the zodiacal cloud should be dominated by comets of initial radii $\sim 50$~km. \par

\subsection{Dust distribution}
\label{subsec:dustres}

Here we present the behaviour of dust produced by comet splittings as it evolves in our collisional evolution model, and the resulting radial and size distributions, as well as its variation in time.

\subsubsection{Optical depth}
\label{subsec:tau}
\begin{figure}
	\centering
	\includegraphics[width=\linewidth]{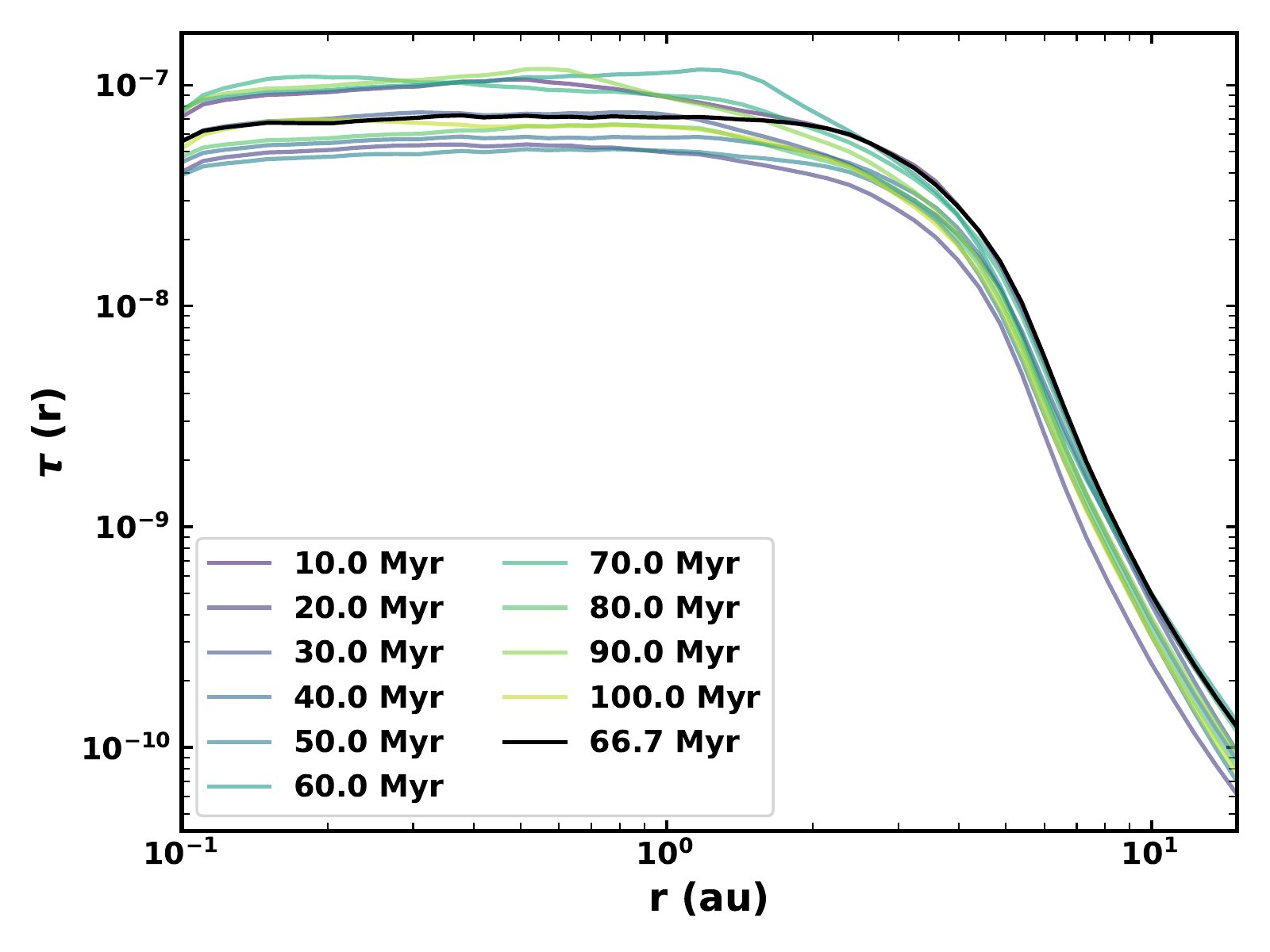}
	\caption{Radial profile of geometrical optical depth at different times for our best fit model, including 66.7~Myr (black) where it best fits the present-day zodiacal cloud as measured by COBE.}
	\label{fig:taur}
\end{figure}

The evolution of the radial profile of geometrical optical depth with time is shown in Figure~\ref{fig:taur}. We also show the radial profile at 66.7~Myr, when the model best fits the profile of the present-day zodiacal cloud, with a value of $7.1 \times 10^{-8}$ at 1~au and a radial slope of -0.34, which both agree well with the COBE/DIRBE measurements. The radial profile is relatively flat inside of 1~au, with a shallow negative slope out to 3~au. The comets act as a distributed source, such that the radial profile continues past $>10$~au, but drops off very sharply outside $\sim 4$~au. Such a sharp drop-off $> 4$~au is not seen in observations of solar system dust \citep[e.g.][]{Poppe19}. However, this is due to the presence of dust from sources other than JFCs that are not included here since they contribute little to the inner few au that is the focus of this work (see also Section~\ref{subsub:sources}). Figure~\ref{fig:taur} also highlights the variations of optical depth with time: the overall level varies depending on how many comets are being scattered in and how massive they are. The shape and slope can also vary based on where comets are depositing the most mass. For example, at 60~Myr a bump is seen at $\sim 1.5$~au, which is likely due to a massive comet depositing a lot of mass there. \par
\begin{figure}
	\centering
	\includegraphics[width=\linewidth]{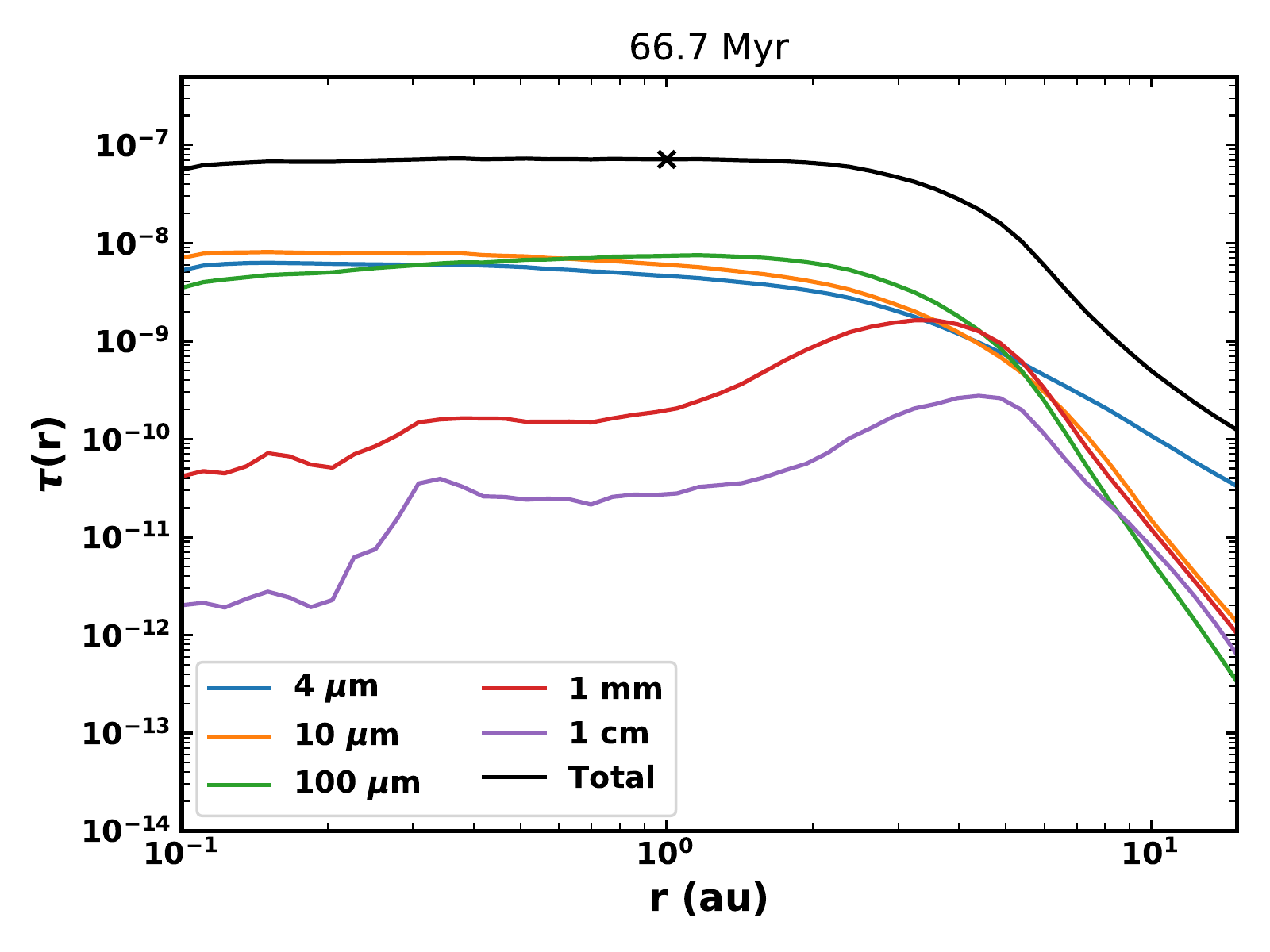}
	\caption{Radial profile of geometrical optical depth from our best fit model for dust grains in size bins centred on the values shown, along with the total optical depth (black). The optical depth of our current zodiacal cloud at 1~au is marked with an x.}
	\label{fig:tauDr}
\end{figure}

It should be noted that the cross-sectional area and optical depth will be dominated by smaller grain sizes, while the largest grains, which dominate the mass, will not contribute significantly to the brightness. The optical depth profiles of various grain sizes are shown in Figure~\ref{fig:tauDr}.  Dust grains which are dominated by P-R drag should migrate inwards to give a flat optical depth profile, while grains which are being destroyed by collisions are expected to be depleted closer in, where collisions are more frequent \citep[e.g.][]{Wyatt05,Rigley20}. The smaller grains which dominate the optical depth ($D \lesssim 100~\mu$m) have flat radial profiles due to P-R drag, causing the overall radial profile to be flat close in. The largest grains (mm- and cm-size), which supply mass to the interplanetary dust complex, are depleted by collisions closer in. The destructive collisions of these grains supply the smaller grains which dominate the zodiacal emission. This is why it is important to model the mass produced by collisions: it describes the shift of mass to smaller grain sizes, allowing us to explain the size distribution and radial profile of the dust. The importance of collisional evolution is discussed further in Section~\ref{subsec:comp_coll}. \par  

\subsubsection{Size distribution}
\label{subsec:sizedist_dust}
\begin{figure}
	\centering
	\includegraphics[width=\linewidth]{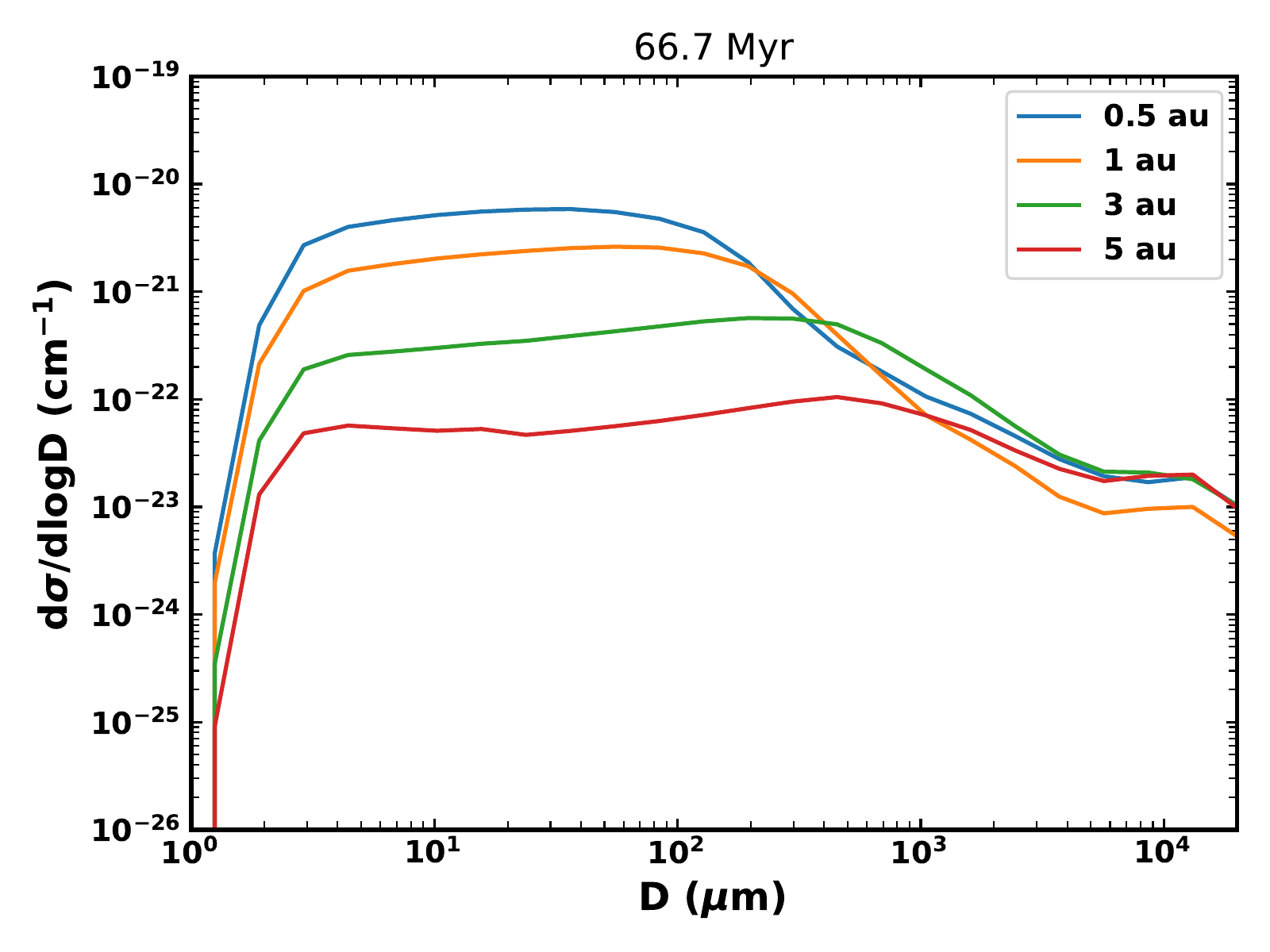}
	\caption{Distribution of the volume density of cross-sectional area per size decade at different heliocentric distances as a function of grain size for our best fit model.}
	\label{fig:dsig_dlogD}
\end{figure}

The size distribution of dust, expressed in terms of the distribution of cross-sectional area, is shown in Figure~\ref{fig:dsig_dlogD} at various locations. While the size distribution of dust input by comets (Figure~\ref{fig:dsig_dlogD_in}) has two peaks and breaks in the distribution, the size distribution of our dust model has been smoothed out. The difference between the size distributions which are input to and resulting from the kinetic model suggests that the final distribution is relatively insensitive to the specifics of the input distribution, and is primarily determined by collisions and drag. At 5~au, the size distribution has a subtle peak at 0.5~mm, which is likely due to the peak there in our input size distribution (Table~\ref{tab:input_dist}). Then moving inwards, the grain size dominating the cross-sectional area decreases. At 1~au the distribution of cross-sectional area peaks at $D\sim 60~\mu$m, though overall it is quite flat in the range $3 \lesssim D \lesssim 100~\mu$m. \par
\begin{figure}
	\centering
	\includegraphics[width=\linewidth]{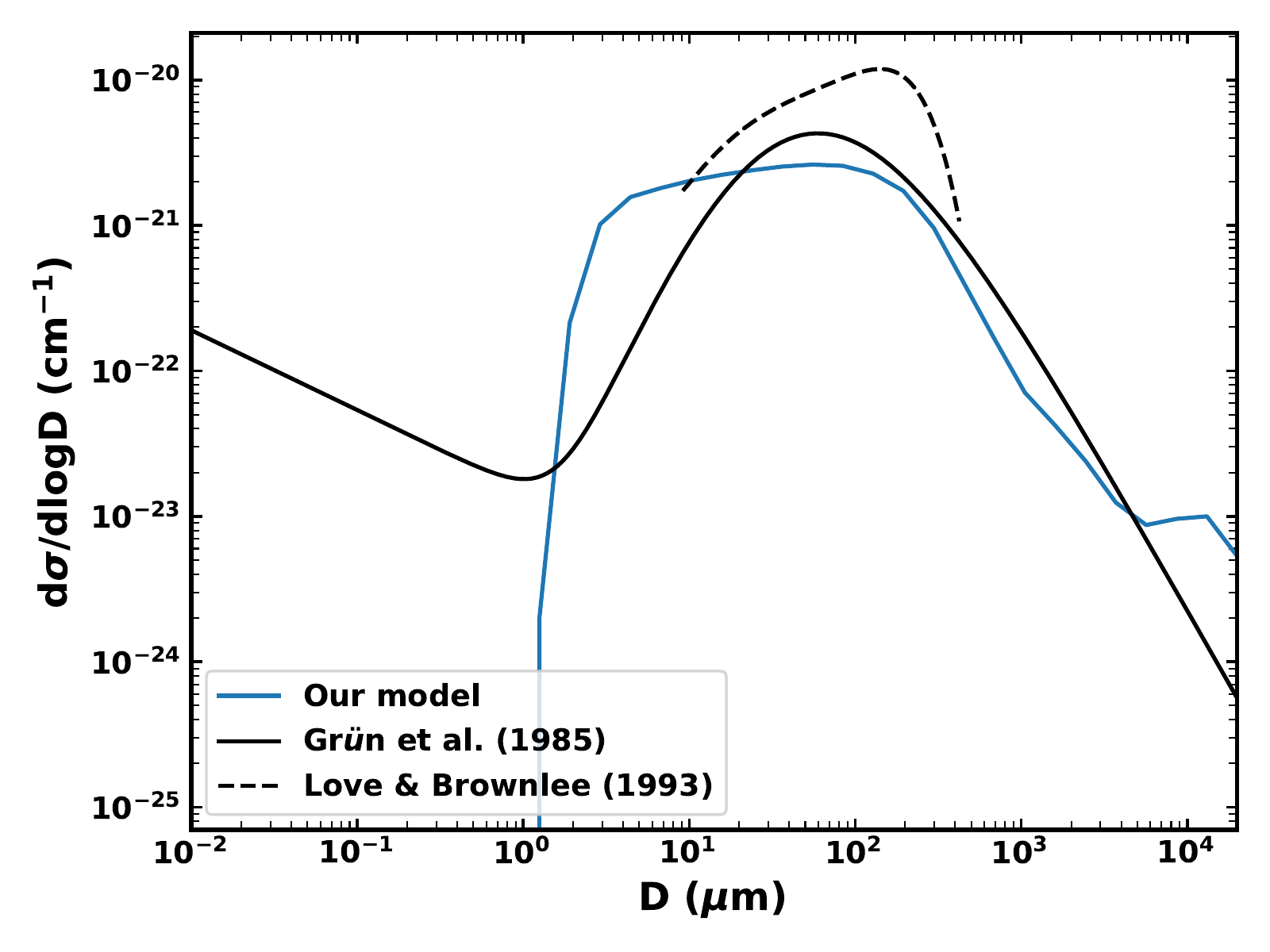}
	\caption{Distribution of cross-sectional area volume density per size decade at 1 au as a function of grain size from our model at 66.7~Myr (blue), the empirical model of \citet[][black, solid]{Grun85}, and measurements from the LDEF satellite \citep[black, dashed]{Love93}.}
	\label{fig:dsig_dlogDg}
\end{figure}

Figure~\ref{fig:dsig_dlogDg} compares the size distribution at 1~au from our model to two measurements of the flux of particles near Earth. \citet{Grun85} developed an empirical model for the distribution of interplanetary dust at 1~au based on lunar microcraters and in situ measurements of interplanetary dust. LDEF \citep{Love93} measured the distribution of dust accreted to Earth over a more limited size range. The cross-sectional area distributions of \citet{Grun85} and \citet{Love93} peak at grain diameters of 60 and 140~$\mu$m respectively, while at this time our distribution peaks at 60$~\mu$m. However, the shapes of the three distributions are slightly different: ours is relatively flat in the region of interest, whereas \citet{Grun85} is more peaked, and LDEF has a peak at a larger grain size. At larger grain sizes our model matches \citet{Grun85} quite well, but for $1 < D < 10~\mu$m we predict a lot more grains than the empirical model. However, it should be noted that the in situ measurements which the empirical model was fitted to cover grain sizes $D \leq 0.9~\mu$m and $D \geq 41~\mu$m, such that there are not direct observations of grains in the range where there is a discrepancy. 

\subsubsection{Variation of the zodiacal cloud}
\label{subsec:variation}
\begin{figure}
	\centering
	\begin{subfigure}[b]{\linewidth}
		\centering
		\includegraphics[width=\linewidth]{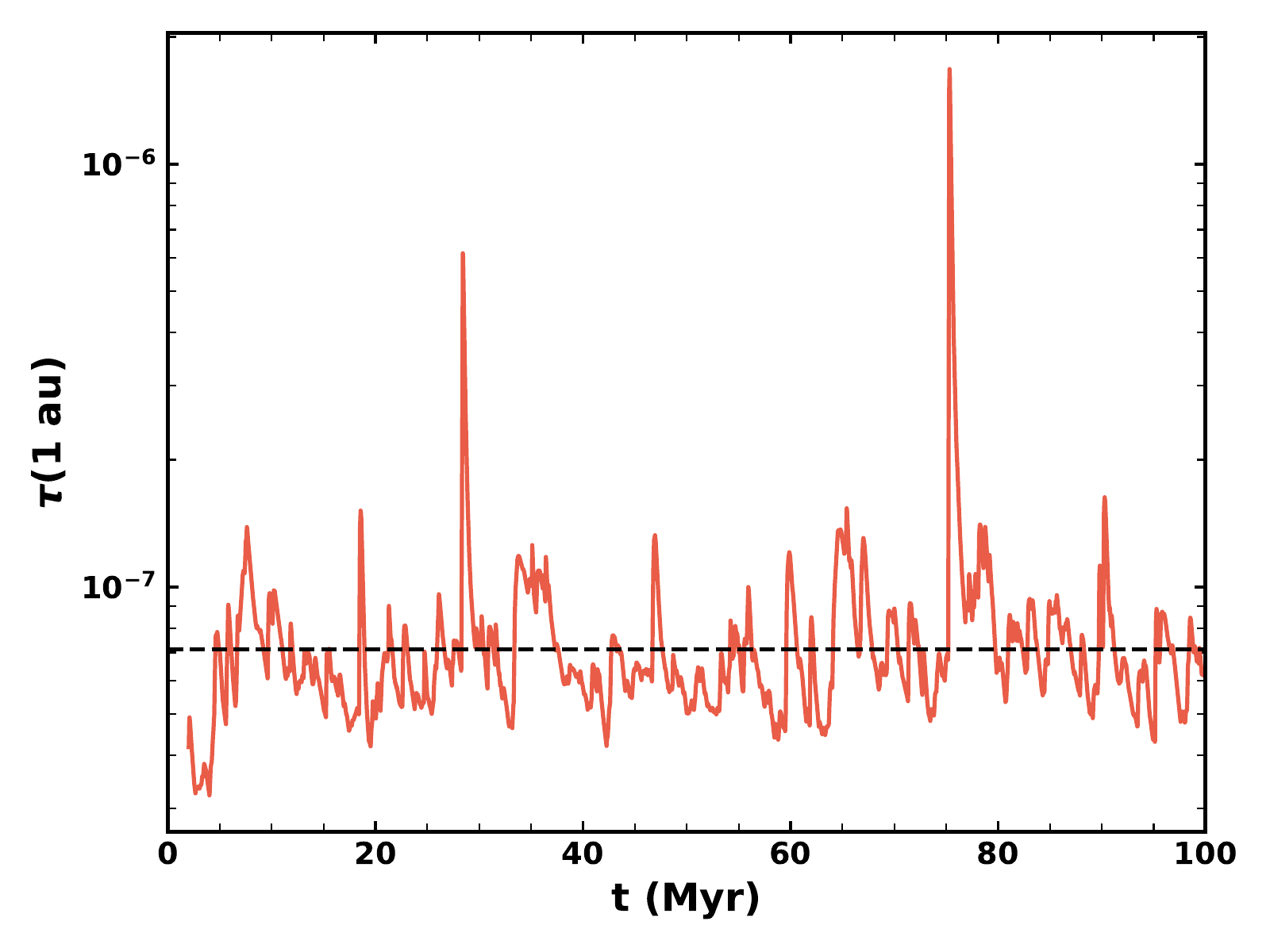}
		\label{fig:tauau}
	\end{subfigure}
	\begin{subfigure}[b]{\linewidth}
		\centering
		\includegraphics[width=\linewidth]{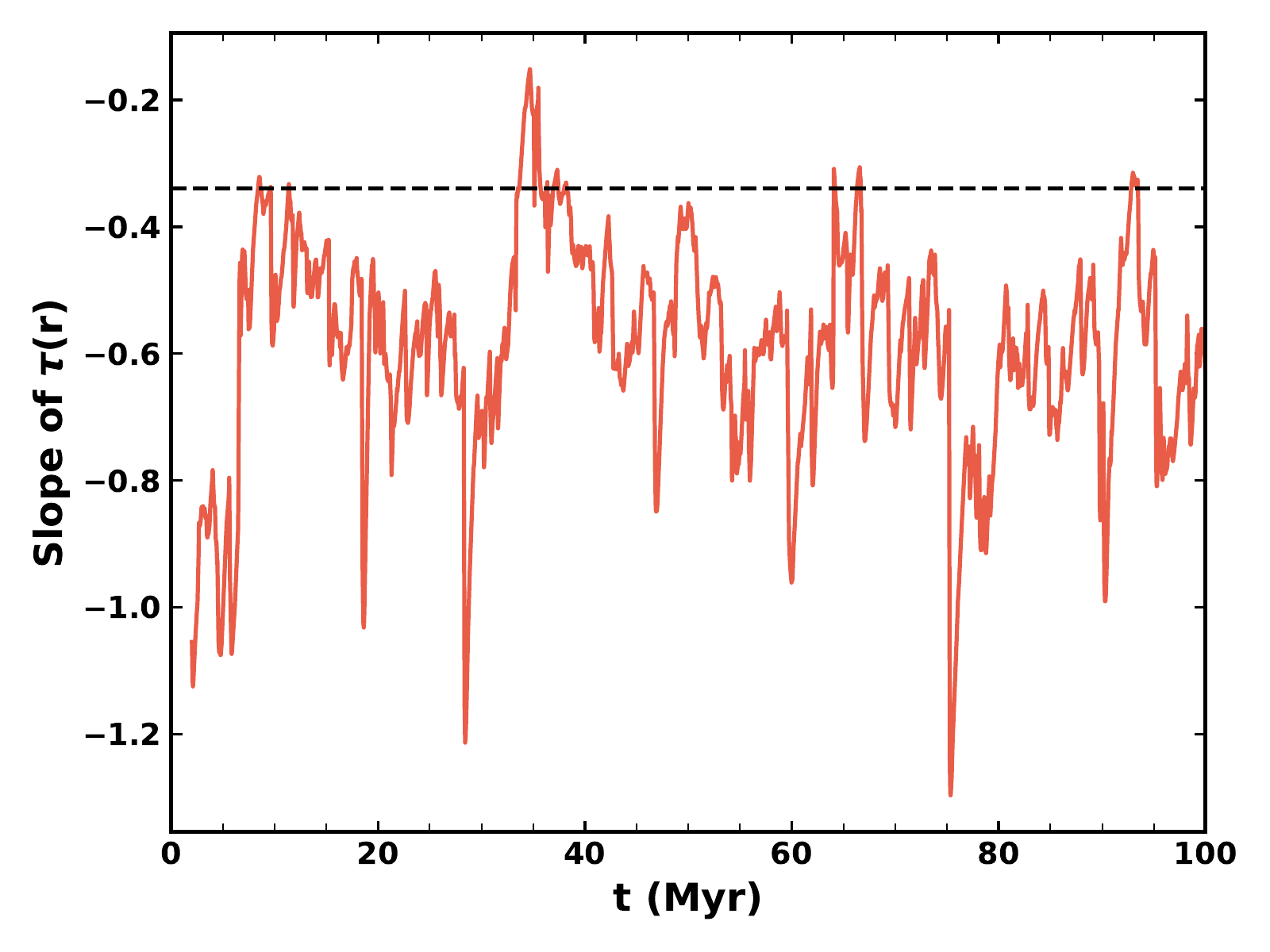}
		\label{fig:tauslope}
	\end{subfigure}
	\begin{subfigure}[b]{\linewidth}
		\centering
		\includegraphics[width=\linewidth]{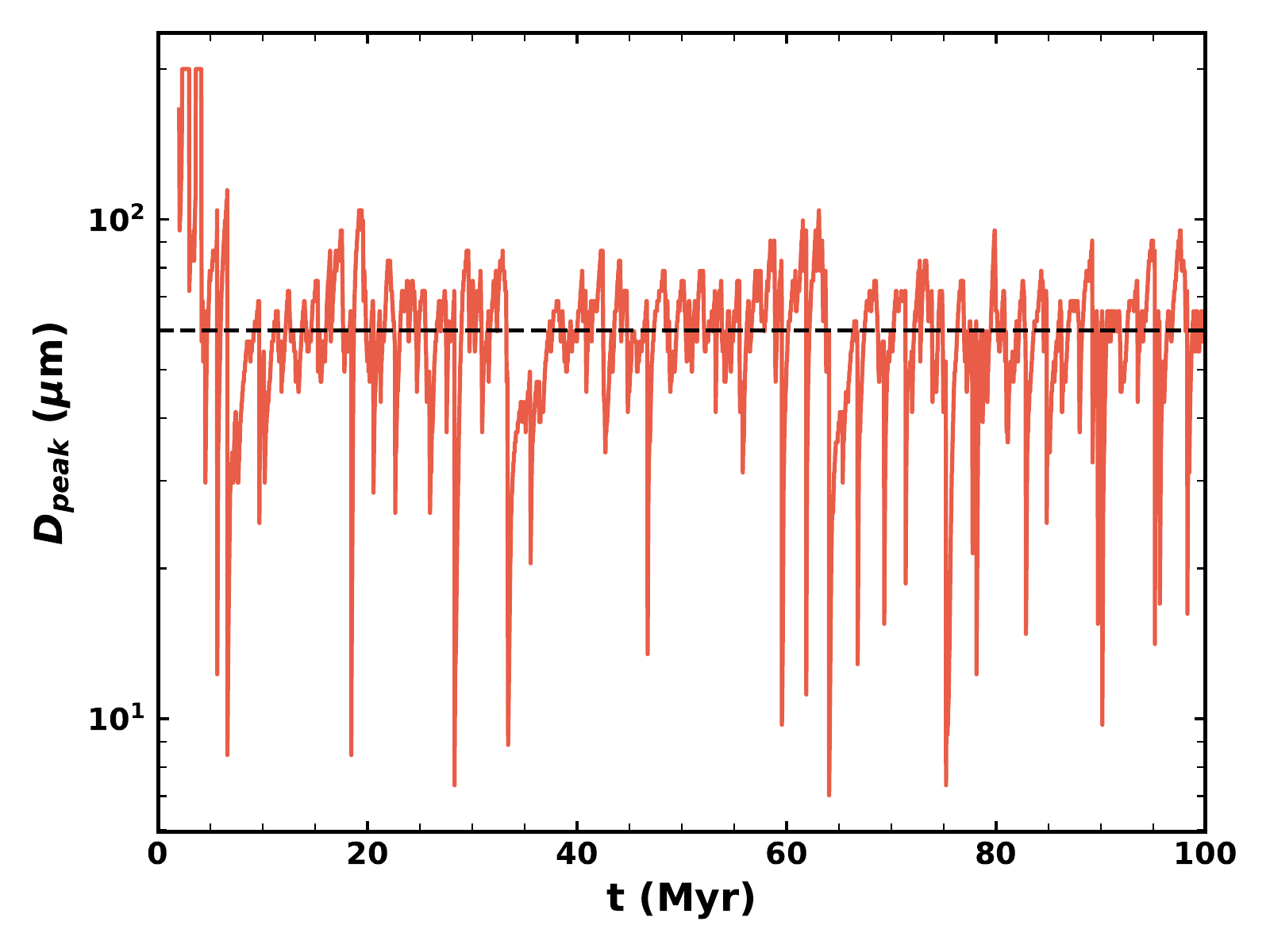}
		\label{fig:Dpeak}
	\end{subfigure}
	\caption{Evolution of our three 'observables' of the zodiacal cloud as a function of time. The values of the present zodiacal cloud are shown with dashed black lines. Top: absolute value of geometrical optical depth at 1~au as a function of time. Middle: slope of the radial profile of optical depth between 1 and 3 au as a function of time. Bottom: grain size which dominates the cross-sectional area of dust grains at 1 au as a function of time. }
	\label{fig:obs_evol}
\end{figure}

As mentioned previously, the distribution of dust in our model is stochastic, depending on which comets are scattered in and where they deposit dust. The variation of our three zodiacal cloud observables with time are shown in Figure~\ref{fig:obs_evol}: the absolute value of optical depth near Earth, $\tau(1~$au), the slope of the radial optical depth profile between 1 and 3 au, and the grain size at which the distribution of cross-sectional area peaks. This demonstrates the relationship between the different variables. While there is variation, the amount of dust as measured by the optical depth at 1~au is roughly constant, with a few large spikes. The radial slope fluctuates, which could be related to where mass is input by the comets: as shown in Figure~\ref{fig:taur}, a large comet depositing a lot of mass in one particular region can cause a change in the shape of the radial profile. The dominant grain size is also highly stochastic. There are a few events where there is a large spike in optical depth, which all correspond to a sharp drop in the slope and the dominant grain size, before evolving back to the quiescent level of dust. The rapid increase in the amount of dust present during a spike likely leads to a much higher collision rate. This would cause the production of small grains and destruction of large grains, shifting the size distribution towards smaller sizes. Similarly, collisions occur more frequently closer in due to the higher relative velocities of particles. The drop in the slope of optical depth could be explained by a higher production rate of small grains by collisions closer in; since it is the smaller grains which will dominate the optical depth, this affects the overall radial slope. \par
While a lot of stochasticity is seen in these variables, it should be noted that the overall level of variation in the optical depth is only a factor of a few, although one spike causes a jump of an order of magnitude. However, this depends on which dynamical paths comets are placed on; when the largest comets have longer dynamical lifetimes, much larger spikes in the optical depth can be seen. \par
\begin{figure}
	\centering
	\includegraphics[width=\linewidth]{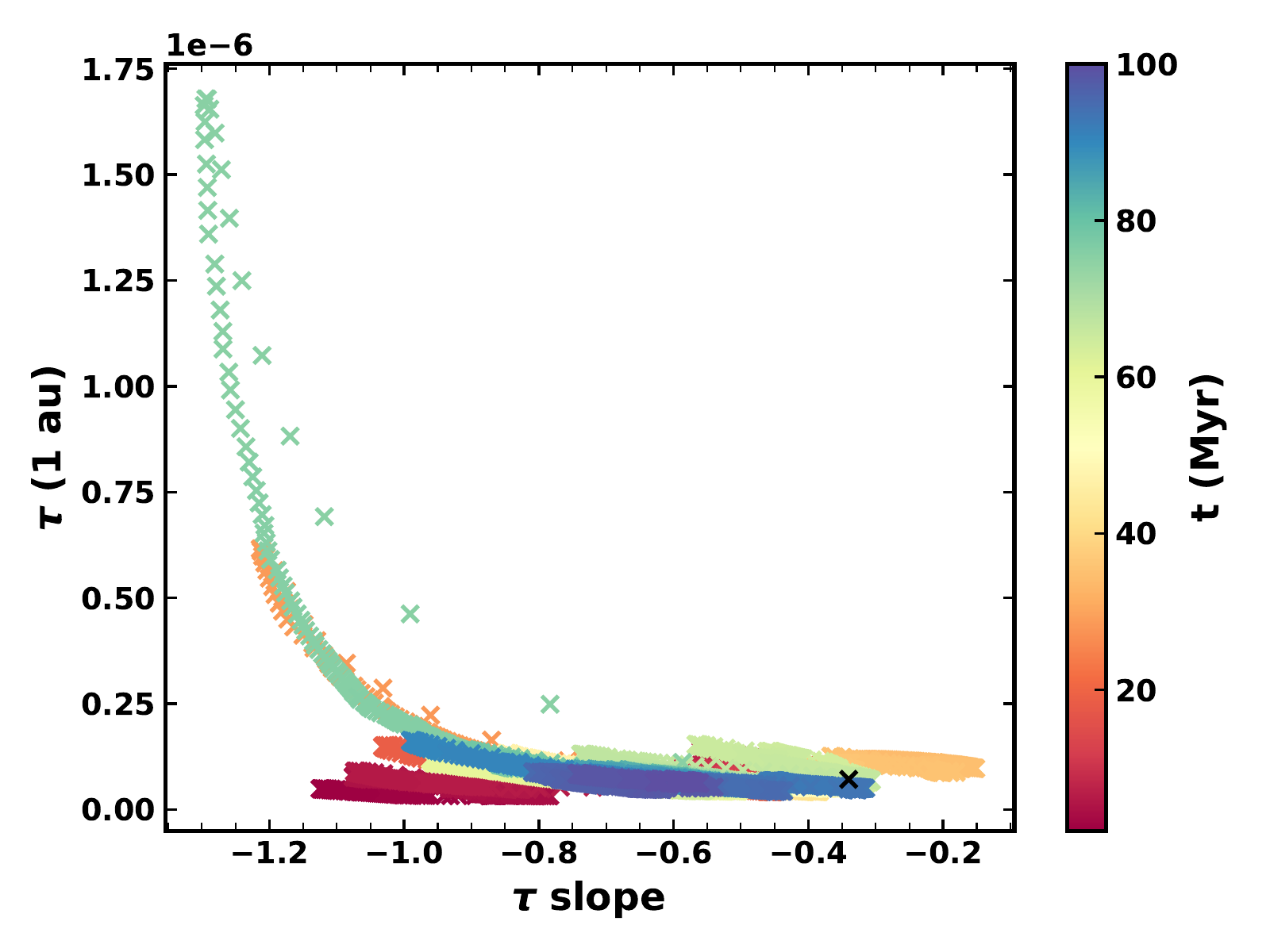}
	\caption{Evolution of the absolute value and radial slope of the geometrical optical depth of our zodiacal cloud model as a function of time. The present-day zodiacal cloud is marked with an x. }
	\label{fig:tauvslope}
\end{figure}

The correlation between the slope and absolute value of the optical depth can be seen more clearly in Figure~\ref{fig:tauvslope}, which shows the evolution of both variables against each other with time. This shows how the slope fluctuates back and forth at the quiescent level of dust depending where comets are inputting mass. Spikes in the amount of dust cause a sharp drop in the radial slope before it returns to the previous level. \par
\begin{figure}
	\centering
	\includegraphics[width=\linewidth]{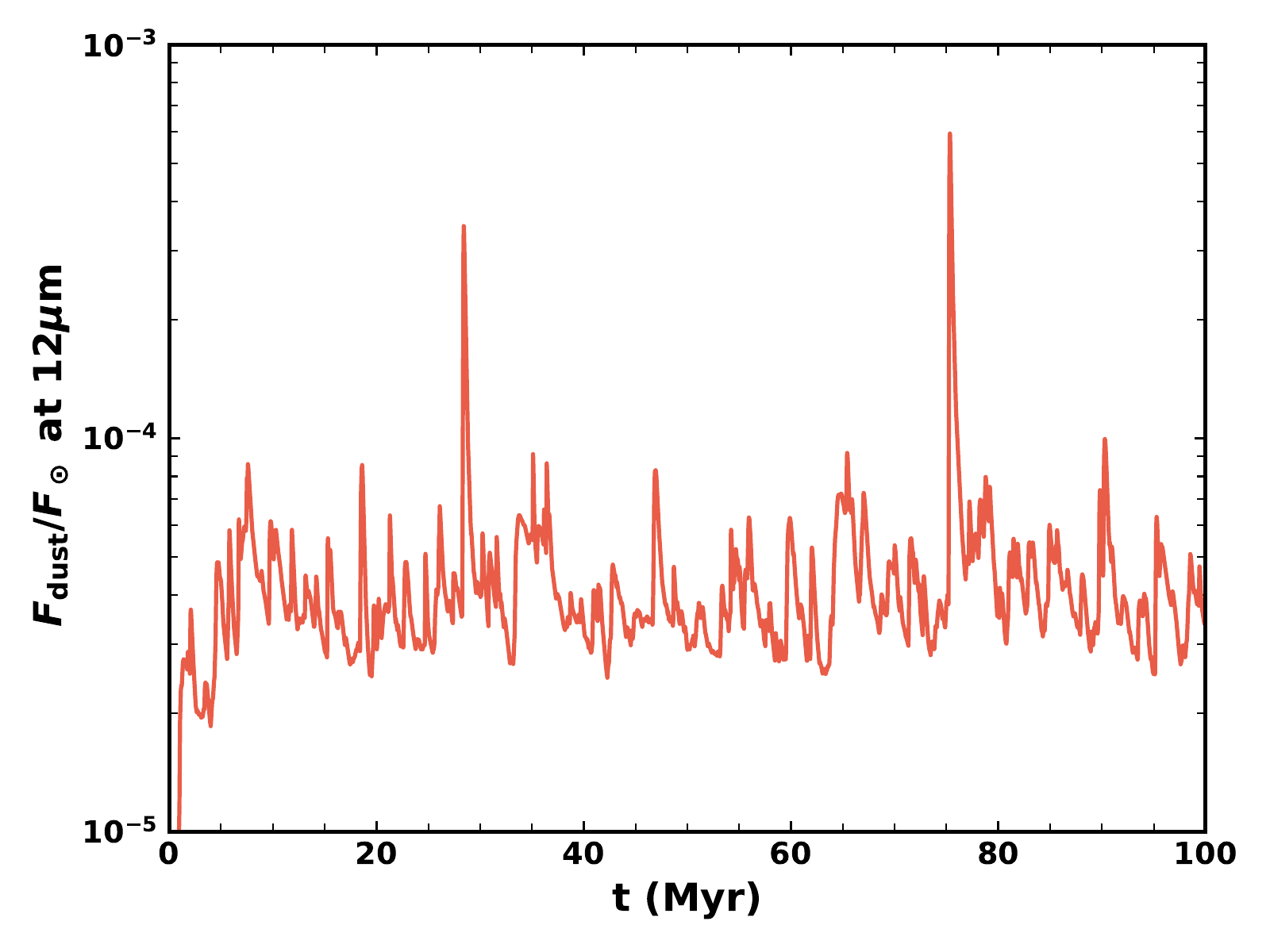}
	\caption{Evolution of the fractional excess of $12~\mu$m emission relative to the Sun which would be seen by a distant observer with time. }
	\label{fig:F12}
\end{figure}

The emission a distant observer would see from the zodiacal cloud will also be highly variable as a result of the stochasticity. Using realistic optical properties (see Section~\ref{subsec:dust_sizedist}), we calculated the emission which would result from our dust model. The fractional $12~\mu$m excess vs. time is shown in Figure~\ref{fig:F12}. This is stochastic and follows the same trends as the overall level of optical depth (Figure~\ref{fig:obs_evol}, top). The $12~\mu$m excess at our best fit time is $4.1 \times 10^{-5}$, with a total cross-sectional area of $8.7 \times 10^{20}$~cm$^2$. However, spikes in the level of dust can cause an excess as high as $6 \times 10^{-4}$, approaching levels detectable with an interferometric instrument such as the Large Binocular Telescope Interferometer \citep{Hinz16}. \par 

\subsubsection{Spike event}
\label{subsec:spike}
\begin{figure}
	\centering
	\begin{subfigure}[b]{\linewidth}
		\centering
		\includegraphics[width=\linewidth]{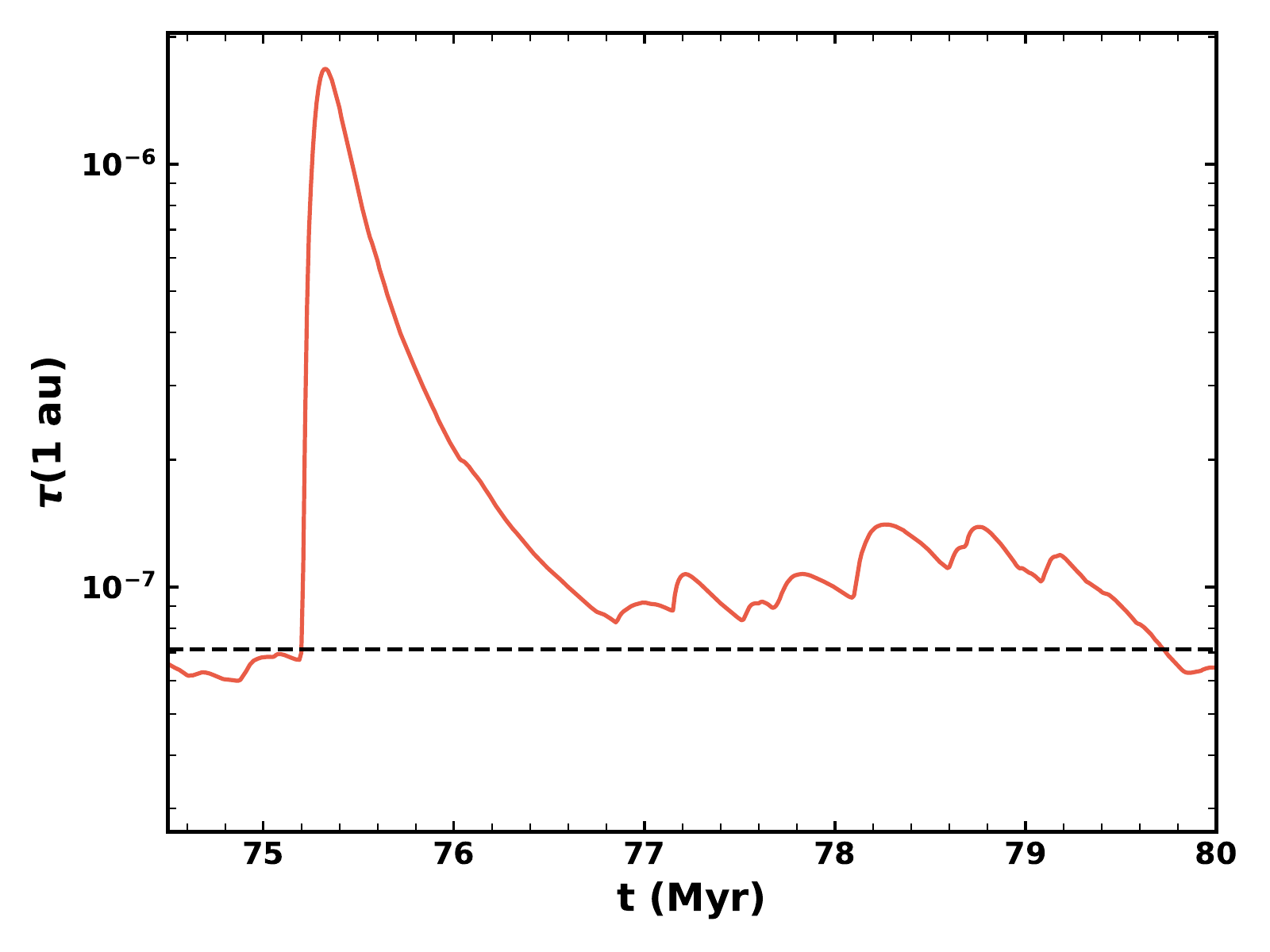}
	\end{subfigure}
	\begin{subfigure}[b]{\linewidth}
		\centering
		\includegraphics[width=\linewidth]{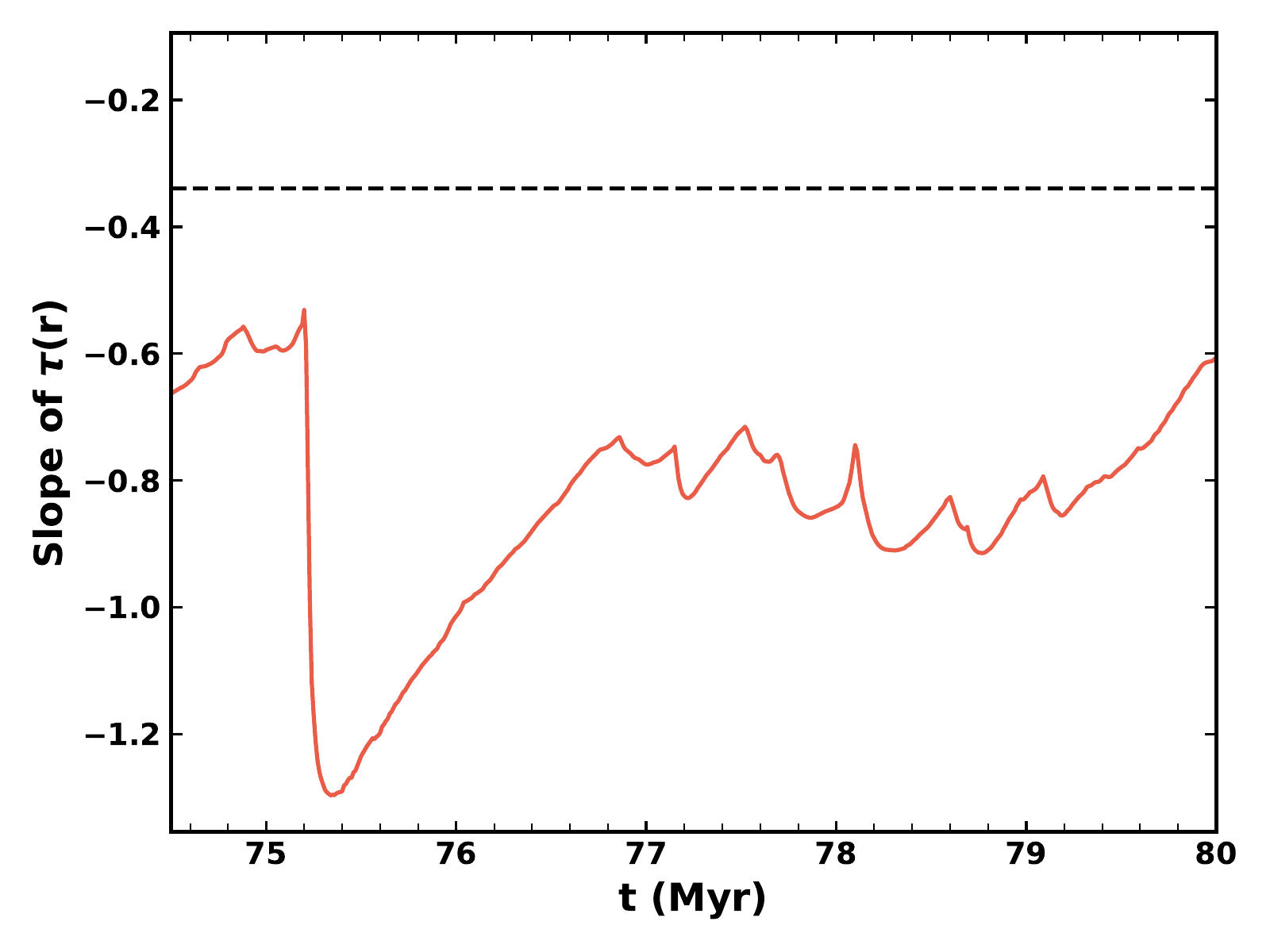}
	\end{subfigure}
	\begin{subfigure}[b]{\linewidth}
		\centering
		\includegraphics[width=\linewidth]{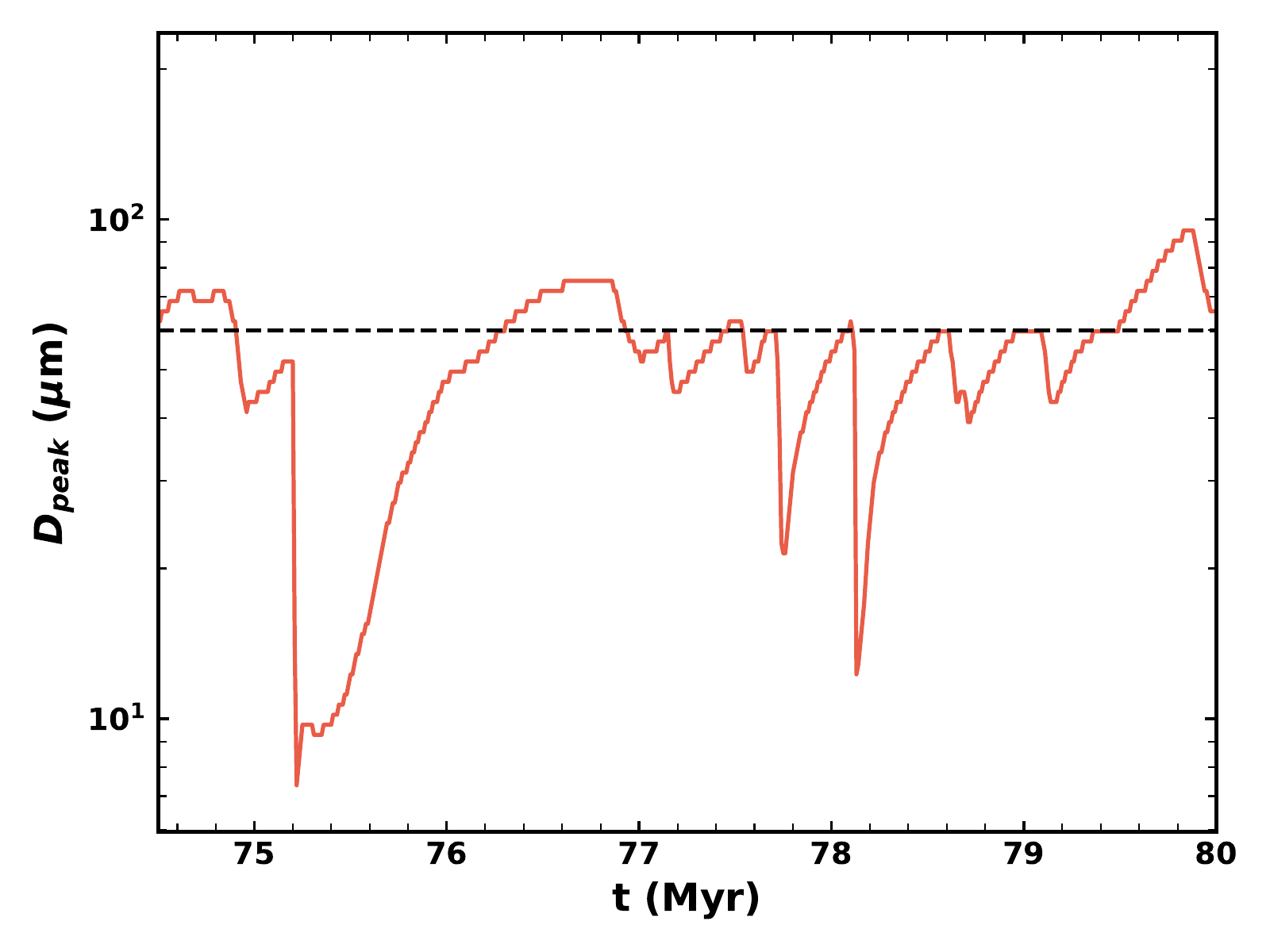}
	\end{subfigure}
	\caption{Evolution of our three 'observables' of the zodiacal cloud as a function of time, the same as Figure~\ref{fig:obs_evol}, but zoomed in on a spike in the amount of dust present. The values of the present zodiacal cloud are shown with dashed black lines.}
	\label{fig:spike}
\end{figure}

As suggested in Section~\ref{subsec:variation}, large comets cause occasional spikes in the level of dust, which correspond to a sharp drop in the radial slope, and a drop in the dominant grain size (see Figure~\ref{fig:obs_evol}). The largest spike occurs at 75.3 Myr, in which the level of dust jumps by an order of magnitude. The three observables are shown again in Figure~\ref{fig:spike}, zoomed in on the evolution of this spike. The optical depth (Figure~\ref{fig:spike}, top) shows that the spike in the level of dust decays after around 1.5~Myr. The slope takes a similar amount of time to return to its previous value. \par
By considering which large ($R \geq$ 100~km) comets are scattered into the inner solar system, this spike may be attributable to a particular comet. It might be expected that the largest spike in mass would be caused by the largest comet scattered in. However, due to the fact that $\gtrsim 100$~km comets will not lose all of their mass to fragmentations in general, the main factor determining whether large comets create massive spikes in the amounts of dust is the length of the dynamical path they are on. For example, there is one 501~km comet, and two 398~km comets scattered in during our simulation. However, these are all on dynamical paths which only spend $<40,000$~yr in the inner solar system. They therefore do not cause particularly large spikes in dust, as they are not present for long enough to lose much mass. The longest-lived large comet is a 125~km comet which is scattered in at 28.2~Myr, and has a dynamical lifetime of 284,700~yr. This seems to correspond to the peak in optical depth which occurs at 28.45~Myr, as while this comet is smaller it has a lot of time in the inner solar system to produce mass. The effects of this particular comet last until around 29~Myr, long after it has left the inner solar system. There is another large comet with a radius of 316~km scattered in at 74.6~Myr with a dynamical lifetime of 108,300~yr. This seems to correspond to the largest spike in optical depth at 75~Myr. Massive comets may therefore have lasting effects on the distribution of zodiacal dust if they spend long enough inside Jupiter's orbit.   \par
This highlights that very large comets may cause huge spikes in the levels of dust, but only if their dynamical lifetimes in the inner solar system are long enough. The highly stochastic nature of dynamical interactions means this may happen occasionally, but often large comets may have shorter lifetimes and therefore not contribute huge amounts of dust. \par

\section{Discussion}
\label{sec:discussion}

\subsection{Stochasticity}
\label{subsec:stochasticity}
In Section~\ref{subsec:variation} we showed that a cometary contribution to the interplanetary dust complex will be highly variable, depending on the sizes of comets which are scattered in and their dynamical lifetimes. In particular, very large comets can cause big spikes in the amount of dust if their dynamical lifetimes are long enough (Section~\ref{subsec:spike}). \par 
In comparison to our model, the parameters of the present-day zodiacal cloud seem to be close to the quiescent levels of optical depth. The radial slope is at the highest end of the range of values in our model. In principle this could suggest that we have not recently had a very large comet with a long dynamical lifetime in the inner solar system to cause a spike. Arguably it may be possible to alter the free parameters of our dust model so that the present-day is in the middle of a spike rather than close to the steady state level. However, it is very difficult to shift the parameters such that the radial slope becomes  more positive. Regardless, if comets dominate the mass input to the zodiacal cloud, it is likely that the zodiacal light has been highly variable over the history of the solar system. \par

\subsection{Other free parameters}
\label{subsec:freepars}
Two free parameters of the model which were not discussed in Section~\ref{subsec:fitting} are the maximum grain size $D_\mathrm{max}$ and the dust density $\rho$. IDPs are generally assumed to have densities $\sim$2\gcm, and the value we assume is 1.9\gcm~based on \citet{Rotundi15}. However, \citet{Fulle16a} measured a density of 0.795\gcm~from Rosetta. We also tried some runs with a density of 0.795\gcm, with the main difference being that the size distribution moved towards much larger grain sizes, giving a poorer fit to the size distribution. However, measurements of IDPs are done in terms of particle mass. \citet{Grun85} and \citet{Love93} assume particle densities of 2.5\gcm, such that the lower density distribution may fit observations better if we considered the size distribution in terms of particle mass, as opposed to particle size. Otherwise the density of dust grains should not affect the model too much. \par
We also tried varying the maximum grain size, $D_\mathrm{max}$, and found that 2~cm best fit the observed properties of the zodiacal cloud. This is the value used in Sections~\ref{sec:fitting} and \ref{sec:results}. With a smaller $D_\mathrm{max}$ of 2~mm, it is possible to fit the radial profile of the zodiacal cloud with our model. However, the size distribution is a worse fit, as the cross-sectional area peaks at smaller grain sizes ($D\sim 20-30~\mu$m). This may be due to collisions of cm-size grains supplying smaller grain sizes. With a much larger (m-size) maximum size, the mass accumulates in the largest particles without being destroyed in collisions or migrating inwards. The largest grain size will dominate the overall mass, so this means that the mass increases such that a quasi-steady state cannot be reached within 100~Myr. Increasing $D_\mathrm{max}$ by a factor of three to 6~cm, the total mass reaches a steady state. However, the radial profile is always far too steep and is not able to match the observed distribution. \par
Observations suggest that cometary dust is dominated by grains of mm to cm-size \citep[e.g.][]{McDonnell93,Green04,Reach07,Rotundi15}, and so ideally $D_\mathrm{max}$ should be at least cm-size.  Further, observations of splitting events suggest that large (>m-size) fragments will have a shallower size distribution \citep{Makinen01,Fuse07,Fernandez09}. These fragments often disappear on short timescales, such that they may undergo further fragmentations themselves. More recently, fireball observations suggest a lack of JFC material in the cm- to m-size range near Earth \citep{Shober21}. Given that fragments $\gtrsim$~cm-size may be able to disrupt via mechanisms other than mutual collisions, and will not contribute significantly to optical depth, we choose to set $D_\mathrm{max}$ to 2~cm. \par 
As discussed in Section~\ref{subsec:dust_sizedist}, there are various size distributions we could have chosen for the dust produced by comets. We chose the size distribution found by \citet{Reach07} when studying images of SPC debris trails, which is a broken power law with three different slopes depending on the grain size. However, we could have instead chosen to use a distribution based on fragments of comet splittings. For example, \citet{Makinen01} found a distribution of fragments with a slope -2.7 fit the splitting of comet C/1999 S4 (LINEAR). By converting the magnitude of 19 fragments of comet 57P, \citet{Fernandez09} found they had a rather shallow slope of -2.3. Meanwhile, the fragmentation of comet 73P/Schwassmann-Wachmann 3 has been widely studied \citep{Boehnhardt02,Sekanina07}. \citet{Fuse07} measured the size distribution of a group of 54 fragments, and derived a slope of -2.1. This suggests that the large fragments of comet splitting may have a different distribution than the dust. While these size distributions are quite different, the distribution resulting from our kinetic model differs significantly from the input distribution. We therefore expect that it is relatively insensitive to the details of the input distribution, and the important part of the size distribution should be which sizes dominate the mass and cross-sectional area. The exact distribution of dust produced by fragmentations is highly uncertain, but hopefully dust trails give a good approximation. \par

\subsection{Dominant comet size}
We showed in Section~\ref{subsec:frag_results} that $R \sim 50$~km comets should dominate the overall mass created by comet fragmentation (Figure~\ref{fig:dm_dlogR}). This conclusion was not changed after removing dust grains which are dominated by dynamical interactions (Figure~\ref{fig:dm_dlogR_nodyn}). While larger ($R > 100$~km) comets will dominate when they are present, they are very rare, and do not lose all of their mass to fragmentations, such that they contribute a smaller fraction of the overall mass to the interplanetary dust complex. Conversely, comets 10s of km in size are always present, and some will fully disrupt. The largest comets seen today have $R \sim 30$~km: the largest JFCs are 29P/Schwassmann-Wachmann 1 (30.2~km) and C/2011 KP36 (27.5~km) (JPL Small-Body Database\footnote{https://ssd.jpl.nasa.gov/sbdb\_query.cgi}). As $R_\mathrm{max} > 50$~km 13 per cent of the time and we clone particles every 12,000~yr, Figure~\ref{fig:maxR} shows that we estimate comets $\gtrsim 50$~km should be scattered in on average every 100,000~yr. The dust from large comets will last for longer, such that the present-day zodiacal cloud should be dominated by the dust from these comets, despite no comets so large being seen by us in the last $\sim 200$~yr.

\subsection{Historical brightness}
\label{subsec:history}
We have assumed a constant scattering rate of comets into the inner solar system, but this is not true over the history of the solar system. While we have shown that stochastic variations should be important over timescales $\sim 100$~Myr, variations in comet input must also be taken into account. For example, the Nice model \citep{Tsiganis05,Gomes05} suggests that there was a phase when many more comets were scattered into the inner solar system at early times. Therefore, the historical brightness of the zodiacal cloud should vary both due to the stochasticity of comets which are scattered in, and also due to variations in the overall influx of comets caused by processes such as dynamical instability. \par

\subsection{Model parameters}
\label{subsec:bestfitpars}
In Section~\ref{subsec:fitting} we fitted the free parameters of our model to match the present-day zodiacal cloud. These parameters are related to the collisional behaviour of dust grains ($Q_0$, $a$, and \ar), and the fraction of mass lost in a comet fragmentation which becomes dust grains ($\epsilon$). \par
\begin{figure}
	\centering
	\includegraphics[width=\linewidth]{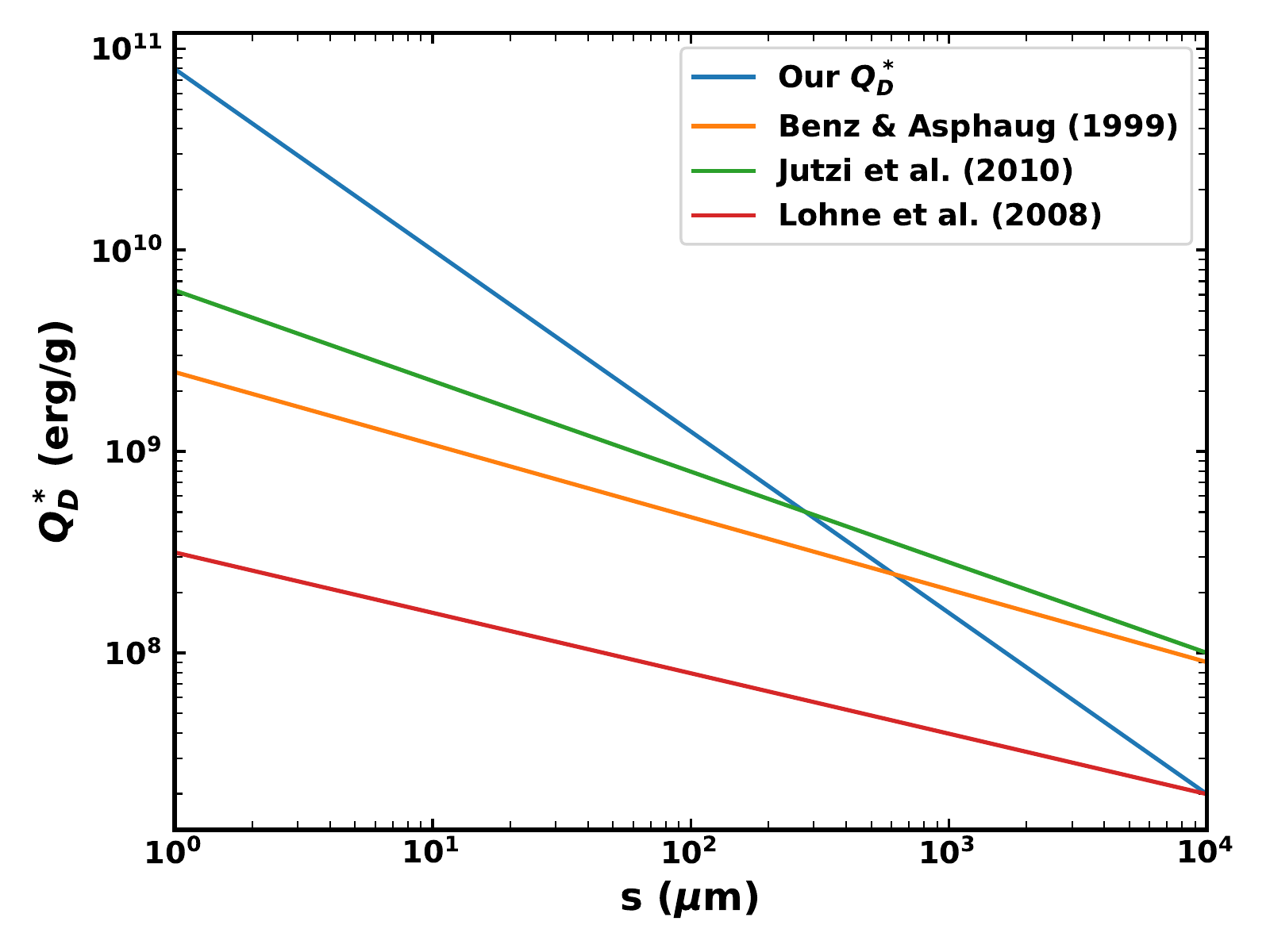}
	\caption{Collisional strength, \QD, of dust grains of different sizes. Four different prescriptions are compared: our best fit model, \citet{Benz99}, \citet{Jutzi10}, and \citet{Lohne08}.}
	\label{fig:QD}
\end{figure}

Laboratory experiments \citep{Fujiwara86} suggest that the redistribution function of collisional fragments, \ar, has a possible range of $2.5 \lesssim~$\ar$~\lesssim 4.0$. Thus our best fit value of 3.75 is reasonable. The collisional strength of dust grains, however, is poorly constrained. Our final values were a normalisation of $Q_0 = 2.0 \times 10^7$~erg/g, and a slope of $a = 0.9$. In Figure~\ref{fig:QD} we compare our model parameters to other prescriptions for the collisional strength of particles. While the normalisation $Q_0$ is quite typical, the slope $a$ is steeper than previous models in the literature. This means that we require the smallest grains ($\mu$m-sized) to be about ten times stronger than other models in the literature. However, the collisional strength is not well known for dust grains, and usually only characterised for particles $> 10~$cm-sized. \par
The final parameter we fitted was the fraction of mass lost in a fragmentation event which becomes dust, $\epsilon$. Our best fit value was 5 per cent. Implicit in this assumption is that fragmentation of comets will also produce m-size fragments which remain without producing dust themselves. However, the exact fraction of mass becoming dust is not well constrained. For example, photometric observations of the disruption of comet C/1999 S4 (LINEAR) \citep{Farnham01} suggested that most of the mass was hidden in fragments 1~mm to 50~m in size. The mass of $<$~mm dust observable was $3 \times 10^8$~kg, and the comet nucleus was estimated to be $4 \times 10^{11}$~kg, suggesting only 0.1 per cent of the initial comet mass was put into dust grains $<$~mm-sized. \par
It should be noted that there is further uncertainty on the value of $\epsilon$ derived from our model due to the fact that we have assumed there are 58 visible JFCs in the range $1 \leq R \leq 10$~km, which is probably a lower limit on the number of comets in this range, as the observed sample is likely incomplete. For example, \citet{DiSisto09} estimated that there are 107 visible JFCs with $R > 1$~km and $q < 2.5$~au, which would increase the normalisation of our size distribution, which scales linearly with the number of visible 1-10~km comets. To compensate for this we would expect the best fit model to require a value of $\epsilon$ that is decreased by a corresponding amount. There may therefore be a factor $\sim 2$ uncertainty in our best value of $\epsilon$. Increasing the normalisation of the mean size distribution would also increase the probability of large ($> 100$~km) comets being scattered in, which may lead to an increase in how frequently large increases in dust mass occur.  \par

\subsection{Model limitations}
\label{subsec:limitations}

\subsubsection{Vertical distribution}
\label{subsub:vertical}
One of the main limitations of our model is that because we are using a kinetic model which assumes a uniform inclination distribution, we cannot follow the evolution of particle inclinations. While this should not be too important for collisions and P-R drag, \citet{Nesvorny10} showed that after being released from a comet, JFC particles are scattered by Jupiter, such that their inclination distribution is broader than that of JFCs. Therefore, by not modelling the dynamical interactions after dust is released from a comet, we are unable to study the inclination distribution of dust. One key metric which many models use to compare to the zodiacal cloud is the profile of thermal emission with ecliptic latitude. We are unable to compare with IRAS based on our model. \par

\subsubsection{Fragmentation prescription}
\label{subsub:frag}
We have modelled comet fragmentations using the model of \citet{DiSisto09}, who modelled comets with $q < 2.5$~au and $1 \leq R \leq 10$~km based on the need for a relatively complete sample of observations to compare to. We have extrapolated this model outside the region of parameter space it was fitted to in terms of both pericentre and comet size. We have extrapolated the pericentres fragmentation occurs at out to 5.2~au. While we would expect the probability of fragmentations to continue decreasing at larger pericentres, there could be a change in fragmentation rate e.g. at 2.5~au due to the onset of water sublimation and increased cometary activity. Fragmentations have been observed much further from the star than 2.5~au \citep[e.g.][]{Fernandez05}, but their frequency is not well constrained. \par
Further, comets much smaller or larger than those modelled by \citeauthor{DiSisto09} may fragment at different rates. According to the model, the fraction of a comet's mass lost in a fragmentation event is inversely proportional to its radius (equation~\ref{eq:sR}). A 10~km comet loses 0.7 per cent of its mass in a fragmentation. This means that very small, sub-km comets lose most of their mass in a single event: a 0.1~km comet will lose 70 per cent of its mass in a single fragmentation, such that it will only survive two fragmentation events. Conversely, larger ($>10$~km) comets require many fragmentations to lose all of their mass. \par
\begin{figure}
	\centering
	\includegraphics[width=\linewidth]{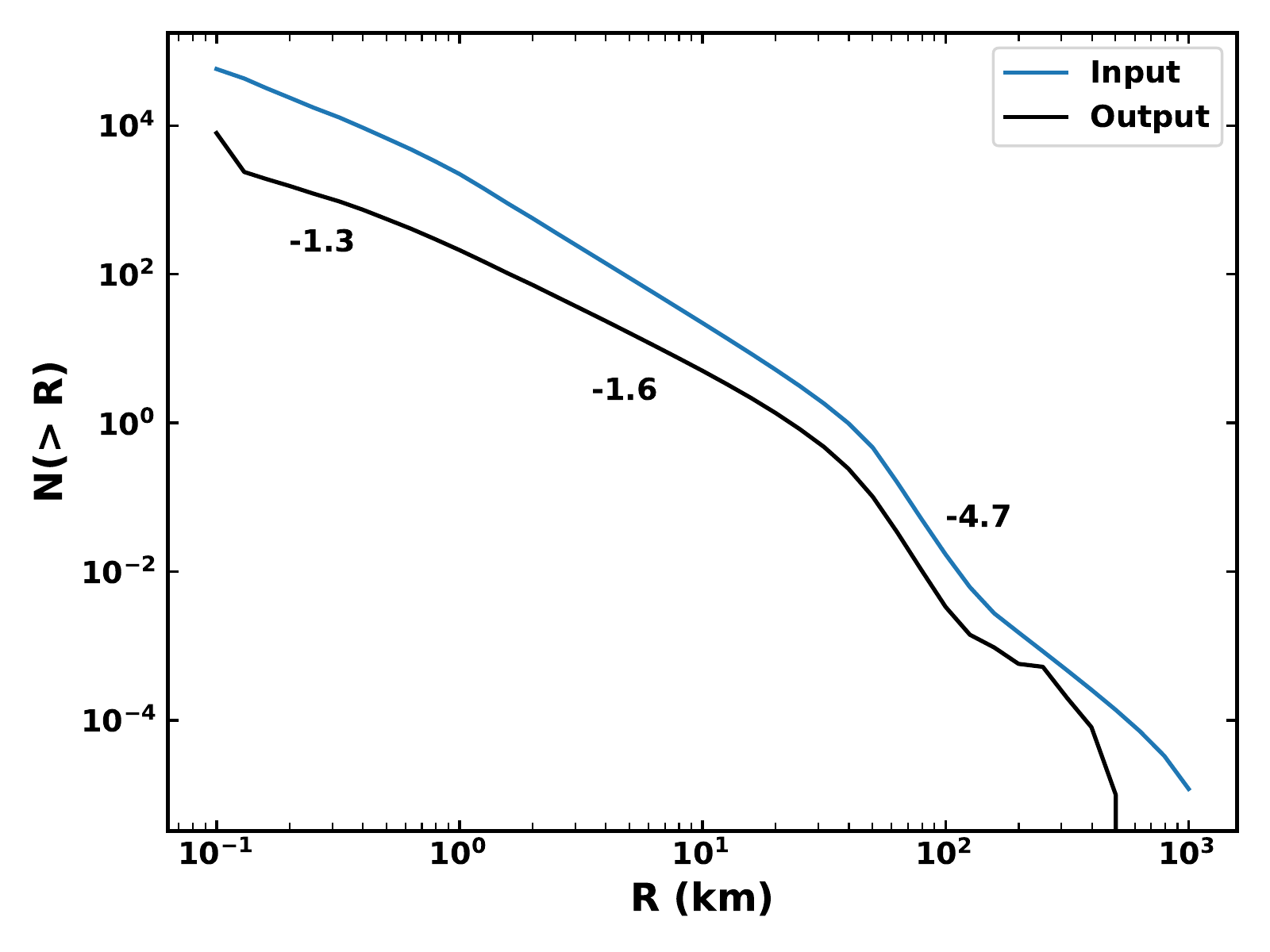}
	\caption{Cumulative size distribution (CSD) of comets which is present on average in a 100~yr period (black) compared with the distribution of comets which is input (blue). The same as Figure~\ref{fig:CSD_com}, but an alternative prescription for the fraction of mass lost in a fragmentation event is used with a weaker dependence on comet size. The slopes of the CSD in each region are labelled by the curve.}
	\label{fig:CSD_sqrt}
\end{figure}

The size dependence of the \citeauthor{DiSisto09} model was based on the fact that the escape velocity of a comet should be proportional to its radius, and was not considered a free parameter of the model. However, we found that the slopes of the comet size distribution resulting from the fragmentation model were too shallow compared to observations of JFCs (Section~\ref{subsec:frag_results}). Restricting the pericentres fragmentations can occur at to $q < 2.5$~au so that more comets survive long enough to reach $< 2.5$~au improved the comet size distribution slopes slightly, but the fit was still poor. Therefore, it is possible that a different size dependence of mass loss in fragmentations is needed. We tried a weaker dependence of the fractional mass loss on size, with $\frac{\Delta M}{M} \propto 1/\sqrt{R}$, with the resulting CSD shown in Figure~\ref{fig:CSD_sqrt}. This gave a better fit to the slopes of the comet distribution for $R \lesssim 10$~km comets. Reducing the fractional mass loss per event extends the lifetime of comets, such that smaller comets had much longer lifetimes than with the $1 / R$ prescription. However, the amount of mass input and the location mass was input to was not significantly changed. The main effect this has is to extend the lifetimes of comets, such that there are roughly twice as many visible comets. Such a prescription would therefore require us to halve our mass input rate of comets, and change $\epsilon$ accordingly to fit the zodiacal cloud. It is therefore possible that the size dependence of comet fragmentation should be further explored in order to match both the input and output size distributions of comets. \par

\subsubsection{Other sources}
\label{subsub:sources}
It is important to acknowledge that while JFCs are believed to dominate, other sources will contribute to the interplanetary dust complex. Asteroids, the ISM, and other families of comets should contribute at least small amounts to our zodiacal cloud. Here we focus on the distribution of dust resulting from comet fragmentation and its variability, but a comprehensive model of the zodiacal cloud requires modelling all potential sources of dust. \par
Recent dynamical models place an upper limit on the asteroidal contribution of 10 per cent \citep{Nesvorny10,Ueda17}. In order to mimic a cometary source with an additional asteroidal contribution, we ran our best fit model again with a mass input rate which was 10 per cent lower. We then included an asteroid belt which had a constant mass input rate of 100~\kgs~(10 per cent of the mean mass input), with eccentricities in the range 0.04-0.27, and pericentres in the range 1.8-3.5~au. Dust from this source was placed in a size distribution with a differential slope of -3.5, the typical value for a collisional cascade \citep{Dohnanyi69}. \par
Including this 'asteroidal' component still allowed us to fit the observed values of the zodiacal cloud relatively well, with a best fit of $7.2\times 10^{-8}$, -0.34, and a peak of $55~\mu$m. The radial slope becomes slightly flatter on average as dust from the asteroid belt will migrate in via P-R drag. Increasing the contribution of this asteroidal source to 30 per cent, we could still obtain a reasonable fit to the zodiacal cloud, but the size distribution peaks at smaller grain sizes. \par
Therefore, with an 'asteroidal' contribution we could still produce a size distribution which is reasonable compared to the zodiacal cloud. However, the limitation of this approximation is that we cannot use different particle inclinations, which is the main difference between asteroidal and cometary grains, and the basis of many arguments for why comets should be the dominant source. Further, asteroidal and cometary grains will likely have different compositions, densities, and collisional strengths, rather than being homogenous. \par
When considering the distribution of dust further out in the solar system, other sources become more important. For example, based on in situ measurements from the New Horizons Student Dust Counter, \citet{Poppe19} modelled the relative contributions of different sources to interplanetary dust in the outer solar system. They found that JFCs should be the dominant source at distances of $\lesssim 10$~au, while further out the dominant sources are the Kuiper Belt and Oort Cloud comets. Our model focuses on the inner few au of the solar system, and so only considers the contribution of JFCs. As such, its predictions for the region $>10$~au are expected to be inaccurate. Indeed, the model may also not include all of the dust expected from JFCs in this outer region, since we only considered comets when they reached within 5.2~au, whereas they could also fragment when further from the Sun. \par

\subsubsection{Dynamical grains}
\label{subsub:lostgrains}
Due to computational limitations we were unable to follow particles which are released by JFCs and dominated by dynamical interactions with Jupiter, instead assuming they are 'lost' on short timescales and therefore do not contribute significantly. These particles are typically the largest sized grains, such that they constitute a large fraction of the mass, but should not contribute significantly to the optical depth of the zodiacal cloud. \par
\begin{figure}
	\centering
	\begin{subfigure}[b]{\linewidth}
		\centering
		\includegraphics[width=\linewidth]{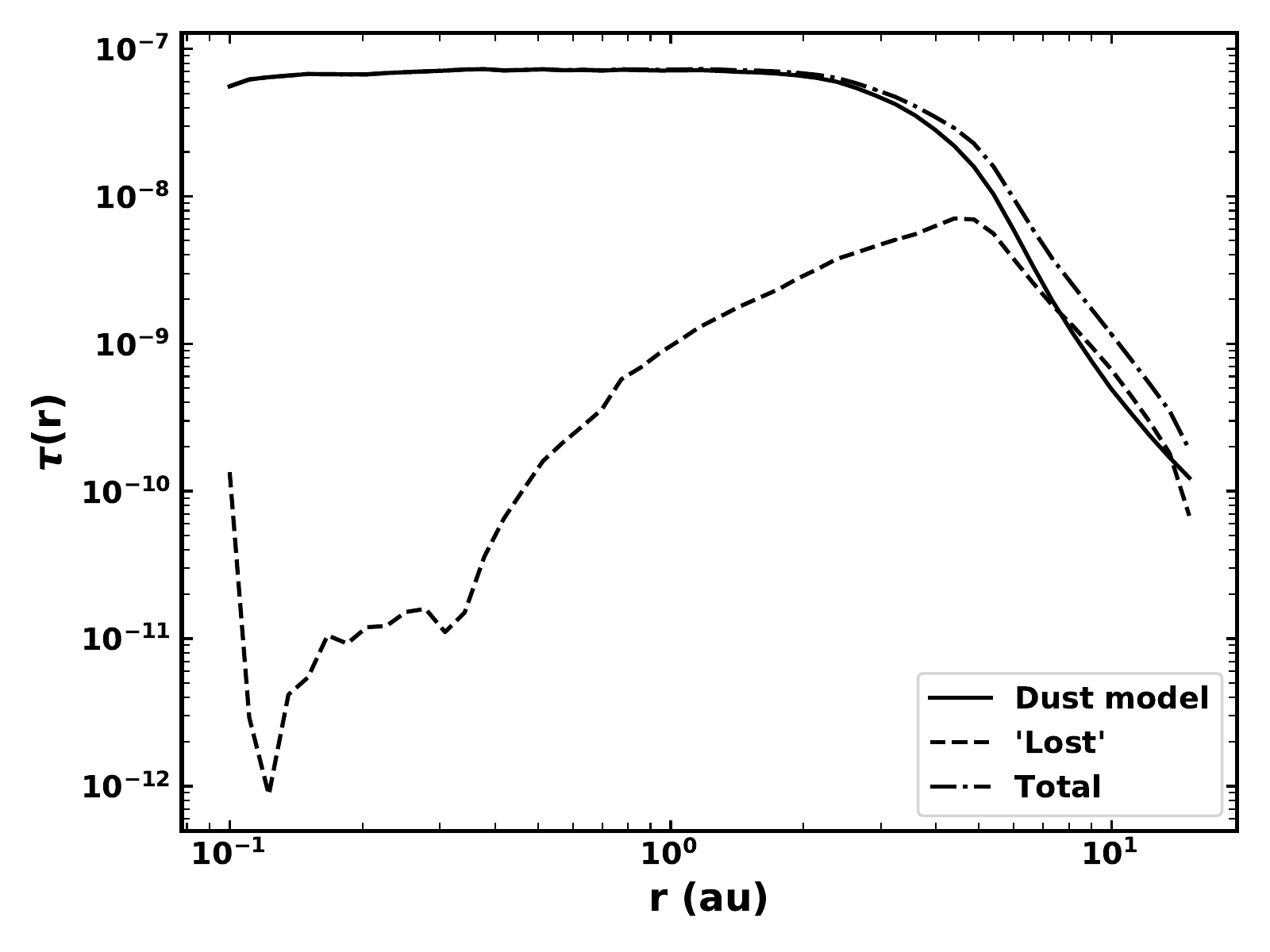}
	\end{subfigure}
	\begin{subfigure}[b]{\linewidth}
		\centering
		\includegraphics[width=\linewidth]{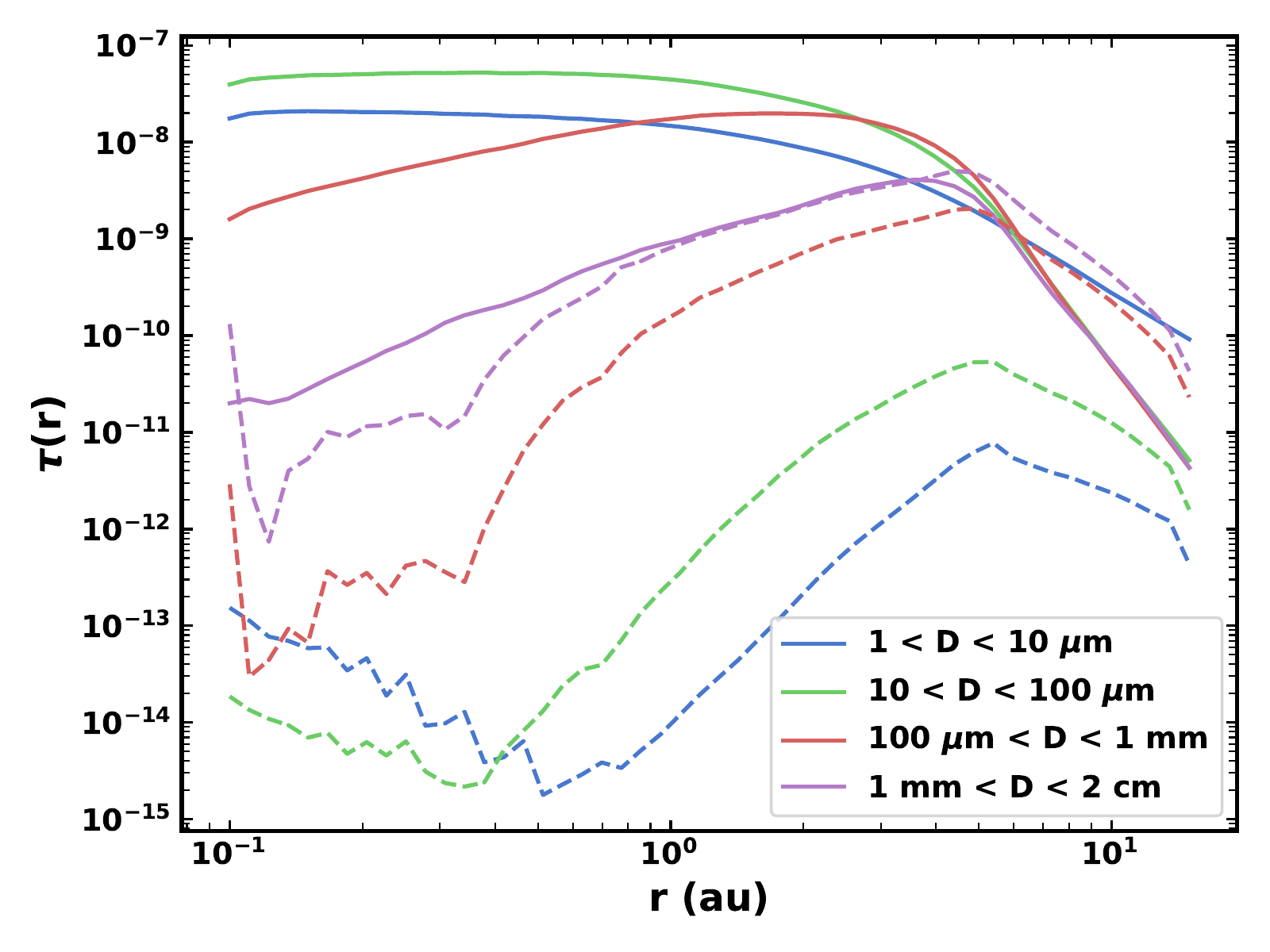}
	\end{subfigure}
	\caption{Top: comparison of the total geometrical optical depth of our model at 66.7 Myr (solid) with the dynamically lost grains (dashed), and the overall profile summing the two contributions (dash-dotted). Bottom: Comparison of geometrical optical depth in our model (solid) with dynamically lost grains (dashed) for different grain sizes, averaged over time.}
	\label{fig:lost}
\end{figure}

In order to estimate the contribution of these lost grains, we recorded the distribution of dust produced by fragmentations which is dominated by dynamics, weighted by the dynamical lifetime divided by the length of our simulation. This gives the 'lost' cross-sectional area, weighted by the fraction of time that the comet spends after fragmentation in the inner solar system, to give the average distribution of dynamical grains. This is an approximation which assumes the grains stay on the orbit of the parent comet when they are produced, when in reality they will bounce around. Our best fit radial profile at 66.7~Myr is compared with the average distribution of dynamical grains in Figure~\ref{fig:lost} (top). The optical depth of dynamical grains is much higher further from the Sun, where P-R drag timescales are longer. The dynamical grains dominate the optical depth at $\gtrsim 8$~au. Superposing these dynamical grains on our best fit model, the slope of the radial profile for $1 \leq r \leq 3$~au goes from -0.34 to -0.27. The optical depth and size distribution at 1~au are not significantly affected. However, as mentioned above this assumes dynamical grains stay where they are produced. Therefore the actual distribution of dynamical grains may be weighted more towards smaller radii as they get scattered inwards, and so may not affect the radial slope so much. However, the exact parameters of our best fit model may be slightly different if dynamical grains could be included fully. \par
Figure~\ref{fig:lost} (bottom) compares the average radial profiles of different grain sizes. For small grains ($D < 100~\mu$m), the dynamical grains are never significant compared to those dominated by drag and collisions. For $100~\mu$m$~<D<1$~mm, dynamical grains dominate the cross-sectional area at $r > 6$~au. However, for $D > 1$~mm grains, the dynamical grains are always comparable to those dominated by drag and collisions. This means that the main effect of not including dynamical grains in our kinetic model is that we are underestimating the number of cm-size grains. In our model cm-size grains do not contribute significantly to the cross-sectional area (see Figure~\ref{fig:dsig_dlogD}), so the main effect of this is that we are underestimating the collision rate of cm-size grains, which supplies the smaller grains. This will thus have an effect further down the size distribution. Dynamical grains will mostly affect the distribution further from the Sun, where the zodiacal cloud is more poorly characterised. Since we are underestimating the collision rate of cm-size grains, the main effect on our parameters would likely be that the collisional strength \QD~would not need to have such a steep slope ($a = 0.9$) if dynamical grains were included, as the collision rate of cm-size grains would be higher due to there being more grains of that size, rather than them having lower collisional strength. \par
Our simplified treatment of dynamics also means we are unable to study fine structure in the zodiacal cloud. For example, particles may get trapped in mean-motion resonances which extend their lifetimes. Measurements with Juno showed that the radial structure of the zodiacal cloud may have fine structure \citep{Jorgensen21}. Our model instead focusses on studying the broad, overall distribution.  \par

\section{Comparison with other models}
\label{sec:comparison}
As discussed in the introduction, there have been many attempts to model the zodiacal cloud. Most of these models either try to fit the thermal emission \citep[e.g.][]{Liou95,Nesvorny10,Rowan-Robinson13}, usually from IRAS or COBE, or the sporadic meteoroid complex \citep[e.g.][]{Wiegert09,Nesvorny11_ZC,Pokorny14}. NASA's Meteoroid Engineering Model \citep[MEM][]{McNamara04,Moorhead20} focuses on modelling the sporadic meteoroid environment, and is tested against meteoroid impact data from the Pegasus satellites and LDEF. They adopt the \citet{Grun85} size distribution for all sources, and follow particles with collisions and drag. ESA's Interplanetary Meteoroid Environment Model \citep[IMEM][]{Dikarev04,Soja19} is a dynamical model which is compared to the COBE latitudinal brightness profile, meteoroids, and lunar microcraters. Most of these models are dynamical, whereas we use a kinetic approach which includes the collisional evolution of dust, including fragments produced in mutual collisions. \par
Since collisional evolution moves mass from larger particles to smaller grains, using a kinetic model allows us to consider the origin of the size distribution in more detail. However, not including dynamical interactions with Jupiter poses its own limitations (see Section~\ref{subsub:lostgrains}). In particular, dynamical models may be better suited to studying the sporadic meteoroids, for which the direction matters and axisymmetry cannot be assumed. Further, meteoroids are larger grains, for which supply by destructive collisions of bigger grains is less important. However, for smaller grains which dominate the thermal emission of the zodiacal cloud, collisions need to be taken into account (see Section~\ref{subsec:comp_coll}). \par

\subsection{Accretion rate onto Earth}
\label{subsec:acc_rate}
Measurements of particle impacts onto the LDEF satellite \citep{Love93} gave an accretion rate of $(40 \pm 20) \times 10^6$~\kgy~onto Earth from dust grains in the mass range $10^{-9} < m < 10^{-4}$~g. Applying the prescription of \citet{Wyatt10} to find the collision rates of particles on different orbits with Earth, and adding an extra factor to take into account gravitational focussing, we find an accretion rate onto Earth from grains of this size range of $7.7 \times 10^6$~\kgy~at the time our distribution best fits the zodiacal cloud. The range of values over the simulation are $4.4 - 57 \times 10^6$~\kgy, with mean value $11 \times 10^6$~\kgy. This is similar to the accretion rate of $15\times 10^6$~\kgy~found by \citet{Nesvorny11_ZC}. While our model predicts a lower accretion rate than the one measured by LDEF, it is in agreement with previous dynamical models. \par

\subsection{Collisional evolution}
\label{subsec:comp_coll}
As mentioned above, previous models of the zodiacal cloud are primarily dynamical. If collisions are considered, they are included with a simplified prescription in which particles are removed after their collisional lifetime, ignoring the products of collisions. We argue that it is important to include the grains produced in such collisions, as these will contribute to the zodiacal light. This is important for modelling both the size distribution of dust and its radial profile. \par
\begin{figure}
	\centering
	\begin{subfigure}[b]{\linewidth}
		\centering
		\includegraphics[width=\linewidth]{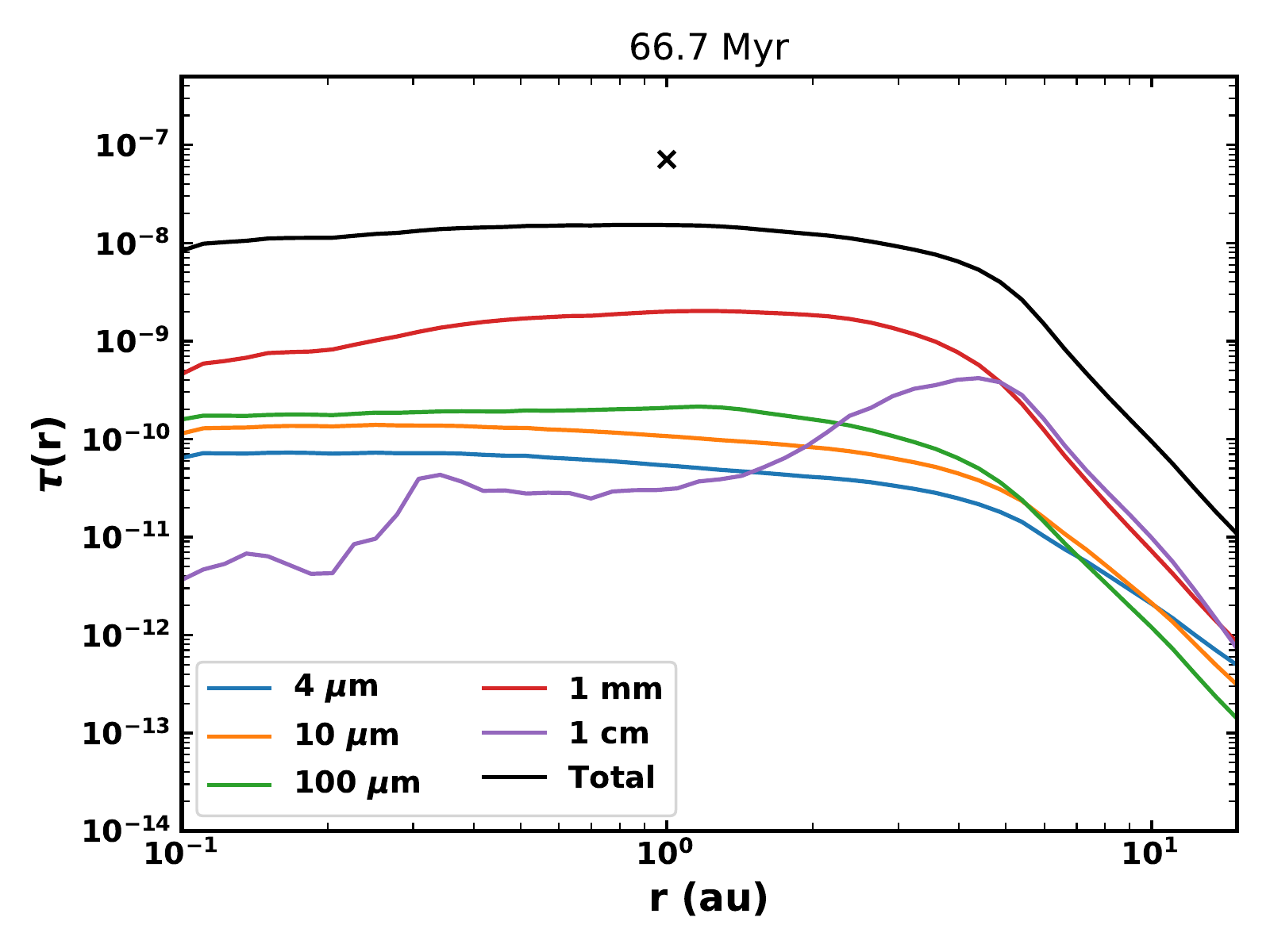}
	\end{subfigure}
	\begin{subfigure}[b]{\linewidth}
		\centering
		\includegraphics[width=\linewidth]{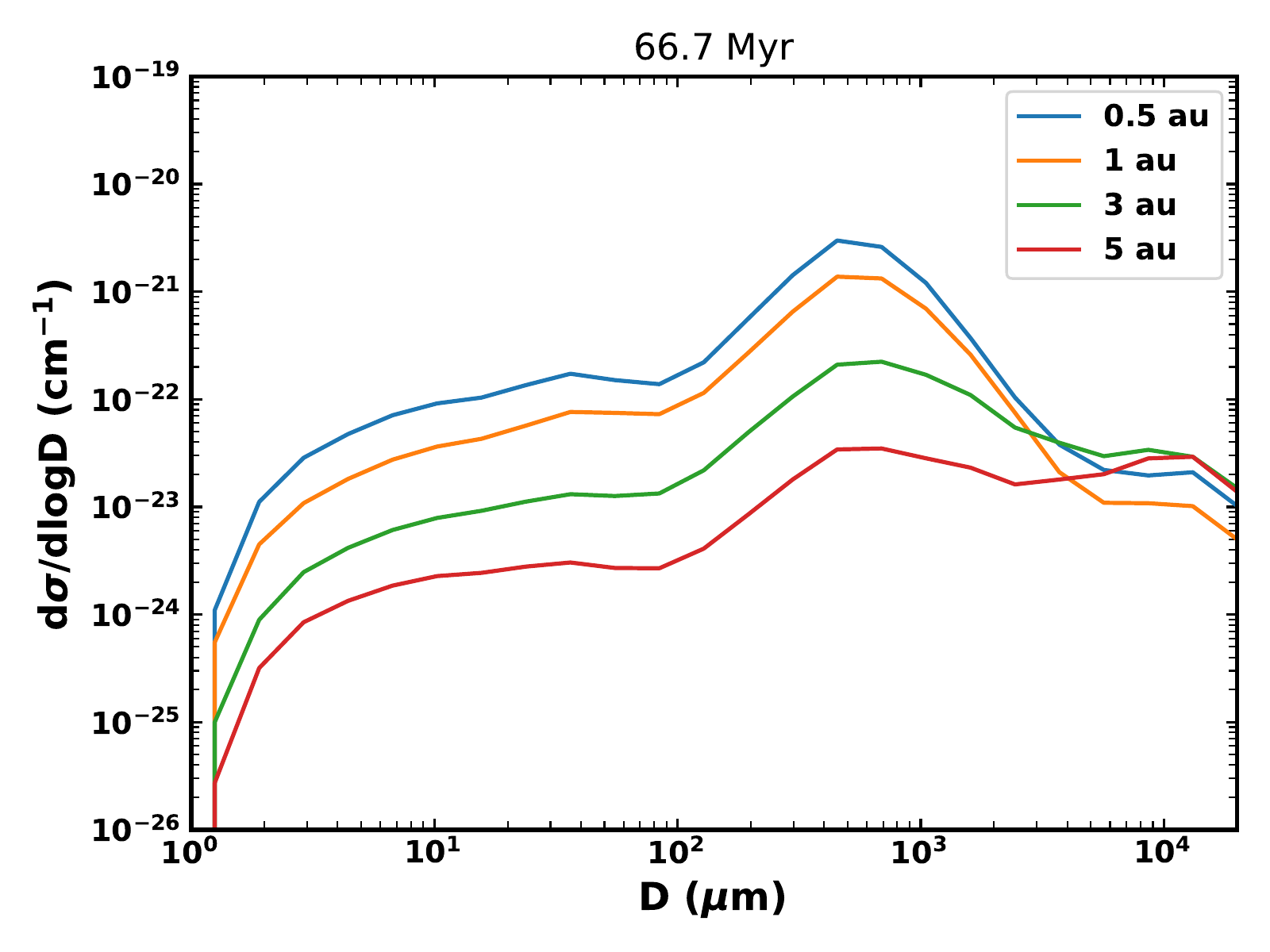}
	\end{subfigure}
	\caption{Top: radial profile of geometrical optical depth in dust grains of different sizes in our model when collisional fragments are not included, to be compared with Figure~\ref{fig:tauDr}. Bottom: size distribution of dust in terms of volume density of cross-sectional area at different heliocentric distances when collisional fragments are not included in our model, to be compared with Figure~\ref{fig:dsig_dlogD}.}
	\label{fig:nofrag}
\end{figure}

In order to ascertain the importance of including collisional fragments in the model, we ran our best fit model again, turning off the part of the code which produces collisional fragments, such that destructive collisions only act as a loss mechanism. The resulting radial profiles are shown in Figure~\ref{fig:nofrag} (top), to compare with Figure~\ref{fig:tauDr}. Overall the radial profile is much flatter than when collisional fragments are included. Small ($D \lesssim 100~\mu$m) grains still have relatively flat radial profiles. Cm-sized grains are still depleted closer in as they are lost to collisions, but mm-sized grains are much flatter. Not including the source of smaller grains from collisions has a significant effect on the size distribution, which is plotted in Figure~\ref{fig:nofrag} (bottom). With collisions only acting as a loss mechanism, the cross-sectional area is now dominated by 600~$\mu$m grains and there are significantly fewer $D < 100~\mu$m grains, whereas with collisional fragments included the size distribution is dominated by particles 10s of $\mu$m in size. Further, the ability of our model to fit the present-day zodiacal cloud depends on the size distribution of collisional fragments \ar~and the collisional strength of particles, \QD. Therefore, collisional evolution is important in order to understand the size distribution, and how dust behaves outside the vicinity of Earth. \par

\section{Conclusions}
\label{sec:conc}

We have developed a model for the distribution of interplanetary dust which would result from comet fragmentations. As comets from N-body simulations bounce around the inner solar system, they undergo recurrent, spontaneous fragmentation events until they either lose all of their mass or get scattered outside Jupiter's orbit. A fraction of the mass lost in these events is converted into dust which supplies the zodiacal cloud. Such dust either stays with the parent comet due to dynamical interactions with the planets, or is input into a kinetic model which follows collisional evolution, P-R drag, and radiation pressure.\par
Comets are drawn from a size distribution based on the Kuiper belt, such that the resulting distribution of dust is highly stochastic, depending on the size and dynamical lifetime of comets which are scattered in. We compare our model to three observables of the present zodiacal cloud: the absolute value and slope of the radial profile of geometrical optical depth, and the grain size which dominates the cross-sectional area at 1~au. While these vary rapidly due to the stochasticity of our model, at two points in the simulation our model fits the present zodiacal cloud. We therefore suggest that comet fragmentation may be able to produce the correct size and spatial distribution of dust to supply the zodiacal cloud. Including a 10 per cent contribution of dust from the asteroid belt does not change our conclusion that the distribution can fit the zodiacal cloud. We also show that the zodiacal cloud should be highly variable over longer (Myr) timescales due to the aforementioned stochasticity. This means that the historical brightness of the zodiacal cloud may have been highly variable. \par
Smaller ($< 10$~km) comets tend to lose all of their mass in successive fragmentations, whereas larger ($\gtrsim 50$~km) comets tend to survive their dynamical lifetime without fully disrupting. Therefore for larger comets, the key factor determining how much mass they supply to the zodiacal cloud is their dynamical lifetime inside Jupiter's orbit. We predict that very large ($>100$~km) comets should only be scattered into the inner solar system rarely, such that the size of comet which should dominate the dust input to the zodiacal cloud should be $\sim 50$~km, as these are more common. We also show that $> 100$~km comets with longer dynamical lifetimes  can cause spikes in the level of zodiacal dust which last for $\sim 1$~Myr, far longer than the dynamical lifetime of the comet itself. Large comets can therefore have a lasting effect on interplanetary dust. \par
Our model is somewhat limited in its treatment of dynamical interactions with planets, such that more work is needed to couple the dynamical and collisional evolution of dust. However, comet fragmentation provides a promising source of interplanetary dust. \par
Comet disruption should also be further explored as a source of exozodiacal dust \citep[e.g.][]{Sezestre19}. The model presented here serves as  good starting point for such an analysis, since its parameters have been tuned to ensure that it reproduces the zodiacal cloud. Further, the stochasticity of our model suggests that if exozodiacal dust is similarly cometary in origin, it may be highly variable on long (Myr) timescales. \par

\section*{Acknowledgements}

We thank the anonymous reviewer for their helpful comments on the manuscript. JKR would like to acknowledge support from the Science and Technology Facilities Council (STFC) towards her doctoral research. We are grateful to David Nesvorn\'y for providing the N-body data used in this paper. 

\section*{Data Availability}
	
The data underlying this article will be shared on reasonable request to the corresponding author.



\bibliographystyle{mnras}
\bibliography{debris} 

\begin{thebibliography}{}
\makeatletter
\relax
\def\mn@urlcharsother{\let\do\@makeother \do\$\do\&\do\#\do\^\do\_\do\%\do\~}
\def\mn@doi{\begingroup\mn@urlcharsother \@ifnextchar [ {\mn@doi@}
  {\mn@doi@[]}}
\def\mn@doi@[#1]#2{\def\@tempa{#1}\ifx\@tempa\@empty \href
  {http://dx.doi.org/#2} {doi:#2}\else \href {http://dx.doi.org/#2} {#1}\fi
  \endgroup}
\def\mn@eprint#1#2{\mn@eprint@#1:#2::\@nil}
\def\mn@eprint@arXiv#1{\href {http://arxiv.org/abs/#1} {{\tt arXiv:#1}}}
\def\mn@eprint@dblp#1{\href {http://dblp.uni-trier.de/rec/bibtex/#1.xml}
  {dblp:#1}}
\def\mn@eprint@#1:#2:#3:#4\@nil{\def\@tempa {#1}\def\@tempb {#2}\def\@tempc
  {#3}\ifx \@tempc \@empty \let \@tempc \@tempb \let \@tempb \@tempa \fi \ifx
  \@tempb \@empty \def\@tempb {arXiv}\fi \@ifundefined
  {mn@eprint@\@tempb}{\@tempb:\@tempc}{\expandafter \expandafter \csname
  mn@eprint@\@tempb\endcsname \expandafter{\@tempc}}}

\bibitem[\protect\citeauthoryear{{Belton}}{{Belton}}{2014}]{Belton14}
{Belton} M. J.~S.,  2014, \mn@doi [\icarus] {10.1016/j.icarus.2013.12.001},
  \href {https://ui.adsabs.harvard.edu/abs/2014Icar..231..168B} {231, 168}

\bibitem[\protect\citeauthoryear{{Benz} \& {Asphaug}}{{Benz} \&
  {Asphaug}}{1999}]{Benz99}
{Benz} W.,  {Asphaug} E.,  1999, \mn@doi [\icarus] {10.1006/icar.1999.6204},
  \href {https://ui.adsabs.harvard.edu/abs/1999Icar..142....5B} {142, 5}

\bibitem[\protect\citeauthoryear{{Boehnhardt}}{{Boehnhardt}}{2004}]{Boehnhardt04}
{Boehnhardt} H.,  2004, in Festou M.C., Keller H.U., Weaver H.A., eds, Comets
  II. University of Arizona Press, Tucson.
p.~301

\bibitem[\protect\citeauthoryear{{Boehnhardt}, {Holdstock}, {Hainaut}, {Tozzi},
  {Benetti}  \& {Licandro}}{{Boehnhardt} et~al.}{2002}]{Boehnhardt02}
{Boehnhardt} H.,  {Holdstock} S.,  {Hainaut} O.,  {Tozzi} G.~P.,  {Benetti} S.,
    {Licandro} J.,  2002, Earth Moon and Planets, \href
  {https://ui.adsabs.harvard.edu/abs/2002EM&P...90..131B} {90, 131}

\bibitem[\protect\citeauthoryear{{Burns}, {Lamy}  \& {Soter}}{{Burns}
  et~al.}{1979}]{Burns79}
{Burns} J.~A.,  {Lamy} P.~L.,   {Soter} S.,  1979, \mn@doi [\icarus]
  {10.1016/0019-1035(79)90050-2}, \href
  {https://ui.adsabs.harvard.edu/abs/1979Icar...40....1B} {40, 1}

\bibitem[\protect\citeauthoryear{{Chen} \& {Jewitt}}{{Chen} \&
  {Jewitt}}{1994}]{Chen94}
{Chen} J.,  {Jewitt} D.,  1994, \mn@doi [\icarus] {10.1006/icar.1994.1061},
  \href {https://ui.adsabs.harvard.edu/abs/1994Icar..108..265C} {108, 265}

\bibitem[\protect\citeauthoryear{{Clube} \& {Napier}}{{Clube} \&
  {Napier}}{1984}]{Clube84}
{Clube} S.~V.~M.,  {Napier} W.~M.,  1984, \mn@doi [\mnras]
  {10.1093/mnras/211.4.953}, \href
  {https://ui.adsabs.harvard.edu/abs/1984MNRAS.211..953C} {211, 953}

\bibitem[\protect\citeauthoryear{{Dermott}, {Nicholson}, {Burns}  \&
  {Houck}}{{Dermott} et~al.}{1984}]{Dermott84}
{Dermott} S.~F.,  {Nicholson} P.~D.,  {Burns} J.~A.,   {Houck} J.~R.,  1984,
  \mn@doi [\nat] {10.1038/312505a0}, \href
  {https://ui.adsabs.harvard.edu/abs/1984Natur.312..505D} {312, 505}

\bibitem[\protect\citeauthoryear{{Dermott}, {Jayaraman}, {Xu}, {Gustafson}  \&
  {Liou}}{{Dermott} et~al.}{1994}]{Dermott94}
{Dermott} S.~F.,  {Jayaraman} S.,  {Xu} Y.~L.,  {Gustafson} B. {\r{A}}.~S.,
  {Liou} J.~C.,  1994, \mn@doi [\nat] {10.1038/369719a0}, \href
  {https://ui.adsabs.harvard.edu/abs/1994Natur.369..719D} {369, 719}

\bibitem[\protect\citeauthoryear{{Dermott}, {Grogan}, {Durda}, {Jayaraman},
  {Kehoe}, {Kortenkamp}  \& {Wyatt}}{{Dermott} et~al.}{2001}]{Dermott01}
{Dermott} S.~F.,  {Grogan} K.,  {Durda} D.~D.,  {Jayaraman} S.,  {Kehoe} T.
  J.~J.,  {Kortenkamp} S.~J.,   {Wyatt} M.~C.,  2001, in Gr\"un E., Gusafson
  B.A.S., Dermott S., Fechtig H., eds, Interplanetary Dust. Springer, Berlin.
p.~569

\bibitem[\protect\citeauthoryear{{Di Sisto}, {Fern{\'a}ndez}  \& {Brunini}}{{Di
  Sisto} et~al.}{2009}]{DiSisto09}
{Di Sisto} R.~P.,  {Fern{\'a}ndez} J.~A.,   {Brunini} A.,  2009, \mn@doi
  [\icarus] {10.1016/j.icarus.2009.05.002}, \href
  {https://ui.adsabs.harvard.edu/abs/2009Icar..203..140D} {203, 140}

\bibitem[\protect\citeauthoryear{{Dikarev}, {Gr{\"u}n}, {Baggaley}, {Galligan},
  {Landgraf}  \& {Jehn}}{{Dikarev} et~al.}{2004}]{Dikarev04}
{Dikarev} V.,  {Gr{\"u}n} E.,  {Baggaley} J.,  {Galligan} D.,  {Landgraf} M.,
  {Jehn} R.,  2004, \mn@doi [Earth Moon and Planets]
  {10.1007/s11038-005-9017-y}, \href
  {https://ui.adsabs.harvard.edu/abs/2004EM&P...95..109D} {95, 109}

\bibitem[\protect\citeauthoryear{{Divine}}{{Divine}}{1993}]{Divine93}
{Divine} N.,  1993, \mn@doi [\jgr] {10.1029/93JE01203}, \href
  {https://ui.adsabs.harvard.edu/abs/1993JGR....9817029D} {98, 17029}

\bibitem[\protect\citeauthoryear{{Dohnanyi}}{{Dohnanyi}}{1969}]{Dohnanyi69}
{Dohnanyi} J.~S.,  1969, \mn@doi [Journal of Geophysical Research]
  {10.1029/JB074i010p02531}, \href
  {https://ui.adsabs.harvard.edu/abs/1969JGR....74.2531D} {74, 2531}

\bibitem[\protect\citeauthoryear{{Duncan} \& {Levison}}{{Duncan} \&
  {Levison}}{1997}]{Duncan97}
{Duncan} M.~J.,  {Levison} H.~F.,  1997, \mn@doi [Science]
  {10.1126/science.276.5319.1670}, \href
  {https://ui.adsabs.harvard.edu/abs/1997Sci...276.1670D} {276, 1670}

\bibitem[\protect\citeauthoryear{{Durda} \& {Dermott}}{{Durda} \&
  {Dermott}}{1997}]{Durda97}
{Durda} D.~D.,  {Dermott} S.~F.,  1997, \mn@doi [\icarus]
  {10.1006/icar.1997.5803}, \href
  {https://ui.adsabs.harvard.edu/abs/1997Icar..130..140D} {130, 140}

\bibitem[\protect\citeauthoryear{{Economou}, {Green}, {Brownlee}  \&
  {Clark}}{{Economou} et~al.}{2013}]{Economou13}
{Economou} T.~E.,  {Green} S.~F.,  {Brownlee} D.~E.,   {Clark} B.~C.,  2013,
  \mn@doi [\icarus] {10.1016/j.icarus.2012.09.019}, \href
  {https://ui.adsabs.harvard.edu/abs/2013Icar..222..526E} {222, 526}

\bibitem[\protect\citeauthoryear{{Espy Kehoe}, {Kehoe}, {Colwell}  \&
  {Dermott}}{{Espy Kehoe} et~al.}{2015}]{EspyKehoe15}
{Espy Kehoe} A.~J.,  {Kehoe} T.~J.~J.,  {Colwell} J.~E.,   {Dermott} S.~F.,
  2015, \mn@doi [\apj] {10.1088/0004-637X/811/1/66}, \href
  {https://ui.adsabs.harvard.edu/abs/2015ApJ...811...66E} {811, 66}

\bibitem[\protect\citeauthoryear{{Farnham}, {Schleicher}, {Woodney}, {Birch},
  {Eberhardy}  \& {Levy}}{{Farnham} et~al.}{2001}]{Farnham01}
{Farnham} T.~L.,  {Schleicher} D.~G.,  {Woodney} L.~M.,  {Birch} P.~V.,
  {Eberhardy} C.~A.,   {Levy} L.,  2001, \mn@doi [Science]
  {10.1126/science.1058886}, \href
  {https://ui.adsabs.harvard.edu/abs/2001Sci...292.1348F} {292, 1348}

\bibitem[\protect\citeauthoryear{{Fern{\'a}ndez}}{{Fern{\'a}ndez}}{2005}]{Fernandez05}
{Fern{\'a}ndez} J.~A.,  2005, {Comets - Nature, Dynamics, Origin and their
  Cosmological Relevance}.
Springer, Dordrecht, \mn@doi{10.1007/978-1-4020-3495-4}

\bibitem[\protect\citeauthoryear{{Fern{\'a}ndez}}{{Fern{\'a}ndez}}{2009}]{Fernandez09}
{Fern{\'a}ndez} Y.~R.,  2009, \mn@doi [\planss] {10.1016/j.pss.2009.01.003},
  \href {https://ui.adsabs.harvard.edu/abs/2009P&SS...57.1218F} {57, 1218}

\bibitem[\protect\citeauthoryear{{Fern{\'a}ndez} \&
  {Morbidelli}}{{Fern{\'a}ndez} \& {Morbidelli}}{2006}]{Fernandez06}
{Fern{\'a}ndez} J.~A.,  {Morbidelli} A.,  2006, \mn@doi [\icarus]
  {10.1016/j.icarus.2006.07.001}, \href
  {https://ui.adsabs.harvard.edu/abs/2006Icar..185..211F} {185, 211}

\bibitem[\protect\citeauthoryear{{Fern{\'a}ndez} et~al.,}{{Fern{\'a}ndez}
  et~al.}{2013}]{Fernandez13}
{Fern{\'a}ndez} Y.~R.,  et~al., 2013, \mn@doi [\icarus]
  {10.1016/j.icarus.2013.07.021}, \href
  {https://ui.adsabs.harvard.edu/abs/2013Icar..226.1138F} {226, 1138}

\bibitem[\protect\citeauthoryear{{Ferr{\'\i}n} \& {Orofino}}{{Ferr{\'\i}n} \&
  {Orofino}}{2021}]{Ferrin21}
{Ferr{\'\i}n} I.,  {Orofino} V.,  2021, \mn@doi [\planss]
  {10.1016/j.pss.2021.105306}, \href
  {https://ui.adsabs.harvard.edu/abs/2021P&SS..20705306F} {207, 105306}

\bibitem[\protect\citeauthoryear{{Fujiwara}}{{Fujiwara}}{1986}]{Fujiwara86}
{Fujiwara} A.,  1986, \memsai, \href
  {https://ui.adsabs.harvard.edu/abs/1986MmSAI..57...47F} {57, 47}

\bibitem[\protect\citeauthoryear{{Fulle} et~al.,}{{Fulle}
  et~al.}{2016a}]{Fulle16a}
{Fulle} M.,  et~al., 2016a, \mn@doi [\mnras] {10.1093/mnras/stw2299}, \href
  {https://ui.adsabs.harvard.edu/abs/2016MNRAS.462S.132F} {462, S132}

\bibitem[\protect\citeauthoryear{{Fulle} et~al.,}{{Fulle}
  et~al.}{2016b}]{Fulle16b}
{Fulle} M.,  et~al., 2016b, \mn@doi [\apj] {10.3847/0004-637X/821/1/19}, \href
  {https://ui.adsabs.harvard.edu/abs/2016ApJ...821...19F} {821, 19}

\bibitem[\protect\citeauthoryear{{Fuse}, {Yamamoto}, {Kinoshita}, {Furusawa}
  \& {Watanabe}}{{Fuse} et~al.}{2007}]{Fuse07}
{Fuse} T.,  {Yamamoto} N.,  {Kinoshita} D.,  {Furusawa} H.,   {Watanabe} J.-I.,
   2007, \mn@doi [\pasj] {10.1093/pasj/59.2.381}, \href
  {https://ui.adsabs.harvard.edu/abs/2007PASJ...59..381F} {59, 381}

\bibitem[\protect\citeauthoryear{{Gomes}, {Levison}, {Tsiganis}  \&
  {Morbidelli}}{{Gomes} et~al.}{2005}]{Gomes05}
{Gomes} R.,  {Levison} H.~F.,  {Tsiganis} K.,   {Morbidelli} A.,  2005, \mn@doi
  [\nat] {10.1038/nature03676}, \href
  {https://ui.adsabs.harvard.edu/abs/2005Natur.435..466G} {435, 466}

\bibitem[\protect\citeauthoryear{{Green} et~al.,}{{Green}
  et~al.}{2004}]{Green04}
{Green} S.~F.,  et~al., 2004, \mn@doi [Journal of Geophysical Research
  (Planets)] {10.1029/2004JE002318}, \href
  {https://ui.adsabs.harvard.edu/abs/2004JGRE..10912S04G} {109, E12S04}

\bibitem[\protect\citeauthoryear{{Grun}, {Zook}, {Fechtig}  \& {Giese}}{{Grun}
  et~al.}{1985}]{Grun85}
{Grun} E.,  {Zook} H.~A.,  {Fechtig} H.,   {Giese} R.~H.,  1985, \mn@doi
  [\icarus] {10.1016/0019-1035(85)90121-6}, \href
  {https://ui.adsabs.harvard.edu/abs/1985Icar...62..244G} {62, 244}

\bibitem[\protect\citeauthoryear{{Gustafson}}{{Gustafson}}{1994}]{Gustafson94}
{Gustafson} B.~A.~S.,  1994, \mn@doi [Annual Review of Earth and Planetary
  Sciences] {10.1146/annurev.ea.22.050194.003005}, \href
  {https://ui.adsabs.harvard.edu/abs/1994AREPS..22..553G} {22, 553}

\bibitem[\protect\citeauthoryear{{Hanner}, {Sparrow}, {Weinberg}  \&
  {Beeson}}{{Hanner} et~al.}{1976}]{Hanner76}
{Hanner} M.~S.,  {Sparrow} J.~G.,  {Weinberg} J.~L.,   {Beeson} D.~E.,  1976,
  in Elsaesser H., Fechtig H., eds, Proc. IAU Colloq. 31, Interplanetary Dust
  and Zodiacal Light. Springer, Berlin.
p.~29, \mn@doi{10.1007/3-540-07615-8_448}

\bibitem[\protect\citeauthoryear{{Hauser} et~al.,}{{Hauser}
  et~al.}{1984}]{Hauser84}
{Hauser} M.~G.,  et~al., 1984, \mn@doi [\apj] {10.1086/184212}, \href
  {https://ui.adsabs.harvard.edu/abs/1984ApJ...278L..15H} {278, L15}

\bibitem[\protect\citeauthoryear{{Hinz} et~al.,}{{Hinz} et~al.}{2016}]{Hinz16}
{Hinz} P.~M.,  et~al., 2016, in {Malbet} F.,  {Creech-Eakman} M.~J.,
  {Tuthill} P.~G.,  eds,  Proc. SPIE Conf. Ser. Vol. 9907, Optical and Infrared
  Interferometry and Imaging V. p. 990704, \mn@doi{10.1117/12.2233795}

\bibitem[\protect\citeauthoryear{{H{\"o}rz} et~al.,}{{H{\"o}rz}
  et~al.}{2006}]{Horz06}
{H{\"o}rz} F.,  et~al., 2006, \mn@doi [Science] {10.1126/science.1135705},
  \href {https://ui.adsabs.harvard.edu/abs/2006Sci...314.1716H} {314, 1716}

\bibitem[\protect\citeauthoryear{{Jorgensen}, {Benn}, {Connerney}, {Denver},
  {Jorgensen}, {Andersen}  \& {Bolton}}{{Jorgensen} et~al.}{2021}]{Jorgensen21}
{Jorgensen} J.~L.,  {Benn} M.,  {Connerney} J.~E.~P.,  {Denver} T.,
  {Jorgensen} P.~S.,  {Andersen} A.~C.,   {Bolton} S.~J.,  2021, \mn@doi
  [Journal of Geophysical Research (Planets)] {10.1029/2020JE006509}, \href
  {https://ui.adsabs.harvard.edu/abs/2021JGRE..12606509J} {126, e06509}

\bibitem[\protect\citeauthoryear{{Jutzi}, {Michel}, {Benz}  \&
  {Richardson}}{{Jutzi} et~al.}{2010}]{Jutzi10}
{Jutzi} M.,  {Michel} P.,  {Benz} W.,   {Richardson} D.~C.,  2010, \mn@doi
  [\icarus] {10.1016/j.icarus.2009.11.016}, \href
  {https://ui.adsabs.harvard.edu/abs/2010Icar..207...54J} {207, 54}

\bibitem[\protect\citeauthoryear{{Kelsall} et~al.,}{{Kelsall}
  et~al.}{1998}]{Kelsall98}
{Kelsall} T.,  et~al., 1998, \mn@doi [\apj] {10.1086/306380}, \href
  {https://ui.adsabs.harvard.edu/abs/1998ApJ...508...44K} {508, 44}

\bibitem[\protect\citeauthoryear{{Krivov}, {Srem{\v{c}}evi{\'c}}  \&
  {Spahn}}{{Krivov} et~al.}{2005}]{Krivov05}
{Krivov} A.~V.,  {Srem{\v{c}}evi{\'c}} M.,   {Spahn} F.,  2005, \mn@doi
  [\icarus] {10.1016/j.icarus.2004.10.003}, \href
  {https://ui.adsabs.harvard.edu/abs/2005Icar..174..105K} {174, 105}

\bibitem[\protect\citeauthoryear{{Krivov}, {L{\"o}hne}  \&
  {Srem{\v{c}}evi{\'c}}}{{Krivov} et~al.}{2006}]{Krivov06}
{Krivov} A.~V.,  {L{\"o}hne} T.,   {Srem{\v{c}}evi{\'c}} M.,  2006, \mn@doi
  [\aap] {10.1051/0004-6361:20064907}, \href
  {https://ui.adsabs.harvard.edu/abs/2006A&A...455..509K} {455, 509}

\bibitem[\protect\citeauthoryear{{Kr{\"u}ger} et~al.,}{{Kr{\"u}ger}
  et~al.}{2010}]{Kruger10}
{Kr{\"u}ger} H.,  et~al., 2010, \mn@doi [\planss] {10.1016/j.pss.2009.11.002},
  \href {https://ui.adsabs.harvard.edu/abs/2010P&SS...58..951K} {58, 951}

\bibitem[\protect\citeauthoryear{{Lamy}, {Toth}, {Fernandez}  \&
  {Weaver}}{{Lamy} et~al.}{2004}]{Lamy04}
{Lamy} P.~L.,  {Toth} I.,  {Fernandez} Y.~R.,   {Weaver} H.~A.,  2004, in
  Festou M.C., Keller H.U., Weaver H.A., eds, Comets II. University of Arizona
  Press, Tucson.
p.~223

\bibitem[\protect\citeauthoryear{{Landgraf}, {Baggaley}, {Gr{\"u}n},
  {Kr{\"u}ger}  \& {Linkert}}{{Landgraf} et~al.}{2000}]{Landgraf00}
{Landgraf} M.,  {Baggaley} W.~J.,  {Gr{\"u}n} E.,  {Kr{\"u}ger} H.,   {Linkert}
  G.,  2000, \mn@doi [\jgr] {10.1029/1999JA900359}, \href
  {https://ui.adsabs.harvard.edu/abs/2000JGR...10510343L} {105, 10343}

\bibitem[\protect\citeauthoryear{{Leinert}, {Richter}, {Pitz}  \&
  {Planck}}{{Leinert} et~al.}{1981}]{Leinert81}
{Leinert} C.,  {Richter} I.,  {Pitz} E.,   {Planck} B.,  1981, \aap, \href
  {https://ui.adsabs.harvard.edu/abs/1981A&A...103..177L} {103, 177}

\bibitem[\protect\citeauthoryear{{Leinert}, {Roser}  \& {Buitrago}}{{Leinert}
  et~al.}{1983}]{Leinert83}
{Leinert} C.,  {Roser} S.,   {Buitrago} J.,  1983, \aap, \href
  {https://ui.adsabs.harvard.edu/abs/1983A&A...118..345L} {118, 345}

\bibitem[\protect\citeauthoryear{{Levison} \& {Duncan}}{{Levison} \&
  {Duncan}}{1997}]{Levison97}
{Levison} H.~F.,  {Duncan} M.~J.,  1997, \mn@doi [\icarus]
  {10.1006/icar.1996.5637}, \href
  {https://ui.adsabs.harvard.edu/abs/1997Icar..127...13L} {127, 13}

\bibitem[\protect\citeauthoryear{{Li} \& {Greenberg}}{{Li} \&
  {Greenberg}}{1997}]{Li97}
{Li} A.,  {Greenberg} J.~M.,  1997, \aap, \href
  {https://ui.adsabs.harvard.edu/abs/1997A&A...323..566L} {323, 566}

\bibitem[\protect\citeauthoryear{{Liou}, {Dermott}  \& {Xu}}{{Liou}
  et~al.}{1995}]{Liou95}
{Liou} J.~C.,  {Dermott} S.~F.,   {Xu} Y.~L.,  1995, \mn@doi [\planss]
  {10.1016/0032-0633(95)00065-D}, \href
  {https://ui.adsabs.harvard.edu/abs/1995P&SS...43..717L} {43, 717}

\bibitem[\protect\citeauthoryear{{L{\"o}hne}, {Krivov}  \&
  {Rodmann}}{{L{\"o}hne} et~al.}{2008}]{Lohne08}
{L{\"o}hne} T.,  {Krivov} A.~V.,   {Rodmann} J.,  2008, \mn@doi [\apj]
  {10.1086/524840}, \href
  {https://ui.adsabs.harvard.edu/abs/2008ApJ...673.1123L} {673, 1123}

\bibitem[\protect\citeauthoryear{{Love} \& {Brownlee}}{{Love} \&
  {Brownlee}}{1993}]{Love93}
{Love} S.~G.,  {Brownlee} D.~E.,  1993, \mn@doi [Science]
  {10.1126/science.262.5133.550}, \href
  {https://ui.adsabs.harvard.edu/abs/1993Sci...262..550L} {262, 550}

\bibitem[\protect\citeauthoryear{{M{\"a}kinen}, {Bertaux}, {Combi}  \&
  {Qu{\'e}merais}}{{M{\"a}kinen} et~al.}{2001}]{Makinen01}
{M{\"a}kinen} J. T.~T.,  {Bertaux} J.-L.,  {Combi} M.~R.,   {Qu{\'e}merais} E.,
   2001, \mn@doi [Science] {10.1126/science.1060858}, \href
  {https://ui.adsabs.harvard.edu/abs/2001Sci...292.1326M} {292, 1326}

\bibitem[\protect\citeauthoryear{{Marboeuf}, {Bonsor}  \&
  {Augereau}}{{Marboeuf} et~al.}{2016}]{Marboeuf16}
{Marboeuf} U.,  {Bonsor} A.,   {Augereau} J.~C.,  2016, \mn@doi [\planss]
  {10.1016/j.pss.2016.03.014}, \href
  {https://ui.adsabs.harvard.edu/abs/2016P&SS..133...47M} {133, 47}

\bibitem[\protect\citeauthoryear{{McDonnell} et~al.,}{{McDonnell}
  et~al.}{1993}]{McDonnell93}
{McDonnell} J.~A.~M.,  et~al., 1993, \mn@doi [\nat] {10.1038/362732a0}, \href
  {https://ui.adsabs.harvard.edu/abs/1993Natur.362..732M} {362, 732}

\bibitem[\protect\citeauthoryear{{McNamara}, {Jones}, {Kauffman}, {Suggs},
  {Cooke}  \& {Smith}}{{McNamara} et~al.}{2004}]{McNamara04}
{McNamara} H.,  {Jones} J.,  {Kauffman} B.,  {Suggs} R.,  {Cooke} W.,   {Smith}
  S.,  2004, \mn@doi [Earth Moon and Planets] {10.1007/s11038-005-9044-8},
  \href {https://ui.adsabs.harvard.edu/abs/2004EM&P...95..123M} {95, 123}

\bibitem[\protect\citeauthoryear{{Meech}, {Hainaut}  \& {Marsden}}{{Meech}
  et~al.}{2004}]{Meech04_size}
{Meech} K.~J.,  {Hainaut} O.~R.,   {Marsden} B.~G.,  2004, \mn@doi [\icarus]
  {10.1016/j.icarus.2004.03.014}, \href
  {https://ui.adsabs.harvard.edu/abs/2004Icar..170..463M} {170, 463}

\bibitem[\protect\citeauthoryear{{Minato}, {K{\"o}hler}, {Kimura}, {Mann}  \&
  {Yamamoto}}{{Minato} et~al.}{2006}]{Minato06}
{Minato} T.,  {K{\"o}hler} M.,  {Kimura} H.,  {Mann} I.,   {Yamamoto} T.,
  2006, \mn@doi [\aap] {10.1051/0004-6361:20054774}, \href
  {https://ui.adsabs.harvard.edu/abs/2006A&A...452..701M} {452, 701}

\bibitem[\protect\citeauthoryear{{Moorhead}, {Kingery}  \& {Ehlert}}{{Moorhead}
  et~al.}{2020}]{Moorhead20}
{Moorhead} A.~V.,  {Kingery} A.,   {Ehlert} S.,  2020, \mn@doi [Journal of
  Spacecraft and Rockets] {10.2514/1.A34561}, \href
  {https://ui.adsabs.harvard.edu/abs/2020JSpRo..57..160M} {57, 160}

\bibitem[\protect\citeauthoryear{{Morbidelli}, {Nesvorny}, {Bottke}  \&
  {Marchi}}{{Morbidelli} et~al.}{2021}]{Morbidelli21}
{Morbidelli} A.,  {Nesvorny} D.,  {Bottke} W.~F.,   {Marchi} S.,  2021, \mn@doi
  [\icarus] {10.1016/j.icarus.2020.114256}, \href
  {https://ui.adsabs.harvard.edu/abs/2021Icar..35614256M} {356, 114256}

\bibitem[\protect\citeauthoryear{{Moreno} et~al.,}{{Moreno}
  et~al.}{2016}]{Moreno16}
{Moreno} F.,  et~al., 2016, \mn@doi [\aap] {10.1051/0004-6361/201527564}, \href
  {https://ui.adsabs.harvard.edu/abs/2016A&A...587A.155M} {587, A155}

\bibitem[\protect\citeauthoryear{{Moro-Mart{\'\i}n} \&
  {Malhotra}}{{Moro-Mart{\'\i}n} \& {Malhotra}}{2003}]{Moro-Martin03}
{Moro-Mart{\'\i}n} A.,  {Malhotra} R.,  2003, \mn@doi [\aj] {10.1086/368237},
  \href {https://ui.adsabs.harvard.edu/abs/2003AJ....125.2255M} {125, 2255}

\bibitem[\protect\citeauthoryear{{Napier}}{{Napier}}{2001}]{Napier01}
{Napier} W.~M.,  2001, \mn@doi [\mnras] {10.1046/j.1365-8711.2001.04020.x},
  \href {https://ui.adsabs.harvard.edu/abs/2001MNRAS.321..463N} {321, 463}

\bibitem[\protect\citeauthoryear{{Napier}}{{Napier}}{2019}]{Napier19}
{Napier} W.~M.,  2019, \mn@doi [\mnras] {10.1093/mnras/stz1769}, \href
  {https://ui.adsabs.harvard.edu/abs/2019MNRAS.488.1822N} {488, 1822}

\bibitem[\protect\citeauthoryear{{Nesvorn{\'y}} \&
  {Vokrouhlick{\'y}}}{{Nesvorn{\'y}} \& {Vokrouhlick{\'y}}}{2016}]{Nesvorny16}
{Nesvorn{\'y}} D.,  {Vokrouhlick{\'y}} D.,  2016, \mn@doi [\apj]
  {10.3847/0004-637X/825/2/94}, \href
  {https://ui.adsabs.harvard.edu/abs/2016ApJ...825...94N} {825, 94}

\bibitem[\protect\citeauthoryear{{Nesvorn{\'y}}, {Bottke}, {Levison}  \&
  {Dones}}{{Nesvorn{\'y}} et~al.}{2003}]{Nesvorny03}
{Nesvorn{\'y}} D.,  {Bottke} W.~F.,  {Levison} H.~F.,   {Dones} L.,  2003,
  \mn@doi [\apj] {10.1086/374807}, \href
  {https://ui.adsabs.harvard.edu/abs/2003ApJ...591..486N} {591, 486}

\bibitem[\protect\citeauthoryear{{Nesvorn{\'y}} et~al.,}{{Nesvorn{\'y}}
  et~al.}{2006}]{Nesvorny06}
{Nesvorn{\'y}} D.,  et~al., 2006, \mn@doi [\aj] {10.1086/505392}, \href
  {https://ui.adsabs.harvard.edu/abs/2006AJ....132..582N} {132, 582}

\bibitem[\protect\citeauthoryear{{Nesvorn{\'y}}, {Bottke}, {Vokrouhlick{\'y}},
  {Sykes}, {Lien}  \& {Stansberry}}{{Nesvorn{\'y}} et~al.}{2008}]{Nesvorny08}
{Nesvorn{\'y}} D.,  {Bottke} W.~F.,  {Vokrouhlick{\'y}} D.,  {Sykes} M.,
  {Lien} D.~J.,   {Stansberry} J.,  2008, \mn@doi [\apjl] {10.1086/588841},
  \href {https://ui.adsabs.harvard.edu/abs/2008ApJ...679L.143N} {679, L143}

\bibitem[\protect\citeauthoryear{{Nesvorn{\'y}}, {Jenniskens}, {Levison},
  {Bottke}, {Vokrouhlick{\'y}}  \& {Gounelle}}{{Nesvorn{\'y}}
  et~al.}{2010}]{Nesvorny10}
{Nesvorn{\'y}} D.,  {Jenniskens} P.,  {Levison} H.~F.,  {Bottke} W.~F.,
  {Vokrouhlick{\'y}} D.,   {Gounelle} M.,  2010, \mn@doi [\apj]
  {10.1088/0004-637X/713/2/816}, \href
  {https://ui.adsabs.harvard.edu/abs/2010ApJ...713..816N} {713, 816}

\bibitem[\protect\citeauthoryear{{Nesvorn{\'y}}, {Janches}, {Vokrouhlick{\'y}},
  {Pokorn{\'y}}, {Bottke}  \& {Jenniskens}}{{Nesvorn{\'y}}
  et~al.}{2011}]{Nesvorny11_ZC}
{Nesvorn{\'y}} D.,  {Janches} D.,  {Vokrouhlick{\'y}} D.,  {Pokorn{\'y}} P.,
  {Bottke} W.~F.,   {Jenniskens} P.,  2011, \mn@doi [\apj]
  {10.1088/0004-637X/743/2/129}, \href
  {https://ui.adsabs.harvard.edu/abs/2011ApJ...743..129N} {743, 129}

\bibitem[\protect\citeauthoryear{{Nesvorn{\'y}}, {Vokrouhlick{\'y}}, {Dones},
  {Levison}, {Kaib}  \& {Morbidelli}}{{Nesvorn{\'y}} et~al.}{2017}]{Nesvorny17}
{Nesvorn{\'y}} D.,  {Vokrouhlick{\'y}} D.,  {Dones} L.,  {Levison} H.~F.,
  {Kaib} N.,   {Morbidelli} A.,  2017, \mn@doi [\apj]
  {10.3847/1538-4357/aa7cf6}, \href
  {https://ui.adsabs.harvard.edu/abs/2017ApJ...845...27N} {845, 27}

\bibitem[\protect\citeauthoryear{{Pokorn{\'y}}, {Vokrouhlick{\'y}},
  {Nesvorn{\'y}}, {Campbell-Brown}  \& {Brown}}{{Pokorn{\'y}}
  et~al.}{2014}]{Pokorny14}
{Pokorn{\'y}} P.,  {Vokrouhlick{\'y}} D.,  {Nesvorn{\'y}} D.,  {Campbell-Brown}
  M.,   {Brown} P.,  2014, \mn@doi [\apj] {10.1088/0004-637X/789/1/25}, \href
  {https://ui.adsabs.harvard.edu/abs/2014ApJ...789...25P} {789, 25}

\bibitem[\protect\citeauthoryear{{Poppe} et~al.,}{{Poppe}
  et~al.}{2019}]{Poppe19}
{Poppe} A.~R.,  et~al., 2019, \mn@doi [\apjl] {10.3847/2041-8213/ab322a}, \href
  {https://ui.adsabs.harvard.edu/abs/2019ApJ...881L..12P} {881, L12}

\bibitem[\protect\citeauthoryear{{Reach}, {Kelley}  \& {Sykes}}{{Reach}
  et~al.}{2007}]{Reach07}
{Reach} W.~T.,  {Kelley} M.~S.,   {Sykes} M.~V.,  2007, \mn@doi [\icarus]
  {10.1016/j.icarus.2007.03.031}, \href
  {https://ui.adsabs.harvard.edu/abs/2007Icar..191..298R} {191, 298}

\bibitem[\protect\citeauthoryear{{Rigley} \& {Wyatt}}{{Rigley} \&
  {Wyatt}}{2020}]{Rigley20}
{Rigley} J.~K.,  {Wyatt} M.~C.,  2020, \mn@doi [\mnras]
  {10.1093/mnras/staa2029}, \href
  {https://ui.adsabs.harvard.edu/abs/2020MNRAS.497.1143R} {497, 1143}

\bibitem[\protect\citeauthoryear{{Rotundi} et~al.,}{{Rotundi}
  et~al.}{2015}]{Rotundi15}
{Rotundi} A.,  et~al., 2015, \mn@doi [Science] {10.1126/science.aaa3905}, \href
  {https://ui.adsabs.harvard.edu/abs/2015Sci...347a3905R} {347, aaa3905}

\bibitem[\protect\citeauthoryear{{Rowan-Robinson} \& {May}}{{Rowan-Robinson} \&
  {May}}{2013}]{Rowan-Robinson13}
{Rowan-Robinson} M.,  {May} B.,  2013, \mn@doi [\mnras] {10.1093/mnras/sts471},
  \href {https://ui.adsabs.harvard.edu/abs/2013MNRAS.429.2894R} {429, 2894}

\bibitem[\protect\citeauthoryear{{Samarasinha}}{{Samarasinha}}{2007}]{Samarasinha07}
{Samarasinha} N.~H.,  2007, \mn@doi [Advances in Space Research]
  {10.1016/j.asr.2004.07.016}, \href
  {https://ui.adsabs.harvard.edu/abs/2007AdSpR..39..421S} {39, 421}

\bibitem[\protect\citeauthoryear{{Sekanina}}{{Sekanina}}{1982}]{Sekanina82}
{Sekanina} Z.,  1982, in {Wilkening} L.~L.,  ed., IAU Colloq. 61: Comet
  Discoveries, Statistics, and Observational Selection. pp 251--287

\bibitem[\protect\citeauthoryear{{Sekanina}}{{Sekanina}}{1997}]{Sekanina97}
{Sekanina} Z.,  1997, \aap, \href
  {https://ui.adsabs.harvard.edu/abs/1997A&A...318L...5S} {318, L5}

\bibitem[\protect\citeauthoryear{{Sekanina}}{{Sekanina}}{1999}]{Sekanina99}
{Sekanina} Z.,  1999, \aap, \href
  {https://ui.adsabs.harvard.edu/abs/1999A&A...342..285S} {342, 285}

\bibitem[\protect\citeauthoryear{{Sekanina}}{{Sekanina}}{2007}]{Sekanina07}
{Sekanina} Z.,  2007, in {Valsecchi} G.~B.,  {Vokrouhlick{\'y}} D.,   {Milani}
  A.,  eds,  Vol. 236, Near Earth Objects, our Celestial Neighbors: Opportunity
  and Risk. pp 211--220, \mn@doi{10.1017/S1743921307003249}

\bibitem[\protect\citeauthoryear{{Sekanina}}{{Sekanina}}{2021}]{Sekanina21}
{Sekanina} Z.,  2021, arXiv e-prints, \href
  {https://ui.adsabs.harvard.edu/abs/2021arXiv210901297S} {p. arXiv:2109.01297}

\bibitem[\protect\citeauthoryear{{Sekanina} \& {Chodas}}{{Sekanina} \&
  {Chodas}}{2002}]{SekaninaChodas02}
{Sekanina} Z.,  {Chodas} P.~W.,  2002, \mn@doi [\apj] {10.1086/344261}, \href
  {https://ui.adsabs.harvard.edu/abs/2002ApJ...581.1389S} {581, 1389}

\bibitem[\protect\citeauthoryear{{Sezestre}, {Augereau}  \&
  {Th{\'e}bault}}{{Sezestre} et~al.}{2019}]{Sezestre19}
{Sezestre} {\'E}.,  {Augereau} J.~C.,   {Th{\'e}bault} P.,  2019, \mn@doi
  [\aap] {10.1051/0004-6361/201935250}, \href
  {https://ui.adsabs.harvard.edu/abs/2019A&A...626A...2S} {626, A2}

\bibitem[\protect\citeauthoryear{{Shober} et~al.,}{{Shober}
  et~al.}{2021}]{Shober21}
{Shober} P.~M.,  et~al., 2021, Planet. Sci. J., \href
  {https://ui.adsabs.harvard.edu/abs/2021PSJ.....2...98S} {2, 98}

\bibitem[\protect\citeauthoryear{{Snodgrass}, {Fitzsimmons}, {Lowry}  \&
  {Weissman}}{{Snodgrass} et~al.}{2011}]{Snodgrass11}
{Snodgrass} C.,  {Fitzsimmons} A.,  {Lowry} S.~C.,   {Weissman} P.,  2011,
  \mn@doi [\mnras] {10.1111/j.1365-2966.2011.18406.x}, \href
  {https://ui.adsabs.harvard.edu/abs/2011MNRAS.414..458S} {414, 458}

\bibitem[\protect\citeauthoryear{{Soja} et~al.,}{{Soja} et~al.}{2015}]{Soja15}
{Soja} R.~H.,  et~al., 2015, \mn@doi [\aap] {10.1051/0004-6361/201526184},
  \href {https://ui.adsabs.harvard.edu/abs/2015A&A...583A..18S} {583, A18}

\bibitem[\protect\citeauthoryear{{Soja} et~al.,}{{Soja} et~al.}{2019}]{Soja19}
{Soja} R.~H.,  et~al., 2019, \mn@doi [\aap] {10.1051/0004-6361/201834892},
  \href {https://ui.adsabs.harvard.edu/abs/2019A&A...628A.109S} {628, A109}

\bibitem[\protect\citeauthoryear{{Sykes}}{{Sykes}}{1988}]{Sykes88}
{Sykes} M.~V.,  1988, \mn@doi [\apjl] {10.1086/185311}, 334, L55

\bibitem[\protect\citeauthoryear{{Sykes} \& {Walker}}{{Sykes} \&
  {Walker}}{1992}]{Sykes92}
{Sykes} M.~V.,  {Walker} R.~G.,  1992, \mn@doi [\icarus]
  {10.1016/0019-1035(92)90037-8}, \href
  {https://ui.adsabs.harvard.edu/abs/1992Icar...95..180S} {95, 180}

\bibitem[\protect\citeauthoryear{{Sykes}, {Lebofsky}, {Hunten}  \&
  {Low}}{{Sykes} et~al.}{1986}]{Sykes86}
{Sykes} M.~V.,  {Lebofsky} L.~A.,  {Hunten} D.~M.,   {Low} F.,  1986, \mn@doi
  [Science] {10.1126/science.232.4754.1115}, \href
  {https://ui.adsabs.harvard.edu/abs/1986Sci...232.1115S} {232, 1115}

\bibitem[\protect\citeauthoryear{{Tancredi}, {Fern{\'a}ndez}, {Rickman}  \&
  {Licandro}}{{Tancredi} et~al.}{2000}]{Tancredi00}
{Tancredi} G.,  {Fern{\'a}ndez} J.~A.,  {Rickman} H.,   {Licandro} J.,  2000,
  \mn@doi [\aaps] {10.1051/aas:2000263}, \href
  {https://ui.adsabs.harvard.edu/abs/2000A&AS..146...73T} {146, 73}

\bibitem[\protect\citeauthoryear{{Tancredi}, {Fern{\'a}ndez}, {Rickman}  \&
  {Licandro}}{{Tancredi} et~al.}{2006}]{Tancredi06}
{Tancredi} G.,  {Fern{\'a}ndez} J.~A.,  {Rickman} H.,   {Licandro} J.,  2006,
  \mn@doi [\icarus] {10.1016/j.icarus.2006.01.007}, \href
  {https://ui.adsabs.harvard.edu/abs/2006Icar..182..527T} {182, 527}

\bibitem[\protect\citeauthoryear{{Tsiganis}, {Gomes}, {Morbidelli}  \&
  {Levison}}{{Tsiganis} et~al.}{2005}]{Tsiganis05}
{Tsiganis} K.,  {Gomes} R.,  {Morbidelli} A.,   {Levison} H.~F.,  2005, \mn@doi
  [\nat] {10.1038/nature03539}, \href
  {https://ui.adsabs.harvard.edu/abs/2005Natur.435..459T} {435, 459}

\bibitem[\protect\citeauthoryear{{Ueda}, {Kobayashi}, {Takeuchi}, {Ishihara},
  {Kondo}  \& {Kaneda}}{{Ueda} et~al.}{2017}]{Ueda17}
{Ueda} T.,  {Kobayashi} H.,  {Takeuchi} T.,  {Ishihara} D.,  {Kondo} T.,
  {Kaneda} H.,  2017, \mn@doi [\aj] {10.3847/1538-3881/aa5ff3}, \href
  {https://ui.adsabs.harvard.edu/abs/2017AJ....153..232U} {153, 232}

\bibitem[\protect\citeauthoryear{{Weissman} \& {Lowry}}{{Weissman} \&
  {Lowry}}{2003}]{Weissman03}
{Weissman} P.~R.,  {Lowry} S.~C.,  2003, in {Mackwell} S.,  {Stansbery} E.,
  eds,  Lunar and Planetary Science Conference Vol. 34, Lunar and Planetary
  Science Conference. p.~2003

\bibitem[\protect\citeauthoryear{{Weissman} \& {Lowry}}{{Weissman} \&
  {Lowry}}{2008}]{Weissman08}
{Weissman} P.~R.,  {Lowry} S.~C.,  2008, \mn@doi [Meteoritics and Planetary
  Science] {10.1111/j.1945-5100.2008.tb00691.x}, \href
  {https://ui.adsabs.harvard.edu/abs/2008M&PS...43.1033W} {43, 1033}

\bibitem[\protect\citeauthoryear{{Wiegert}, {Vaubaillon}  \&
  {Campbell-Brown}}{{Wiegert} et~al.}{2009}]{Wiegert09}
{Wiegert} P.,  {Vaubaillon} J.,   {Campbell-Brown} M.,  2009, \mn@doi [\icarus]
  {10.1016/j.icarus.2008.12.030}, \href
  {https://ui.adsabs.harvard.edu/abs/2009Icar..201..295W} {201, 295}

\bibitem[\protect\citeauthoryear{{Wyatt}}{{Wyatt}}{2005}]{Wyatt05}
{Wyatt} M.~C.,  2005, \mn@doi [\aap] {10.1051/0004-6361:20042073}, \href
  {https://ui.adsabs.harvard.edu/abs/2005A&A...433.1007W} {433, 1007}

\bibitem[\protect\citeauthoryear{{Wyatt} \& {Dent}}{{Wyatt} \&
  {Dent}}{2002}]{Wyatt02}
{Wyatt} M.~C.,  {Dent} W.~R.~F.,  2002, \mn@doi [\mnras]
  {10.1046/j.1365-8711.2002.05533.x}, \href
  {https://ui.adsabs.harvard.edu/abs/2002MNRAS.334..589W} {334, 589}

\bibitem[\protect\citeauthoryear{{Wyatt} \& {Whipple}}{{Wyatt} \&
  {Whipple}}{1950}]{Wyatt50}
{Wyatt} S.~P.,  {Whipple} F.~L.,  1950, \mn@doi [\apj] {10.1086/145244}, \href
  {https://ui.adsabs.harvard.edu/abs/1950ApJ...111..134W} {111, 134}

\bibitem[\protect\citeauthoryear{{Wyatt}, {Booth}, {Payne}  \&
  {Churcher}}{{Wyatt} et~al.}{2010}]{Wyatt10}
{Wyatt} M.~C.,  {Booth} M.,  {Payne} M.~J.,   {Churcher} L.~J.,  2010, \mn@doi
  [\mnras] {10.1111/j.1365-2966.2009.15930.x}, \href
  {https://ui.adsabs.harvard.edu/abs/2010MNRAS.402..657W} {402, 657}

\bibitem[\protect\citeauthoryear{{Yang} \& {Ishiguro}}{{Yang} \&
  {Ishiguro}}{2015}]{Yang15}
{Yang} H.,  {Ishiguro} M.,  2015, \mn@doi [\apj] {10.1088/0004-637X/813/2/87},
  \href {https://ui.adsabs.harvard.edu/abs/2015ApJ...813...87Y} {813, 87}

\bibitem[\protect\citeauthoryear{{Yoshida} \& {Terai}}{{Yoshida} \&
  {Terai}}{2017}]{Yoshida17}
{Yoshida} F.,  {Terai} T.,  2017, \mn@doi [\aj] {10.3847/1538-3881/aa7d03},
  \href {https://ui.adsabs.harvard.edu/abs/2017AJ....154...71Y} {154, 71}

\bibitem[\protect\citeauthoryear{{van Lieshout}, {Dominik}, {Kama}  \&
  {Min}}{{van Lieshout} et~al.}{2014}]{vLieshout14}
{van Lieshout} R.,  {Dominik} C.,  {Kama} M.,   {Min} M.,  2014, \mn@doi [\aap]
  {10.1051/0004-6361/201322090}, \href
  {https://ui.adsabs.harvard.edu/abs/2014A&A...571A..51V} {571, A51}

\makeatother
\end{thebibliography}




%
%


\bsp	
\label{lastpage}
\end{document}